\documentclass[aps,twocolumn,superscriptaddress,fleqn]{revtex4-1} 

\usepackage{graphicx}
\usepackage{epsfig}

\usepackage{amsmath}
\usepackage{amssymb}
\graphicspath{{figures/}}
\usepackage{float}
\usepackage[colorlinks, linkcolor=red,anchorcolor=green,citecolor=blue]{hyperref}

\begin{document}

\title{Heavy quark potential in quark-gluon Plasma: Deep neural network meets lattice quantum chromodynamics}

\date{\today}

\affiliation{Department of Physics, McGill University, Montreal, Quebec H3A 2T8, Canada.}
\affiliation{Frankfurt Institute for Advanced Studies, Ruth Moufang Strasse 1, D-60438, Frankfurt am Main, Germany.}
\affiliation{Department of Physics, Tsinghua University, Beijing 100084, China.}
\affiliation{Physics Department, Brookhaven National Laboratory, Upton, New York 11973, USA.}

\author{Shuzhe Shi}
\email{shuzhe.shi@mcgill.ca}
\affiliation{Department of Physics, McGill University, Montreal, Quebec H3A 2T8, Canada.}

\author{Kai Zhou}
\email{zhou@fias.uni-frankfurt.de}
\affiliation{Frankfurt Institute for Advanced Studies, Ruth Moufang Strasse 1, D-60438, Frankfurt am Main, Germany.}

\author{Jiaxing Zhao}
\affiliation{Department of Physics, Tsinghua University, Beijing 100084, China.}

\author{Swagato Mukherjee}
\affiliation{Physics Department, Brookhaven National Laboratory, Upton, New York 11973, USA.}

\author{Pengfei Zhuang}
\affiliation{Department of Physics, Tsinghua University, Beijing 100084, China.}

\begin{abstract}
Bottomonium states are key probes for experimental studies of the quark-gluon plasma (QGP) created in high-energy nuclear collisions. Theoretical models of bottomonium productions in high-energy nuclear collisions rely on the in-medium interactions between the bottom and antibottom quarks. The latter can be characterized by the temperature ($T$) dependent potential, with real ($V_R(T,r)$) and imaginary ($V_I(T,r)$) parts, as a function of the spatial separation ($r$). Recently, the masses and thermal widths of up to $3S$ and $2P$ bottomonium states in QGP were calculated using lattice quantum chromodynamics (LQCD). Starting from these LQCD results and through a novel application of deep neural network, here, we obtain $V_R(T,r)$ and $V_I(T,r)$ in a model-independent fashion. The temperature dependence of $V_R(T,r)$ was found to be very mild between $T\approx0-334$~MeV. For $T=151-334$~MeV, $V_I(T,r)$ shows a rapid increase with $T$ and $r$, which is much larger than the perturbation-theory-based expectations.
\end{abstract}
\maketitle
%
%
\section{Introduction} 
In-medium modifications of 
quarkonium states, i.e. bound states of a heavy charm or bottom quark and its antiquark, are sensitive probes of the quark-gluon plasma (QGP) produced in high-energy nuclear collisions~\cite{Matsui:1986dk,Karsch:1987pv,Blaizot:1996nq,BraunMunzinger:2000px,Digal:2001ue,Grandchamp:2003uw,Song:2011xi,Du:2015wha,Liu:2010ej,Zhou:2014kka,Katz:2015qja,Yao:2020xzw,Islam:2020gdv}. Sequential suppression patterns among the $\Upsilon(1S)$,  $\Upsilon(2S)$, and $\Upsilon(3S)$ states have been observed in heavy ion collision experiments~\cite{Chatrchyan:2011pe,Chatrchyan:2012lxa,Khachatryan:2016xxp,Sirunyan:2017lzi}. Theoretical understanding of these experimental observations relies on effective field theories (EFTs), which naturally lead to an open-quantum-system-based treatment of both open and hidden bottom states in QGP (for a recent review, see \cite{Yao:2020xzw}). Owing to the large mass ($m_b$) and small relative velocity ($v$) of the bottom quark, there exists a hierarchy of scales at high temperature: $m_b\gg m_bv\gg m_bv^2$. Sequentially integrating out the scales larger than $m_b$ and $m_bv$ from the QCD Lagrangian, one, respectively, arrives at the nonrelativistic QCD(NRQCD)~\cite{Caswell:1985ui} and potential nonrelativistic QCD(pNRQCD)~\cite{Brambilla:1999xf} EFTs. If interactions between the color-singlet and color-octet states are neglected then the pNRQCD reduces to a theoretical description of quarkonia solely based on a potential between the heavy quark and antiquark. A potential based description allows studies of quarkonia by employing Schr\"odinger-type equations~\cite{Satz:2005hx,Zhao:2020jqu,Crater:2008rt,Guo:2012hx}. One-loop hard thermal loop (HTL) perturbative QCD calculations~\cite{Laine:2006ns,Beraudo:2007ky}, and later on pNRQCD calculations~\cite{Brambilla:2008cx,Brambilla:2010vq}, show that at finite temperatures heavy quark potential becomes complex with a nonvanishing imaginary part. However, it is difficult to provide satisfactory descriptions of bound states arising out of strong interactions solely using perturbative expansions and a nonperturbative treatment, such as the lattice quantum chromodynamics (LQCD), is called for.

A bound state of strong interaction is a nonperturbative problem, which is difficult to be completely or relevantly treated in a conventional perturbation theory. Therefore, a critical step in all EFT based studies is to relate parameters of the EFT to the underlying fundamental theory, i.e. a model-independent determination of the heavy quark potential starting from nonperturbative QCD. In the static limit, the heavy quark potential can be extracted from the spectral functions of the thermal Wilson loop using nonperturbative LQCD calculations~\cite{Rothkopf:2011db,Burnier:2014ssa,Burnier:2015tda,Bala:2019cqu}. On the other hand, recent LQCD studies have led to quantification of the masses, thermal widths, and Bethe--Salpeter amplitudes (BSA) of up to $3S$ and $2P$ bottomonium states in QGP~\cite{Larsen:2019bwy,Larsen:2019zqv,Larsen:2020rjk}. While the lattice QCD results of Refs.~\cite{Rothkopf:2011db,Burnier:2014ssa,Burnier:2015tda,Bala:2019cqu} were obtained based on thermal correlation functions that provide a  well-defined static quark potential in some specific limits, \textit{a priori}, there is no obvious reason to expect that the LQCD results~\cite{Larsen:2019bwy,Larsen:2019zqv,Larsen:2020rjk} on the properties of in-medium bottomonia admit any consistent in-medium potential-based description. Even the fact that the vacuum bottomonia masses below the threshold can be reasonably well-described by Cornell-type potential is an empirical observation, and cannot be rigorously proven from first-principle QCD. In this work we will empirically investigate whether the LQCD results of Refs.~\cite{Larsen:2019bwy,Larsen:2019zqv,Larsen:2020rjk} can be consistently described by an in-medium potential, $V_R(T,r)$ and $V_I(T,r)$. Furthermore, even if it turns out that LQCD results of Refs.~\cite{Larsen:2019bwy,Larsen:2019zqv,Larsen:2020rjk} can be consistently described using some $V_R(T,r)$ and $V_I(T,r)$, there is no theoretical reason for these to agree with the static quark potential obtained from correlation functions of thermal Wilson loops~\cite{Rothkopf:2011db,Burnier:2014ssa,Burnier:2015tda,Bala:2019cqu}. Comparisons among these different in-medium potentials is an interesting study by itself.

As we shall see later, one-loop HTL-motivated functional forms of $V_R(T,r)$ and $V_I(T,r)$ are not compatible with these LQCD results. This observation calls for a model-independent treatment
of the in-medium heavy quark potential. 
In this work, we introduce a model-independent deep-neural-network-based (DNN-based) method and determine the $r$ and $T$-dependence of the in-medium heavy quark potential starting from the LQCD results~\cite{Larsen:2019zqv} for the masses and thermal widths of up to $3S$ and $2P$ bottomonium states at various temperatures. 
The underlying idea is as follows: At a fixed $T$, various bottomonium states differ in sizes and their wavefunctions probe different distances. Knowledge of the masses and thermal widths of multiple bottomonium states, thereby, provide constraints on not only the strength of the real and imaginary parts of the bottom-antibottom interactions in QGP but also its $r$-dependence. Thus, LQCD results for the masses and thermal widths of multiple bottomonium states at different temperatures can be used to extract $V_R(T,r)$ and $V_I(T,r)$ and, presently, DNN is probably the best tool to achieve this in an unbiased fashion. To this goal, we develop a new method to optimize the Deep Neural Network coupled with the Schr\"odinger equation. 

This manuscript is organized as follows.
We discuss the details of vacuum potential determination in Sec.~\ref{sec.schroedinger}, followed by
a detailed description of the potential extraction method in Sec.~\ref{sec.methodology}.
Then, we show our results of complex-valued heavy flavor potential using DNNs that depend on both distance and temperature in Sec.~\ref{sec.potential_dnn}. Finally we perform two consistency tests: (i) compare the DNN potential to the potentials using two other parameterization schemes --- temperature-independent DNNs and polynomials (Sec.~\ref{sec.consistency_potential}); (ii) compare the eigenstate wave-functions with the corresponding lattice QCD results~\cite{Larsen:2020rjk} of Bethe--Salpeter amplitude (Sec.~\ref{sec.consistency_wavfunc}).
After the Summary in Sec.~\ref{sec.conclusion}, we provide supplemental materials (App.~\ref{sec.spectral_function}) to discuss the connection between the imaginary energy and the width extracted in lattice QCD.

\section{Schr\"odinger Equation and Vacuum Potential}\label{sec.schroedinger}
\begin{figure}[!hbtp]\centering
\includegraphics[width=0.35\textwidth]{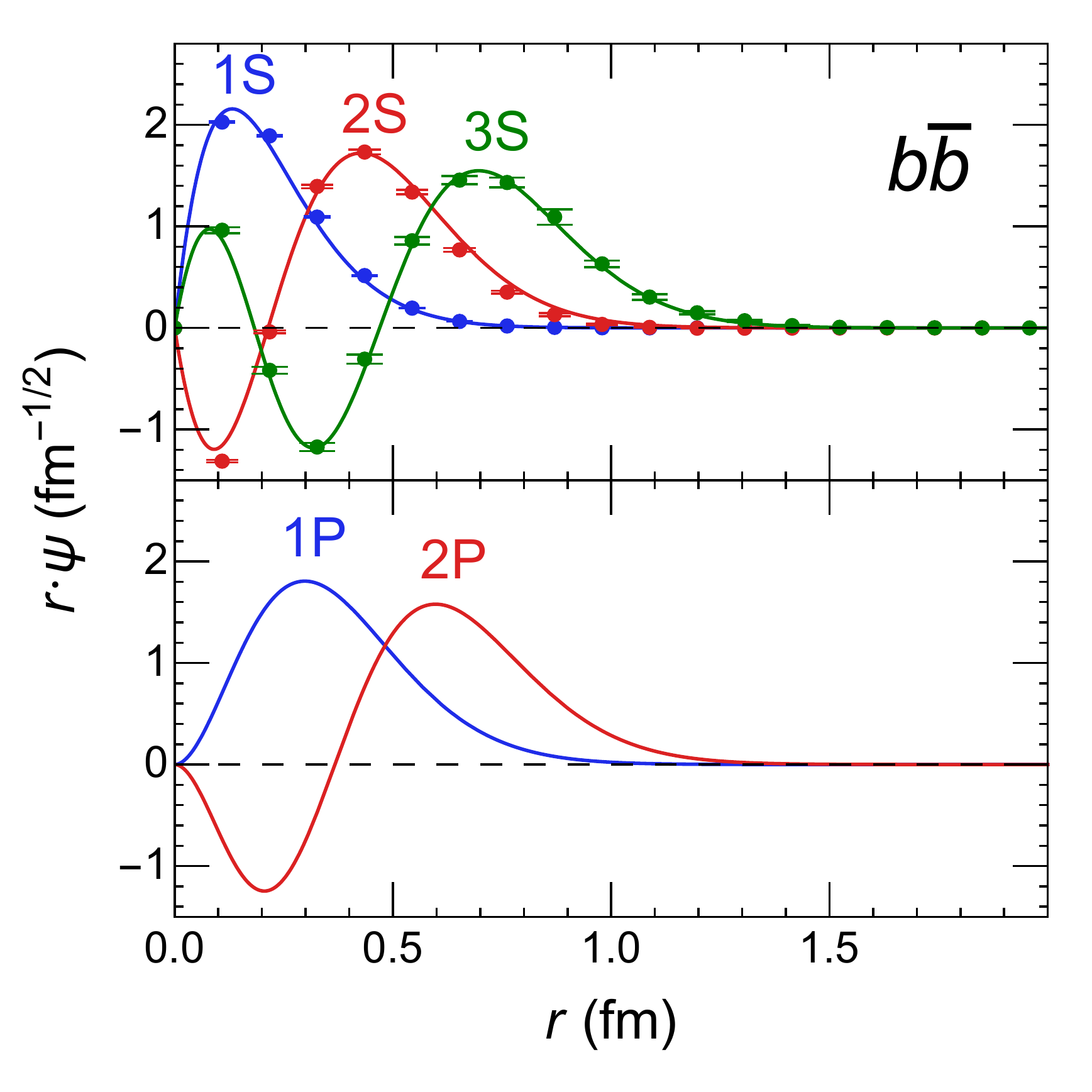}
\caption{Bottomonia wave-functions in vacuum.
Symbols in the upper panel are the Bethe--Salpeter amplitudes from lattice QCD calculation~\cite{Larsen:2020rjk}.
\label{fig.vacuum_wf}}
\end{figure}

Bottomonium states can be described well by the reduced two-body time-independent Schr\"odinger equation, \footnote{It might be worth noting that the potential here is the effective potential for a Schr{\"o}dinger equation.}
\begin{equation}
 -\frac{\nabla^2}{m_b} \psi_n + \left[ V_R(T,r) + i \; V_I(T,r) \right] \psi_n = E_n \psi_n \,.
 \label{eq:SE}
\end{equation}
Here, the heavy quark potential $V(T,r) = V_R(T,r) + i \; V_I(T,r)$ is complex-valued. Accordingly, the wavefunction $\psi_n$ and the energy eigenvalues $E_n$ for in-medium bottomonia are also complex-valued. Further, $V_I(T=0,r)=0$, $\mathrm{Re}[E_n] = m_n-2m_b$ and $\mathrm{Im}[E_n] = -\Gamma_n$, where $m_n$ and $\Gamma_n$ are the mass and thermal width of the $n^\mathrm{th}$ bottomonium state, respectively. (see \protect{App.~\ref{sec.spectral_function}} for detailed discussions.)
Vacuum properties of up to $3S$ and $2P$ bottomonium states~\cite{Tanabashi:2018oca} were found to be reproduced well by the Cornell potential 
\begin{equation}
V_R(T=0,r) = -\frac{\alpha}{r} + \sigma \, r\, + B,
\end{equation}
with $m_b = 6.00~\text{GeV}$, strong coupling $\alpha=0.406$, string tension $\sigma=0.221~\text{GeV}^2$, bag constant $B=-2.53~\text{GeV}$. These parameters are determined by fitting the vacuum masses of the bottomonium states reported in the Particle Data Group booklet~\cite{Tanabashi:2018oca} and the wave-function, to be compared with the BSA from lattice QCD calculations~\cite{Larsen:2020rjk}. The latter is computed in a consistent way as to evaluate the mass and width~\cite{Larsen:2019zqv}. Due to its $r$-dependence, the BS amplitudes are more decisive, compared to the mass spectrum, in determining the interaction potential.
Considering that the lattice results are for spin-averaged states --- e.g. there is no distinction between $\eta_b$(nS) and $\Upsilon$(nS) --- 
we fit the parameters in potential model by fitting the spin-averaged Bottomonia mass spectrum
\begin{equation}
m_\mathrm{averaged} \equiv \frac{1}{4}m_\mathrm{singlet} + \frac{3}{4}m_\mathrm{triplet} \,,
\label{eq.avg_mass}
\end{equation}
as well as the BS amplitude for 1S, 2S, and 3S states. The global parameter tuning technique employed here will be described later in Sec.~\ref{sec.parameter_tuning_wf}.
We present the bottomonium wave-functions in Fig.~\ref{fig.vacuum_wf}, and correspondingly the mass in Table~\ref{tab.vacuum_mass}. One can see the potential model agrees very well with the experimental results (maximum absolute difference being 14~MeV), as well as with the lattice results on the wave-functions (BSAs).

\begin{table}[!hbtp]\centering
\begin{tabular}{c|ccccc}
\hline\hline
	& 1S	& 2S	& 3S	& 1P	& 2P \\
\hline
experiment (MeV) &
	9445 & 10017 & 10352 & 9891 & 10254 \\
\hline
model (MeV) & 
	9449 & 10003 & 10356 & 9893 & 10258 \\
\hline
difference (MeV) & 
	$+4$ & $-14$ & $+4$ & $+2$ & $+4$ \\
\hline \hline
\end{tabular}
\caption{Best fit of spin-averaged bottomonium mass spectrum~\protect{\eqref{eq.avg_mass}}.
\label{tab.vacuum_mass}}
\end{table}

\begin{figure*}[!hbtp]
\includegraphics[width=0.8\textwidth]{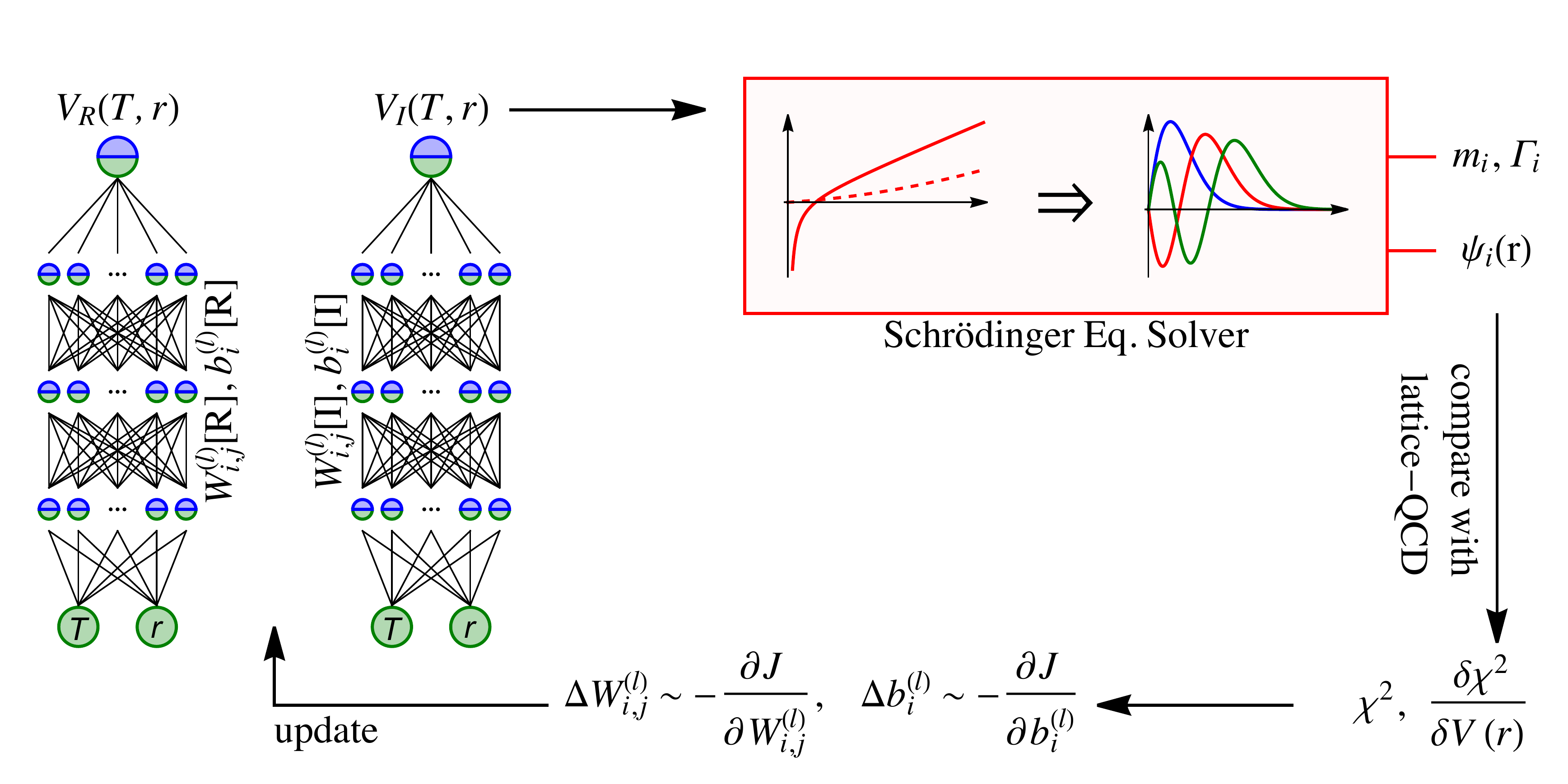}
\caption{Flow chart of the potential reconstruction scheme --- using generalized back-propagation to optimize parameters in the deep neural networks coupled with a Schr\"odinger equation.
\label{fig.flow_chart}}
\end{figure*}
\section{Methodology: How a DNN Learns the Potential from the Spectrum}\label{sec.methodology}
We will move on to discuss our model-independent methodology of the potential reconstruction with DNNs coupled to a Schr\"odinger equation~\eqref{eq:SE}, which is summarized in the flow chart Fig.~\ref{fig.flow_chart}.
We represent the real and imaginary potentials by DNN --- a multi-step iterative function composition scheme to approximate mapping between two functions in a smooth and unbiased manner~\cite{Leshno93multilayerfeedforward, Kratsios_2021}. 
We optimize the network parameters by minimizing the chi-square function, i.e. uncertainty-weighted distance,
\begin{equation}
\chi^2  = \sum_{T,n} \frac{(m_{T,n}-m_{T,n}^\text{LQCD})^2}{(\delta m_{T,n}^\text{LQCD})^2} + \frac{(\Gamma_{T,n}-\Gamma_{T,n}^\text{LQCD})^2}{(\delta \Gamma_{T,n}^\text{LQCD})^2}\,.
\label{eq.chi_square}
\end{equation}
The summation runs over six temperature points, $T\in\{0, 151, 173, 199, 251, 334\}~\text{MeV}$, and five bottomonium states, $n \in \{\text{1S}, \text{2S}, \text{3S}, \text{1P}, \text{2P}\}$ and the LQCD values were taken from Ref.~\cite{Larsen:2019zqv}. We used gradient descent with Back-Propagation optimization technique, which is based on the derivatives of the cost function with respect to the network parameters. This technique requires knowledge of the explicit functional relationship between the cost function and the DNN output. However, such a relationship in our problem is implicit. We overcame this challenge of gradients evaluation through perturbative solution of the Schr\"odinger equation with respect to small change of $V(T,r)$. Moreover, we invoked Bayesian inference for uncertainty quantification, whereby the posterior distribution of the network parameters was evaluated. To the best of our knowledge, the current method is developed for the first time here. In the rest of this section, we provide all the details about this method, including a general introduction to DNN (Sec.~\ref{sec.methodology_general}), parameter optimization algorithm (Sec.~\ref{sec.methodology_optimization} and~\ref{sec.parameter_tuning_wf}), uncertainty estimation using Bayesian Inference (Sec.~\ref{sec.methodology_bayesian}), and a closure test to justify our methodology and assess its reliability (Sec.~\ref{sec.methodology_closure}).

\subsection{General Introduction of Deep Neural Network}\label{sec.methodology_general}
According to the universal approximation theorem~\cite{Leshno93multilayerfeedforward, Kratsios_2021}, DNN can generally provide an unbiased, yet flexible enough, parameterization to approximate arbitrary functional relations.
Algorithms based on DNNs have been deployed to address various physics problems, e.g., determining the parton distribution function~\cite{Forte_2002, Collaboration_2007}, reconstructing the spectral function~\cite{Kades:2019wtd, Zhou:2021bvw,Chen:2021giw}, identifying phase transition~\cite{Carrasquilla_2017,Pang:2016vdc, Wang:2020tgb, Jiang:2021gsw}, assisting lattice field theory calculation~\cite{PhysRevD.100.011501, Boyda:2020hsi, Kanwar:2020xzo, Albergo:2019eim}, evaluating centrality for heavy ion collisions~\cite{OmanaKuttan:2020brq, Thaprasop:2020mzp, Li:2020qqn}, estimating parameter under detector effects~\cite{Andreassen:2020gtw, Kuttan:2021npg}, and speeding up hydrodynamics simulation~\cite{Huang:2018fzn}.

To express, approximately, an $\mathbb{R}_n \to \mathbb{R}_m$ function mapping between independent variables ${\bf x} = \{x_1, \cdots, x_n\}$ and dependent variables ${\bf y}=\{y_1,\cdots,y_m\}$, ${\bf y} = {\bf y}({\bf x})$, the DNN constructs functional form by composing iteratively $N$ simple building blocks (also called \textit{layer} representing a vector-to-vector function). Each layer performs a linear transformation on the output from the preceding layer, followed by an element-wise non-linear transformation dictated by the \textit{activation function} $\sigma^{(l)}(z)$:
\begin{equation}
a^{(l)}_i = \sigma^{(l)}(z^{(l)}_{i}),\qquad z^{(l)}_{i} \equiv {b}^{(l)}_{i} + \sum_{j} {W}^{(l)}_{ij} {a}^{(l-1)}_{j}\,, \label{eq.dnn.iteration}
\end{equation}
for $i=1,\cdots,n^{(l)}$ and $l=1,\cdots,N$, while $a_i^{(0)} \equiv {x_i}$ stands for the input variables.
The final layer gives the model output, which composes all the successive layers and defines the approximate function formula, $\widetilde{\bf y}({\bf{x};\{W_{ij}^{(l)}},b_i^{(l)}\}) = {\bf a}^{(N)}$.
This compositional way of parameterization renders DNN an universal function approximator to be able to fit any continuous function to arbitrary accuracy given enough hidden units. 

\begin{table}
\begin{tabular}{l|l|l}
\hline
\hline
Name & Description &  $\sigma(z)$ \\
\hline
\texttt{linear} & linear& $z$, \\
\texttt{tanh} & hyperbolic tangent & $\tanh(z)$,	\\
\texttt{relu} & rectified linear units& $\max(0,z)$, \\
\texttt{elu} & exponential linear units&
	$\left\{\begin{array}{ll}
	\exp(z)-1 & z<0 \,,\\
	z & z\geq0\,.\\
	\end{array}\right.$\\
\hline
\hline
\end{tabular}
\caption{Common choices of activation functions.\label{tab.activation}}
\end{table}

In the language of deep learning, $N$ is referred to as the \textit{depth} of the network, ${\bf x}$ the \textit{input layer}, ${\bf y}$ the \textit{output layer}, ${\bf a}^{(l)}$ the $l$-th layer, $n^{(l)}$ the \textit{width} of the $l$-th layer. The intermediate layers with $1\le l < N$ are called \textit{hidden layers}. The activation functions are non-linear functions that modulate the function behavior. We list the common choices of activation functions in Table~\ref{tab.activation}. The choice of $N$, $n^{(l)}$, and $\sigma^{(l)}$ are {\it hyper-parameters} of the model.
Meanwhile, $W^{(l)}_{ij}$ and $b^{(l)}_i$ are respectively called \textit{weights} and \textit{biases}. 
They are parameters to be tuned during training by minimizing the \textit{cost function}, which characterize the distance between the approximation formula $\widetilde{\bf y}({\bf x})$ and the corresponding true function ${\bf y}({\bf x})$.
The process of parameter optimization is called \textit{model training} in deep learning.

The power of the DNN comes from its advanced parameter training method, called \textit{gradient descent} via \textit{back-propagation}. 
It updates the parameters according to the gradient of the cost function:
\begin{equation}
\Delta \boldsymbol{\theta} \equiv \boldsymbol{\theta}^{[k+1]} - \boldsymbol{\theta}^{[k]} 
\sim -\boldsymbol{\nabla}_{\boldsymbol{\theta}} J(\boldsymbol{\theta})
\end{equation}
where $\boldsymbol{\theta}$ is the abbreviation of all the parameters, i.e. $\boldsymbol{\theta}\equiv\{W^{(l)}_{ij}, b^{(l)}_{i}\}$, and the superscript ${[k]}$ stands for the $k$-th training step.
While different optimization schemes take different exact relations between $\Delta \boldsymbol{\theta}$ and $\boldsymbol{\nabla}_{\boldsymbol{\theta}} J$, there is a common feature --- noting that $\boldsymbol{\nabla}_{\boldsymbol{\theta}} J = 0$ when the cost function reaches its minimum, the training iterations eventually stop when closing to such a point.
In this work, we adopt the \texttt{Adam} optimization method~\cite{2014arXiv1412.6980K}, an acceleration method based on gradient descent.

In the most typical case for regression problems, the cost function is defined as the summation of the mean-square-error and a regularizer, where the latter is introduced to avoid overfitting manifested with unreasonably large values of parameters:
\begin{equation}
J(\boldsymbol{\theta}) = 
	 \frac{1}{2} \sum_{{\bf x} \in \text{data set}} \big|\widetilde{\bf y}(\boldsymbol{\theta},{\bf x}) - {\bf y}({\bf x}) \big|^2 + \frac{\lambda}{2} \boldsymbol{\theta \cdot \theta} ,
\end{equation}
and 
\begin{equation}
\frac{\partial J}{\partial \theta_i} = \sum_{{\bf x} \in \text{data set}}
	\big( \widetilde{\bf y}(\boldsymbol{\theta},{\bf x}) - {\bf y}({\bf x}) \big) \cdot \frac{\partial \widetilde{\bf y} (\boldsymbol{\theta},{\bf x})}{\partial{\theta_i}} +  \lambda \theta_i,
\end{equation}
where $(\cdot)$ is the inner product of all dimensions of ${\bf y}$.

The calculation of $\boldsymbol{\nabla}_{\boldsymbol{\theta}} J$ can be computationally very expensive in general, or even undoable, for arbitrary parameterized functions.
However, computing $\boldsymbol{\nabla}_{\boldsymbol{\theta}} J$ is straightforward and efficient for the DNN via back-propagation algorithm, thanks to its simple functional building block~\eqref{eq.dnn.iteration}. This is one of the major advantages of DNNs.
We denote that
\begin{align}
u^{(l)}_i =&\; \sigma'^{(l)}(z^{(l)}_{i}),\qquad z^{(l)}_{i} \equiv {b}^{(l)}_{i} + \sum_{j} {W}^{(l)}_{ij} {a}^{(l-1)}_{j},
\end{align}
with $\sigma'(z) \equiv \frac{\mathrm{d}\sigma(z)}{\mathrm{d}z}$ to be the derivative of the activation function with respect to its argument.
From Eq.~\eqref{eq.dnn.iteration} one can obtain the derivatives, for any input data point ${\bf x}$, as
\begin{align}
&\frac{\partial a^{(l)}_i}{\partial b^{(l)}_{i}} = u_i^{(l)}\,,\\
&\frac{\partial a^{(l)}_i}{\partial W^{(l)}_{i,j}} = a^{(l-1)}_j \; u_i^{(l)}\,,\\
&\frac{\partial a^{(l)}_i}{\partial a^{(l-1)}_{j}} = W^{(l)}_{i,j} \;u_i^{(l)} \,.
\end{align}
Then, the derivative at each layer can be calculated using the back-propagation iterations:
\begin{align}
&\frac{\partial y_k}{\partial b^{(l)}_{i}} 
    = \frac{\partial y_k}{\partial a^{(l)}_{i}} \; u^{(l)}_{i} \,,\\
&\frac{\partial y_k}{\partial W^{(l)}_{i,j}}
	= a^{(l-1)}_{j} \frac{\partial y_k}{\partial b^{(l)}_{i}}  \,,\\
&\frac{\partial y_k}{\partial a^{(l-1)}_{i}}  
	= \sum_j W^{(l)}_{j,i} \; \frac{\partial y_k}{\partial b^{(l)}_{j}}  \,.
\end{align}
where the iteration begins by ${\partial y_k / \partial a^{(N+1)}_{i}} = \delta_{i,k}$.
With these, one can obtain the $\boldsymbol{\nabla}_{\boldsymbol{\theta}} J$ and then update the parameters accordingly.

In practice, we employed two four-layered networks to represent $V_R(T,r)$ and $V_I(T,r)$. Each network contains a two-dimensional input layer, $\boldsymbol{a}^{(0)} = \{T,r\}$, and a one-dimensional output layer, ${a}^{(4)}=V_{R/I}$. The intermediate hidden layers $\boldsymbol{a}^{(1)}, \cdots, \boldsymbol{a}^{(3)}$ were chosen to consist of $\{32, 16, 32\}$ and $\{16, 16, 16\}$ neurons for the networks corresponding to $V_R(T,r)$ and $V_I(T,r)$, respectively. We adopted the \texttt{elu} activation function, i.e. $\sigma(z)=\exp(x)-1$ for $z<0$ and $\sigma(z)=z$ for $z\ge0$, 
for all the hidden layers, and a linear identity function, i.e. $\sigma(z)=z$, in the output layer. With the $V_R(T,r)$ and $V_I(T,r)$ as inputs, represented by the DNNs described above, as input we solved Eq.~\eqref{eq:SE} to obtain the masses and thermal widths.
We take the regularizers to be $\lambda_R=10^{-3}$ and $\lambda_I=10^{-2}$. By dividing or multiplying $\lambda$ by a factor of two, we have tested that results are insensitive to the choice of $\lambda$.

\subsection{Back-Propagation for DNN Coupled with Schr\"odinger Equations}\label{sec.methodology_optimization}
In this work, however, we employ the DNNs to approximate the functional relation between $(T,r)$ --- as the input layer --- and $(V_R, V_I)$ --- as the output layer, without knowing the ``true'' values of  $V_R$ and $V_I$ to train the parameters.
Instead, we further invoke Schr\"odinger equation solver to convert the DNN constructed potentials $V_R(T,r)$ and $V_I(T,r)$ into the corresponding mass and width of different bound states since their availability from lattice QCD. The cost function is set to be
\begin{equation}
J(\boldsymbol{\theta}) =  \frac{1}{2} \chi^2(\boldsymbol{\theta}) 
    + \frac{\lambda}{2} \boldsymbol{\theta \cdot \theta},
\label{eq.cost_function}
\end{equation}
to train the parameters of the DNNs, where the chi square function $\chi^2$ is the uncertainty-weighted summation of the squared difference between 
mass and width from Schr\"odinger equation, $m_{T,i}$ and $\Gamma_{T,i}$, and those from lattice QCD, $m_{T,i}^\mathrm{LQCD}$ and $\Gamma_{T,i}^\mathrm{LQCD}$.
In the most generic form, $\chi^2$ can be expressed as
\begin{align}\begin{split}
\chi^2 =&\; \sum_{T,i,j} \Big( R^{(T)}_{ij} \Delta m_{T,i} \Delta m_{T,j} 
+ I^{(T)}_{ij} \Delta \Gamma_{T,i} \Delta \Gamma_{T,j} \\&\;
+2M^{(T)}_{ij} \Delta m_{T,i} \Delta \Gamma_{T,j} \Big) \,,
\end{split}\label{eq.chi_square_general}\end{align}
where $\Delta m_{T,i} \equiv m_{T,i} - m_{T,i}^\mathrm{LQCD}$ is the difference between potential model and lattice result for the mass of the $i$-th state at temperature $T$, and likewise for the width $\Delta \Gamma_{T,i} \equiv \Gamma_{T,i} - \Gamma_{T,i}^\mathrm{LQCD}$.
$R_{ij}$, $I_{ij}$, and $M_{ij}$ are the symmetric covariance matrices.
In this work, we neglect the correlation between different quantities, hence $R_{ij}^{(T)} = (\delta m_{T,i}^\mathrm{LQCD})^{-2} \cdot  \delta_{ij}$, 
$I_{ij}^{(T)} = (\delta \Gamma_{T,i}^\mathrm{LQCD})^{-2} \cdot  \delta_{ij}$, and $M_{ij}^{(T)} = 0$.

Computing the parameter gradient is generally complicated if one is not able to find the explicit functional form between $\boldsymbol{\theta}$ and the cost function.
In this system, however, the gradient $\boldsymbol{\nabla}_{\boldsymbol{\theta}} \chi^2$ can be computed explicitly via perturbation treatment on the Schr\"odinger equation.
One can express the eigenvalue problems, before and after a perturbation of the potential, respectively as
\begin{align}
\begin{split}
&	\Big(\frac{\widehat{p}^2}{2m} + V(r) \Big) |\psi_i\rangle = E_i  |\psi_i\rangle,\\
&	\Big(\frac{\widehat{p}^2}{2m} + V(r) + \delta V(r) \Big) |\psi'_i\rangle = (E_i + \delta E_i)  |\psi'_i\rangle.
\end{split}\label{eq.perturbation}
\end{align}
Up to the leading order in $\delta V$ expansion, perturbation theory yields that
\begin{equation}
\delta E_i = \langle \psi_i | \delta V(r) |\psi_i\rangle,
\end{equation}
and
\begin{equation}
| \psi'_i\rangle = | \psi_i\rangle 
+ \sum_{j\neq i} \frac{\langle \psi_j| \delta V(r) | \psi_i\rangle}{E_i - E_j} | \psi_j\rangle.
\end{equation}
The former relation is also referred to as the Hellmann--Feynman theorem.
Noting that both $E_i$ and $V(r)$ can be complex, we separate the real and imaginary parts:
\begin{align}
\begin{split}
\delta m_i &= \langle \psi_i | \delta V_R(r) |\psi_i\rangle,  \\
\delta \Gamma_i &= -\langle \psi_i | \delta V_I(r) |\psi_i\rangle.
\end{split}
\end{align}
In particular, for local perturbations
\begin{equation}
\delta V(r) = v \, \delta(r-r_k),
\end{equation}
one can obtain
\begin{equation}
\frac{\delta E_i}{\delta v} = |\psi_i(r_k)|^2\,,
\end{equation}
which leads to the functional derivative of complex eigenvalues with respect to the complex potential :
\begin{align}
\begin{split}
 \frac{\delta m_i}{\delta V_R(r)} =&\; -\frac{\delta \Gamma_i}{\delta V_I(r)} = |\psi_i(r)|^2 \,, \\
 \frac{\delta m_i}{\delta V_I(r)} =&\; \frac{\delta \Gamma_i}{\delta V_R(r)} = 0\,.
\end{split}
\end{align}

With such relations, we obtain the gradients of the $\chi^2$ 
\begin{align}\begin{split}
\frac{\partial \chi^2}{\partial \theta_{R,n}} 
=&\; \sum_{T,i,k} \frac{\partial \chi^2}{\partial m_{T,i}} 
    \frac{\partial V_R(T,r_k)}{\partial \theta_{R,n}} |\psi_{i}(T,r_k)|^2 \mathrm{d}r \,,\\
\frac{\partial \chi^2}{\partial \theta_{I,n}} 
=&\; -\sum_{T,i,k} \frac{\partial \chi^2}{\partial \Gamma_{T,i}} \frac{\partial V_I(T,r_k)}{\partial \theta_{I,n}} |\psi_{i}(T,r_k)|^2 \mathrm{d}r \,,
\end{split}\end{align}
and of the cost function 
\begin{align}\begin{split}
\frac{\partial J}{\partial \theta_{R,n}} =&\;
	\sum_{T,i} \bigg\{
	\Big[\sum_k \frac{\partial V_R(T,r_k)}{\partial \theta_{R,n}} |\psi_{i}(T,r_k)|^2 \mathrm{d}r \Big] \times
\\&\quad
	 \sum_j \Big[R^{(T)}_{i,j} \Delta m_{T,j} + M^{(T)}_{ij} \Delta \Gamma_{T,j}  \Big]  \bigg\} 
	+ \lambda \theta_{R,n} \,, \\
\frac{\partial J}{\partial \theta_{I,n}} =&\;
	-\sum_{T,i} \bigg\{
	\Big[\sum_k \frac{\partial V_I(T,r_k)}{\partial \theta_{I,n}} |\psi_{i}(T,r_k)|^2 \mathrm{d}r \Big] \times 
\\&\quad
	\sum_j \Big[I^{(T)}_{i,j}  \Delta \Gamma_{T,j} + M^{(T)}_{ij} \Delta m_{T,j}  \Big]  \bigg\} 
	+ \lambda \theta_{I,n}  \,,
\end{split}\label{eq.parameter_gradient}\end{align}
where $\mathrm{d}r$ is the discrete step size in distance $r$. 
Eventually, we develop the back-propagation scheme for the DNNs coupled with a Schr\"odinger equation.

While DNN here can be viewed to provide an unbiased and robust special parameterization for the potentials, one could in principle take any other arbitrary parameterization scheme. The above perturbative analyses for the gradient evaluation could hold for an arbitrary parameterized form of potentials.
Suppose the potentials are functions of parameter $\boldsymbol{\theta}$, $V_{R/I}(\boldsymbol{\theta}; T,r)$; then Eq.~\eqref{eq.parameter_gradient} would remain valid.

\subsection{Fitting Quark Mass and Vacuum Potential}\label{sec.parameter_tuning_wf}
The $b$-quark mass and bottomonia vacuum potential are determined by fitting both the experimental results on the bottomonia masses and their Bethe--Salpeter amplitudes from the lattice calculation~\cite{Larsen:2020rjk}.
To match the lattice result of both Bethe--Salpeter amplitudes and mass spectrum, one can set loss function as 
\begin{align}\begin{split}
J =&\; \frac{\mu}{2} \sum_i w_i (m_i - m^\mathrm{exp}_i)^2 \\
&\;	+ \frac{\nu}{2} \sum_{i,j} w_{ij} \Big(\psi_i(r_j) - \psi^\mathrm{BS}_i(r_j) \Big)^2 ,
\end{split}\end{align}
where local weights $w_i$ and $w_{ij}$ account for the data uncertainties, $\mu$ and $\nu$ set the relative weight between the mass-difference and the wave-function difference in the fitting.
In practice, we employ a global weighting with $w_i = (10~\text{MeV})^{-2}$ and $w_{ij} = (10~\text{MeV})^{-1}$, and take $\mu=\nu=1$.
From the perturbation theory we find the derivative of the loss function with respect to a potential parameter $\theta_n$:
\begin{align}\begin{split}
\frac{\partial J}{\partial \theta_n} =&\;
	 \mu\sum_i w_i (m_i - m^\mathrm{exp}_i) 
	 \sum_k \frac{\partial V(r_k) }{\partial \theta_n} 
	 |\psi_i(r_k)|^2 \mathrm{d}r \\
&\;	+ \nu \sum_{i,j} \bigg[ w_{ij} \Big(\psi_i(r_j) - \psi^\mathrm{BS}_i(r_j)\Big) \times \\
&\;\qquad
	 \sum_{i'\neq i} \sum_k \frac{\partial V(r_k)}{\partial \theta_n} 
	 \frac{\psi_i(r_k) \psi_{i'}(r_k) }{m_i-m_{i'}} \psi_{i'}(r_j) \mathrm{d}r \bigg]\,.\\
\end{split}\end{align}
Similarly, one can analyze the linear response against a perturbation in $m_b$ ass
\begin{align}
\begin{split}
&	\Big(\frac{\widehat{p}^2}{m_b} + V(r) \Big) |\psi_i\rangle = E_i  |\psi_i\rangle,\\
&	\Big(\frac{\widehat{p}^2}{m_b+\delta m_b } + V(r) \Big) |\psi'_i\rangle = (E_i + \delta E_i)  |\psi'_i\rangle,
\end{split}
\end{align}
which leads to
\begin{align}
&\delta E_i = \frac{\delta m_b}{m_b} \Big(\int V(r) |\psi_i(r)|^2 \mathrm{d}r - E_i \Big)\,, \\
&| \psi'_i\rangle = | \psi_i\rangle 
+ \frac{\delta m_b}{m_b} \sum_{j\neq i} \frac{\langle \psi_j| V(r) | \psi_i\rangle}{E_i - E_j} | \psi_j\rangle,
\end{align}
and further arrives the $m_b$-derivative of the loss function,
\begin{align}\begin{split}
\frac{\partial J}{\partial m_b} =&\;
\frac{\mu}{m_b} \sum_i \Big[ w_i (m_i - m^\mathrm{exp}_i) 
\\&\qquad
	\times \Big(4m_b - m_i + \sum_k V(r_k) |\psi_i(r_k)|^2 \mathrm{d}r \Big) \Big] \\
+&\;	 \frac{\nu}{m_b} \sum_{i,j} \bigg\{ w_{ij} 
	 \sum_{i'\neq i} \bigg[ \frac{\sum_k V(r_k) \psi_i(r_k) \psi_{i'}(r_k) \mathrm{d}r}{m_i-m_{i'}}
\\&\qquad\times
	  \psi_{i'}(r_j)\Big(\psi_i(r_j) - \psi^\mathrm{BS}_i(r_j)\Big) \bigg] \bigg\} \,.\\
\end{split}\end{align}
We note that the wave-function driven potential extraction method is also discussed in~Ref.\cite{2020PhRvB.101x5139X}.

\subsection{Uncertainty Quantification with Bayesian Inference}\label{sec.methodology_bayesian}
We invoke the Bayesian inference to estimate the uncertainties of the DNN reconstructed potentials. 
Bayesian inference is a statistical paradigm that utilizes the (experimental) data to constrain model parameters using probability statements.
Based on the Bayes' theorem, the posterior distribution over the model parameters (conditional on the observed data) is proportional to the product of the likelihood given the observed data and the prior distribution of the parameters,
\begin{align}\begin{split}
&\;	\text{Posterior}(\boldsymbol{\theta}|\text{data}) \propto L(\boldsymbol{\theta}|\text{data}) \times \text{Prior}(\boldsymbol{\theta}).
\end{split}\end{align}
The likelihood function of the parameters given the observed data specifies the chance that those data appear under the model with the taken parameters, which due to the central limit theorem can be expressed as Gaussian form with the chi-square shown naturally:
\begin{align}
L(\boldsymbol{\theta}|\text{data}) = P(\text{data}|\boldsymbol{\theta}) \propto \exp[-\chi^2(\boldsymbol{\theta})/2].
\end{align}
The prior distribution, in principle, reflects our naive belief in the model parameters, while in practice we take a Gaussian prior distribution
accounting for the quadratic regularizers introduced in the cost~\eqref{eq.cost_function},
\begin{align}
 \text{Prior}(\boldsymbol{\theta}) \propto \exp[-\frac{\lambda}{2}\boldsymbol{\theta}\cdot\boldsymbol{\theta}],
\end{align}
which represents our relative ``ignorance'' about $\boldsymbol{\theta}$ and also acts as a non-local regulator to account for correlated distributions of the to-be-determined target (potential) values in the language of Bayesian statistics.

With the above we thus obtain the posterior distribution over the parameters to be :
\begin{align}\begin{split}
\text{Posterior}(\boldsymbol{\theta}|\text{data})
= N_0 \exp\Big[-\frac{\chi^2(\boldsymbol{\theta})}{2} -\frac{\lambda}{2}\boldsymbol{\theta}\cdot\boldsymbol{\theta}\Big],
\end{split}\end{align}
with $N_0$ being a constant normalization factor to ensure $\int  \text{Posterior}(\boldsymbol{\theta}) \mathrm{d}^N\boldsymbol{\theta} = 1$.
To estimate the uncertainty of $V(T,r)$ for any given $T$ and $r$, we allow the variation of parameters away from their optimal values, and the probability of accepting such a parameter set, as well as the corresponding potential, is determined by the posterior function
\begin{align}
&\;P(V_{\boldsymbol{\theta},T,r}) \mathrm{d}V = \text{Posterior}(\boldsymbol{\theta}|\text{data}) \mathrm{d}^N\boldsymbol{\theta} \,.
\end{align}
With a sufficient number of potential samples following the above distribution, we can estimate the credible interval of the potential at each of the $(T,r)$ points. 

Sampling high-dimensional parameters is tricky by itself. 
In a most direct way, one sample $M$ points in the parameter space according to a flat distribution, denoted as $\{\boldsymbol{\theta}_i\}$, so that each of them corresponds to the volume element $\mathrm{d}^N\boldsymbol{\theta}_i = M^{-1}$, and the corresponding potential represents a point in the histogram with weighting
\begin{align}
w_i = P(V_{\boldsymbol{\theta_i},T,r}) \mathrm{d}V_i = \frac{\text{Posterior}(\boldsymbol{\theta}_i|\text{data})}{M}.
\label{eq.Vdist_0}
\end{align}
However, there are $\sim10^3$ parameters $\boldsymbol{\theta}$ in DNN, and a majority of points in the parameter space correspond to a vanishing posterior. Thus, computing the likelihood distribution according to Eq.~\eqref{eq.Vdist_0} is computationally expensive.

In principle, the most efficient way would be to sample $\{\boldsymbol{\theta}_i\}$ according to $\text{Posterior}(\boldsymbol{\theta}|\text{data})$, hence the volume element $\mathrm{d}^N\boldsymbol{\theta}_i =   M^{-1} \text{Posterior}^{-1}(\boldsymbol{\theta_i}|\text{data})$, and the corresponding potential is of the weighting
\begin{align}
w_i = P(V_{\boldsymbol{\theta_i},T,r}) \mathrm{d}V_i = \frac{1}{M}.
\label{eq.Vdist_1}
\end{align}
Nevertheless, the posterior function is generally unknown or unable to be represented in a simple way, hence it is not possible to sample according to the posterior function.
A practical method is importance sampling (see e.g.~\cite{rubinstein2016simulation}), which samples $\{\boldsymbol{\theta}_i\}$ according to a reference distribution $\widetilde{P}(\boldsymbol{\theta})$, hence $\mathrm{d}^N\boldsymbol{\theta}_i =  M^{-1} \widetilde{P}^{-1}(\boldsymbol{\theta_i})$, and perform re-weighting on each sample by assigning an extra weight determined by the ratio of the posterior to the reference, $w_i = \text{Posterior}(\boldsymbol{\theta}_i) / \widetilde{P}(\boldsymbol{\theta}_i)/M$.

The computational efficiency would be higher when the reference distribution $\widetilde{P}(\boldsymbol{\theta})$ is close to the Posterior. 
For general systems, one usually invoke variational inference~\cite{graves2011practical} or Bayesian Neural Network~\cite{blundell2015weight} to find $\widetilde{P}(\boldsymbol{\theta})$. 
In the work, however, we are able to make use of the underlying physics to construct the reference distribution.
According to the first-order perturbation theory, the posterior is a non-diagonal normal distribution around the optimal parameter set ($\boldsymbol{\theta}^\text{opt}$), and we adopt it to be the reference distribution,
\begin{align}\begin{split}
\widetilde {P} (\boldsymbol{\theta}) =&\; (2\pi)^{-N_\theta /2} \sqrt{\det[\Sigma^{-1}]} \; \times \\
&\;	 \exp\Big[ -\frac{\Sigma_{ab}^{-1}}{2} (\theta_a - \theta^\text{opt}_a)( \theta_b- \theta^\text{opt}_b) \Big],\label{eq.reference}
\end{split}\end{align}
with the inverse covariance matrix given by
\begin{align}
\Sigma_{ab}^{-1} \equiv \frac{\partial^2 J(\boldsymbol{\theta})}{\partial \theta_a \partial \theta_b} 
	= \lambda \delta_{ab} + \frac{1}{2} \frac{\partial^2 \chi^2(\boldsymbol{\theta})}{\partial \theta_a \partial \theta_b} \,,
\end{align}
where
\begin{align}
\frac{1}{2}\frac{\partial^2 \chi^2(\boldsymbol{\theta})}{\partial \theta_{R,a} \partial \theta_{R,b}} 
= \sum_{T,i,j} R^{(T)}_{i,j} \frac{\partial m_{T,i}}{\partial \theta_{R,a}} \frac{\partial m_{T,j}}{\partial \theta_{R,b}}\,,
\end{align}
and similarly for $\frac{1}{2}\frac{\partial^2 \chi^2(\boldsymbol{\theta})}{\partial \theta_{R,a} \partial \theta_{I,b}} $ and $\frac{1}{2}\frac{\partial^2 \chi^2(\boldsymbol{\theta})}{\partial \theta_{I,a} \partial \theta_{I,b}}$. 

To sample parameters according to the reference distribution $\widetilde {P} (\boldsymbol{\theta})$, one needs to solve the eigenvalues and eigenstates of the inverse covariance matrix for the parameters, $\Sigma_{ab}^{-1}$.
There are $\sim10^3$ parameters in the DNNs, and solving the eigenvalues and eigenvectors for $\Sigma_{ab}^{-1}$ are generally expensive.
Fortunately, $\frac{\partial^2 \chi^2(\boldsymbol{\theta})}{\partial \theta_a \partial \theta_b}$ are large matrices (dimension $\sim10^3$) constructed by multiplying low-rank [$\text{rank}\leq30=6\text{(temperatures)}\times5\text{(states)}$] matrices. One can conclude there are, at most, $30$ non-vanishing eigenvalues for such a large matrix. We obtain these non-vanishing eigenvalues and the corresponding eigenvectors by employing the power method, while the rest eigenvectors, corresponding to the highly-degenerated zero eigenvalue, are obtained by applying the Gram--Schmidt orthogonalization procedure. Such a procedure is also referred to as the \textit{principal component analysis} (PCA) in deep learning.

In our method, the Aleatoric (statistical) uncertainty is naturally encoded inside the posterior, since the $\chi^2$ considered the lattice data error and/or correlations. Meanwhile the Epistemic (systematic) uncertainty is also manifested, on one hand, our ``ignorance'' of the network parameters is reflected in the prior which is consistent with the regularizer used in the cost. On the other hand, the limits of the model --- limited energy levels for a quantum system would only manifest partial information of the interaction --- can also retain in our methodology via the generalized Back-Propagation through the Schr\"odinger equation.

\begin{figure}[!hbtp]\centering
\includegraphics[width=0.23\textwidth]{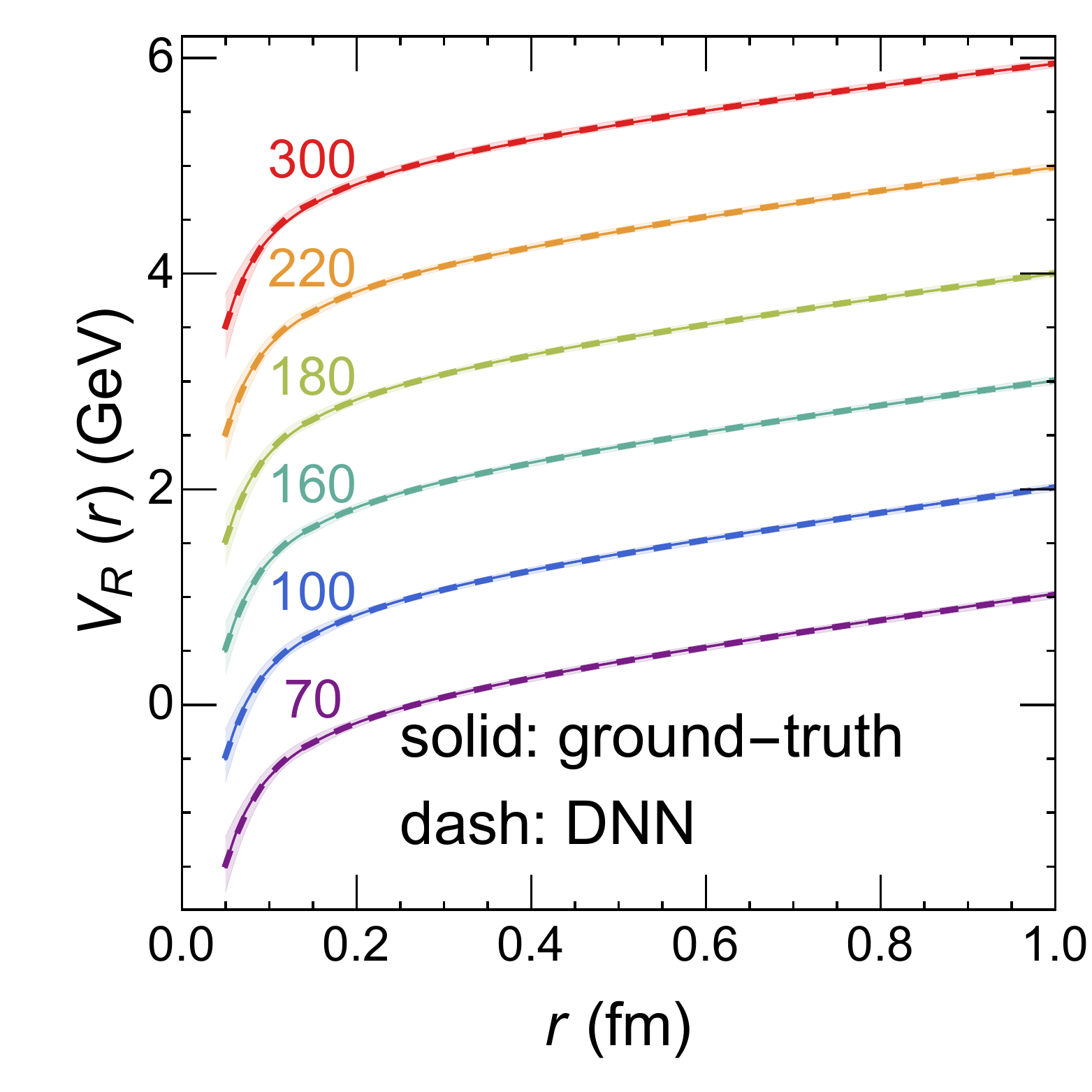}
\includegraphics[width=0.23\textwidth]{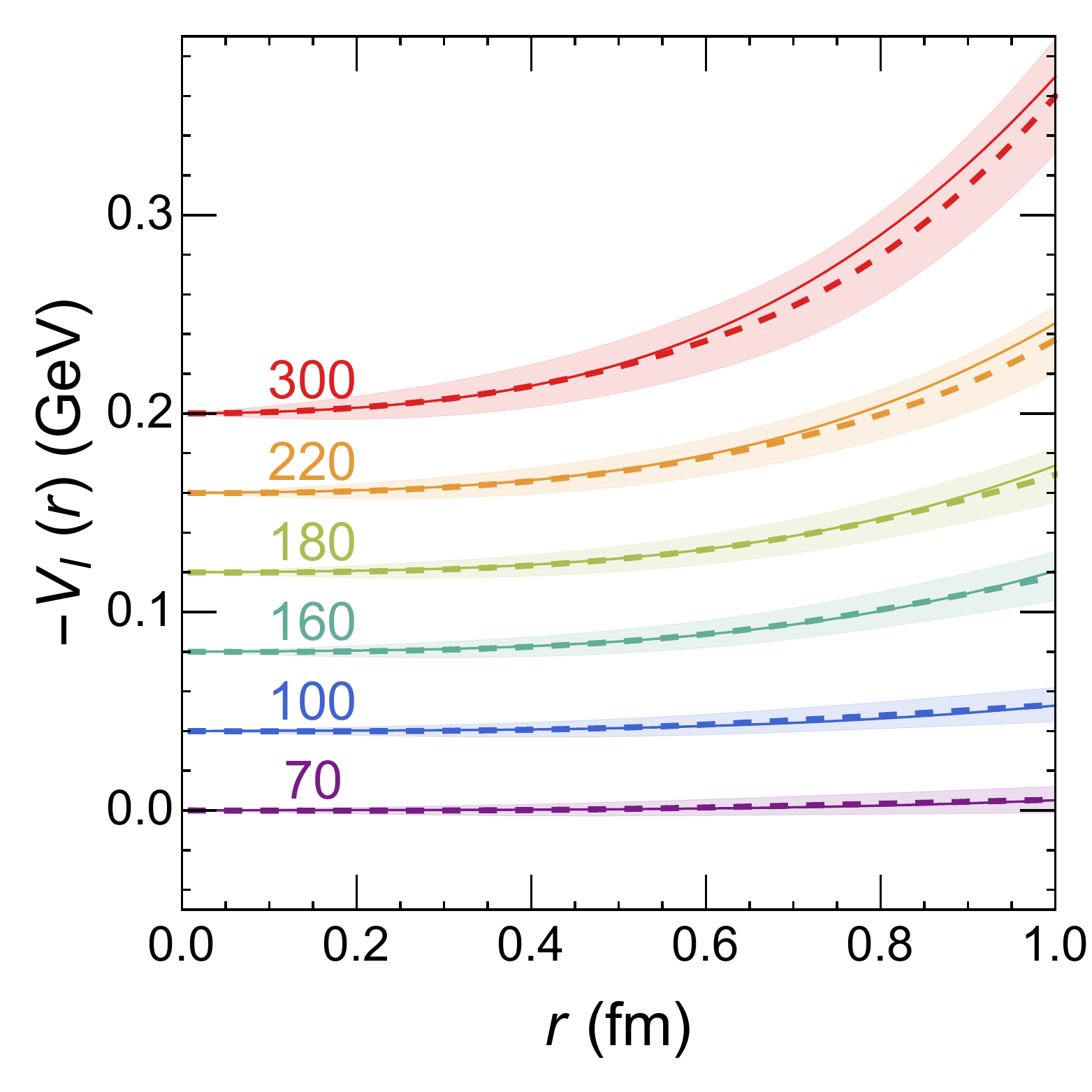} \\
\includegraphics[width=0.23\textwidth]{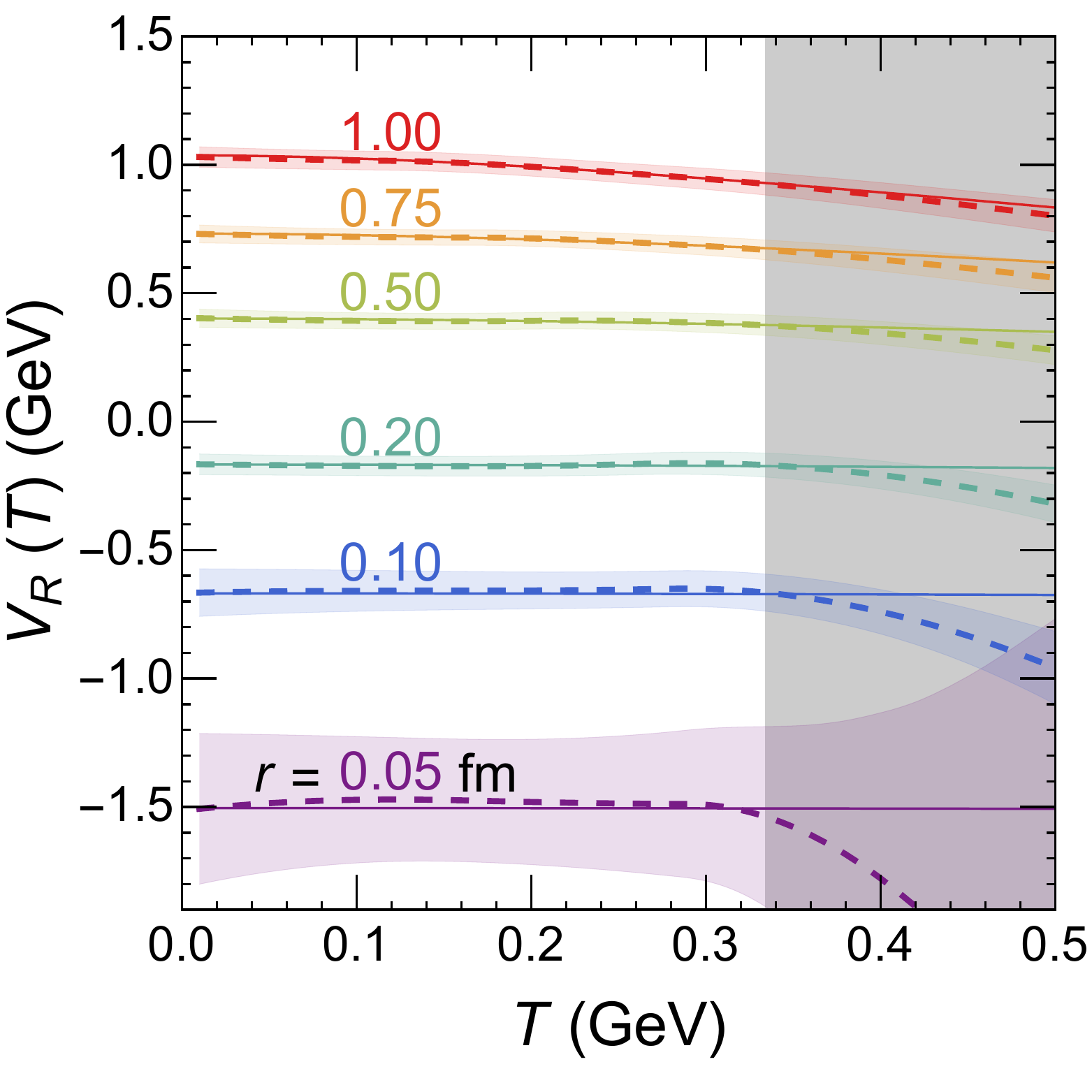}
\includegraphics[width=0.23\textwidth]{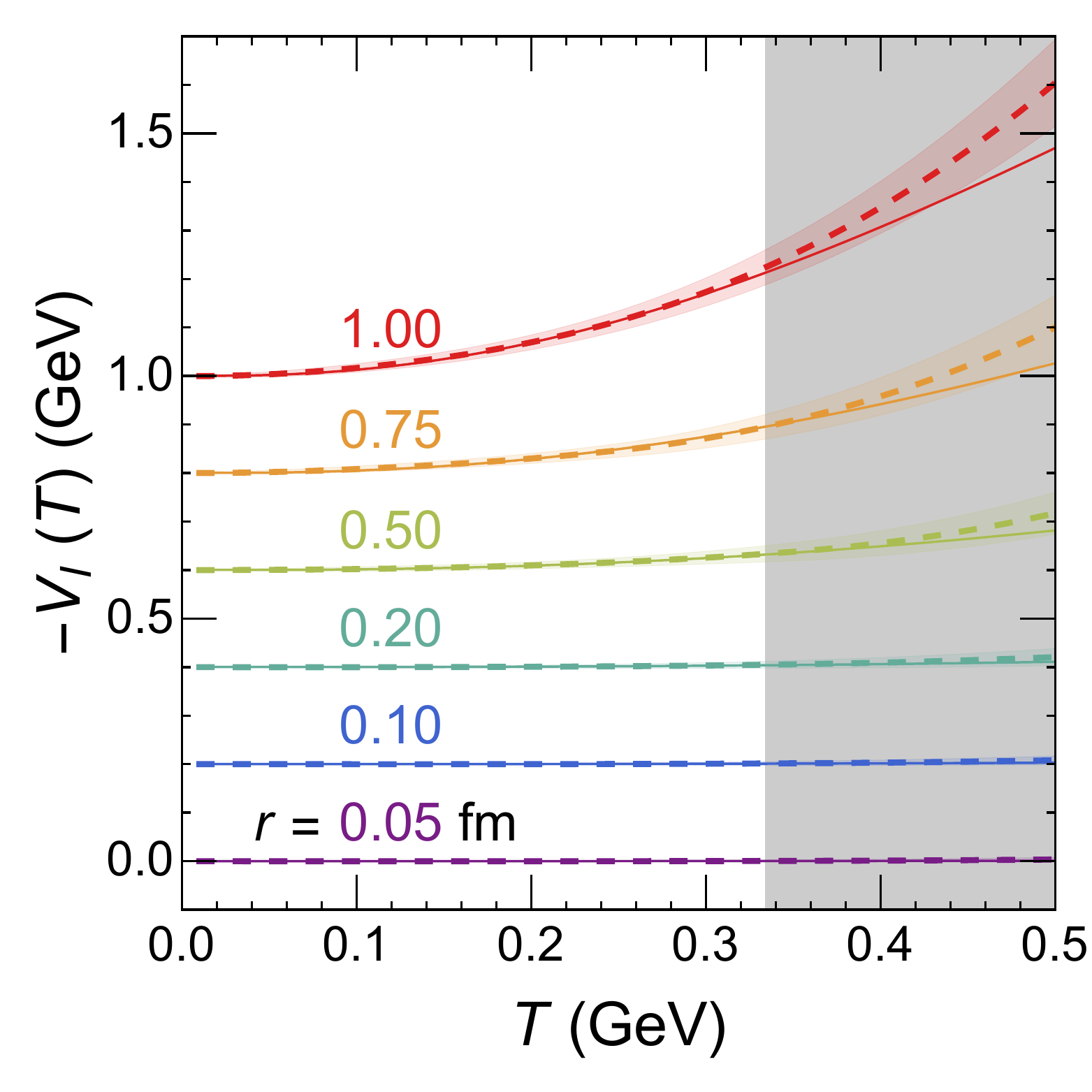}
\caption{Closure test on applying Deep Neural Networks to reconstruct potential functions.
The upper panels show the $r$-dependence at different temperature points, while the lower panels show the $T$-dependence at different distances.
The left(right) panels plot the real(imaginary) potential.
The solid lines represent the ``ground-truth'' formulae \protect{(\ref{eq.VR_screening_closure}-\ref{eq.VI_screening_closure})}, while the dashed lines with the uncertainty band are the potential reconstructed by DNN given only the mass and width for the first five bottomonia states. 
The gray shaded area in the lower panels indicates the extrapolation region.
\label{fig.closure}}
\end{figure}

\subsection{Method Validation: Closure Test With Known Potentials}\label{sec.methodology_closure}
To justify the above described method of potential reconstruction with DNNs, in this subsection we perform closure tests to assess the reliability of the methodology.
We start by assuming a ``ground-truth'' potential taking the known HTL formulae 
\begin{align}
\begin{split}
V_R (T,r) =&\; \frac{\sigma}{\mu_D}\Big(
	2 - (2+\mu_D r) e^{-\mu_D r}
	\Big) 
\\&\;
	- \alpha \Big( \mu_D + \frac{e^{-\mu_D r}}{r} \Big)  + B\,,
\label{eq.VR_screening_closure}
\end{split}\\
\begin{split}
V_I(T,r) =&\; 
	- \frac{\sqrt{\pi}}{4}\mu_D T \sigma r^3 G^{2,2}_{2,4}\Big(^{-\frac{1}{2},-\frac{1}{2}}_{\frac{1}{2},\frac{1}{2},-\frac{3}{2},-1}\Big|\frac{\mu_D^2 r^2}{4}\Big)
\\&\;
	- \alpha\, T\,\phi(\mu_D r) 
	\,,
\label{eq.VI_screening_closure}
\end{split}
\end{align}
where the Debye-screening mass, $\mu_D$, is a function of temperature, 
$G$ the Meijer-$G$ function, and
\begin{equation}
\phi(x) = 2\int_0^\infty \frac{z \, \mathrm{d}z}{(1+z^2)^2} \Big(1-\frac{\sin(x\,z)}{x\,z} \Big) \,.
\end{equation}
We note that the imaginary potential is consistent with the Hard Thermal Loop calculation~\cite{Laine:2006ns,Burnier:2009bk}.
Adopting such a potential, we further solve the Schr\"odinger equations at six different temperatures, $\{0, 151, 173, 199, 251, 334\}$~MeV, to generate a set of pseudo-data --- the mass and width for 1S, 2S, 3S, 1P, and 2P states. This set of pseudo-data is then fed into the above described method to train the DNN for potential reconstruction. With the comparison to the assumed ``ground-truth'' input potential, one can therefore assess the reliability and robustness of the method. 
In Fig.~\ref{fig.closure}, we compare the DNN reconstructed potential with the ``ground-truth'' values systematically. The comparison is shown for different temperatures (not limited to pseudo-data generation temperatures) and distances.

From Fig.~\ref{fig.closure}, it is evident that the DNN reconstructed potentials (dashed lines) are almost identical to the ``ground-truth'' potentials (solid lines) over the physically relevant range in $r$.
Slight deviation and relatively large uncertainty band at $r=0.05$~fm are related to facts that there the wave-functions are small and the energy eigenvalues are insensitive to the potential values at this small range. In particular, note that although the DNN is trained by fitting only the energy eigenvalues (pseudo-data) generated at aforementioned discretized temperature points, the DNN reconstructed potentials show smooth behavior along with temperatures and agree nicely with the ``ground-truth'' potential values in the interpolation region. Not surprisingly, the DNN potential starts to diverge from the ``ground-truth'' values in the extrapolation region ($T>334$~MeV), which is also captured by the increasing uncertainty band from the Bayesian analysis, as indicated by the gray shaded area in the lower panels of Fig.~\ref{fig.closure}.

\begin{figure*}[!hbtp]
\centering
\includegraphics[width=0.32\textwidth]{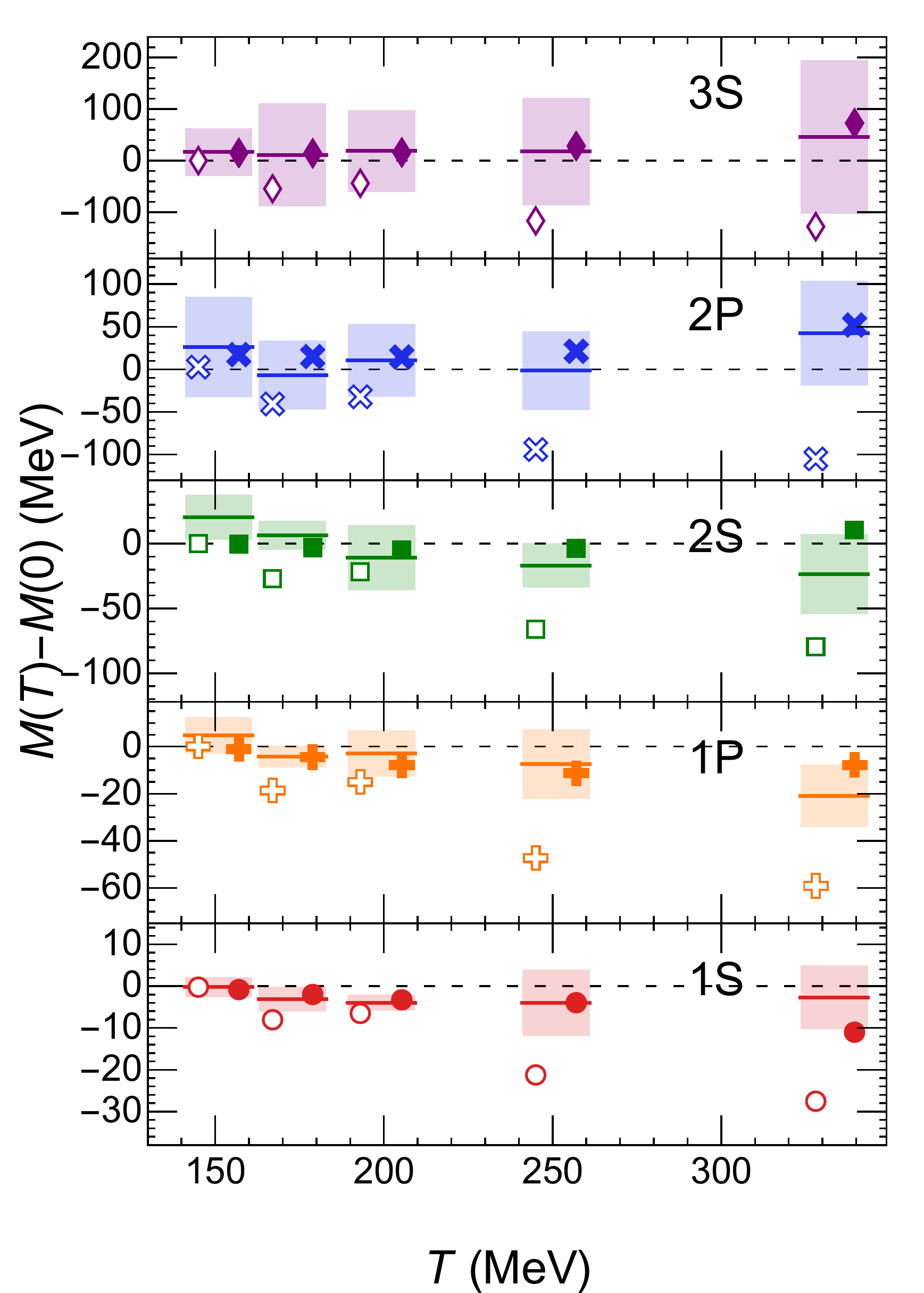}
\includegraphics[width=0.32\textwidth]{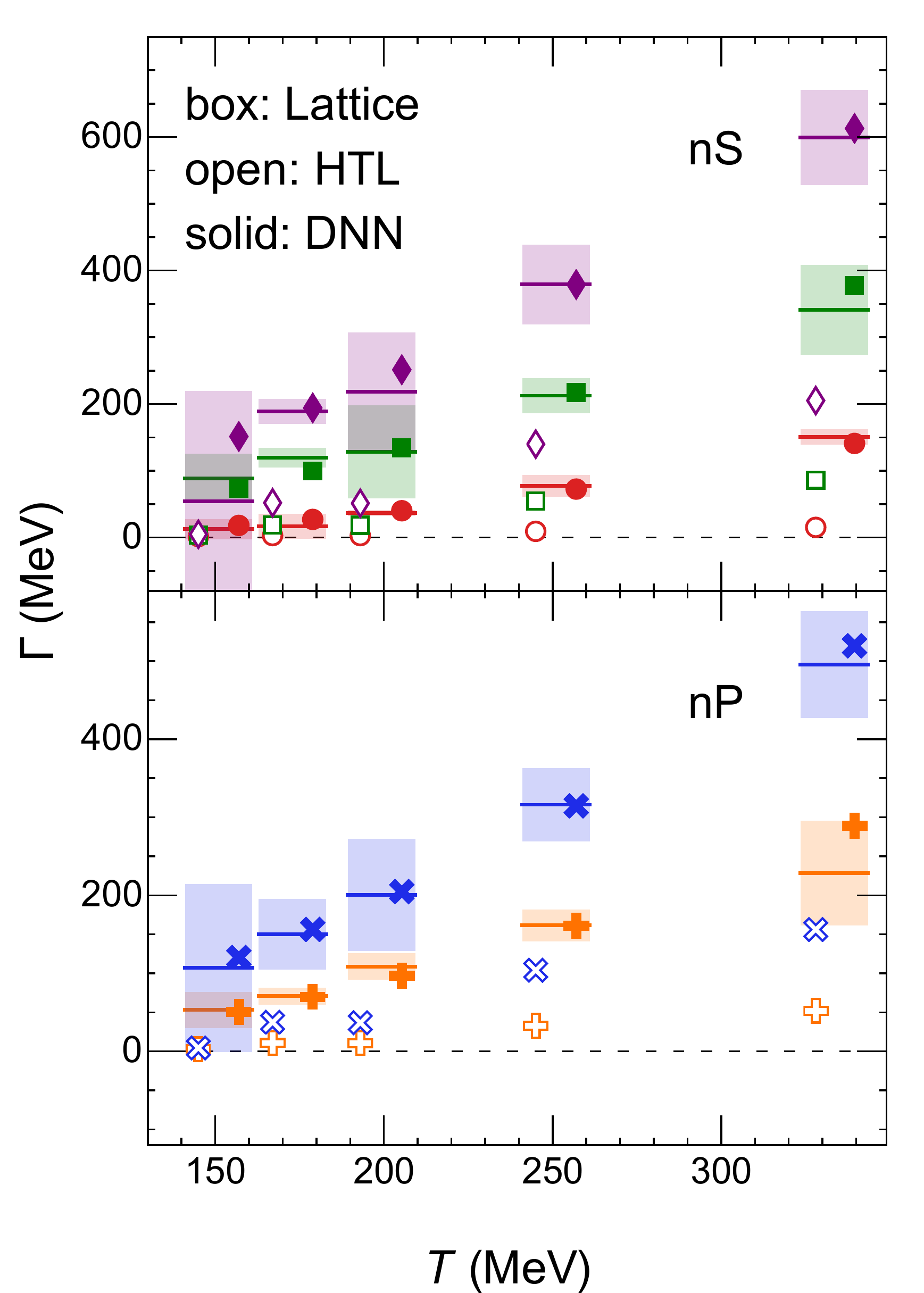}
\includegraphics[width=0.32\textwidth]{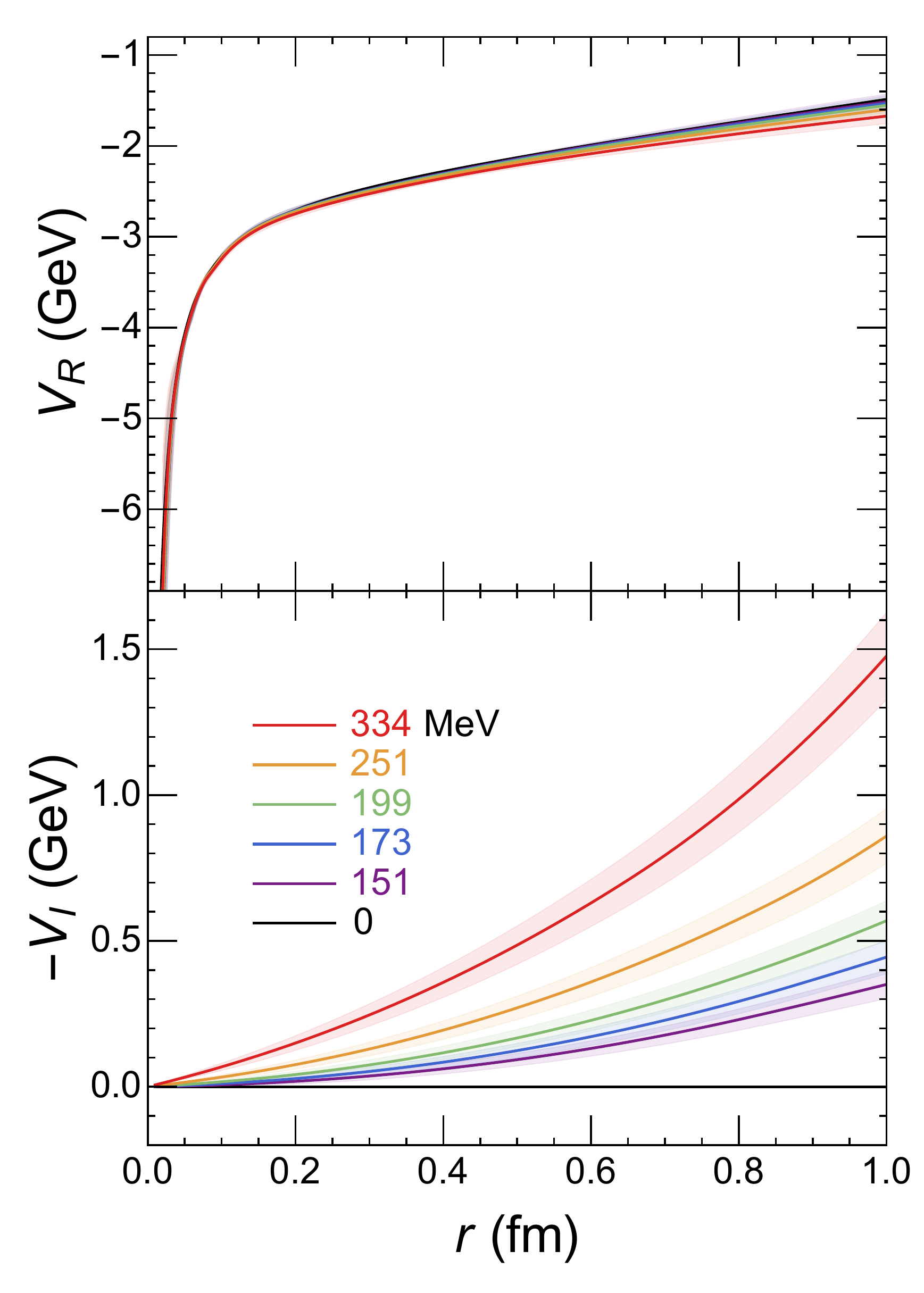}
\caption{Left and middle: In-medium mass shifts with respect to the vacuum mass (left) and the thermal widths (middle) of different bottomonium states obtained from fits to LQCD results of Ref.~\cite{Larsen:2019zqv} (lines and shaded bands) using weak-coupling motivated functional forms~\cite{Lafferty:2019jpr} (open symbols) and DNN based optimization (solid symbols). The points are shifted horizontally for better visualization. 
$\Upsilon(1S)$, $\chi_{b_0}(1P)$, $\Upsilon(2S)$, $\chi_{b_0}(2P)$, and $\Upsilon(3S)$ states are represented by red circles, orange pluses, green squares, blue crosses and purple diamonds, respectively. Right: The DNN reconstructed real (top) and imaginary (bottom) parts of the heavy quark potential at temperatures $T = 0$(black), $151$(purple), $173$(blue), $199$(green), $251$(orange), and $334$(red)~MeV. The uncertainty bands represent the $68\%(1\sigma)$ confident region.
\label{fig.mass}}
\end{figure*}

\section{Heavy Quark Potential from DNN}\label{sec.potential_dnn}
With the computation framework established and tested, we move on to discuss the extraction of heavy quark potential from LQCD results.
We begin with pointing out the inadequacy of the weak-coupling motivated functional form of the potential to consistently describe the LQCD results for bottomonia masses and thermal widths~\cite{Larsen:2019zqv}. For this purpose, we chose the functional form proposed in Ref.~\cite{Lafferty:2019jpr}. This incorporates one-loop HTL based functional forms of $V_I$ and of color-electric screening, in addition to a vacuum potential satisfying Gauss's law [see Eqs. \eqref{eq.VR_screening_closure} and \eqref{eq.VI_screening_closure}]. Taking this functional form for the potential, we fix $\alpha$, $\sigma$, and $B$ by their vacuum values, and tune $\mu_D$ at different temperatures to fit the finite-temperature bottomonia masses and widths. We find the most optimal values to be $\mu_D = 0.01$, $0.19$, $0.17$, $0.32$, $0.37$~GeV for $T = 151$, $173$, $199$, $251$, and $334$~MeV, respectively, with the corresponding $\chi^2$-per-data being $\{13.5, 159, 111, 154, 244\}/5$.
As shown by the open symbols in Fig.~\ref{fig.mass} (left and middle), one-loop HTL motivated functional form of $V_I$ and color-electric screening in $V_R$ fail to simultaneously reproduce the LQCD results for the mass shifts and the thermal widths of bottomonium. This failure might be due to the missing contributions from the color-magnetic scale, which is normally beyond the scope of a conventional perturbation theory~\cite{Fukushima:2013xsa}. 

\begin{figure}[!hbtp]\centering
\includegraphics[width=0.45\textwidth]{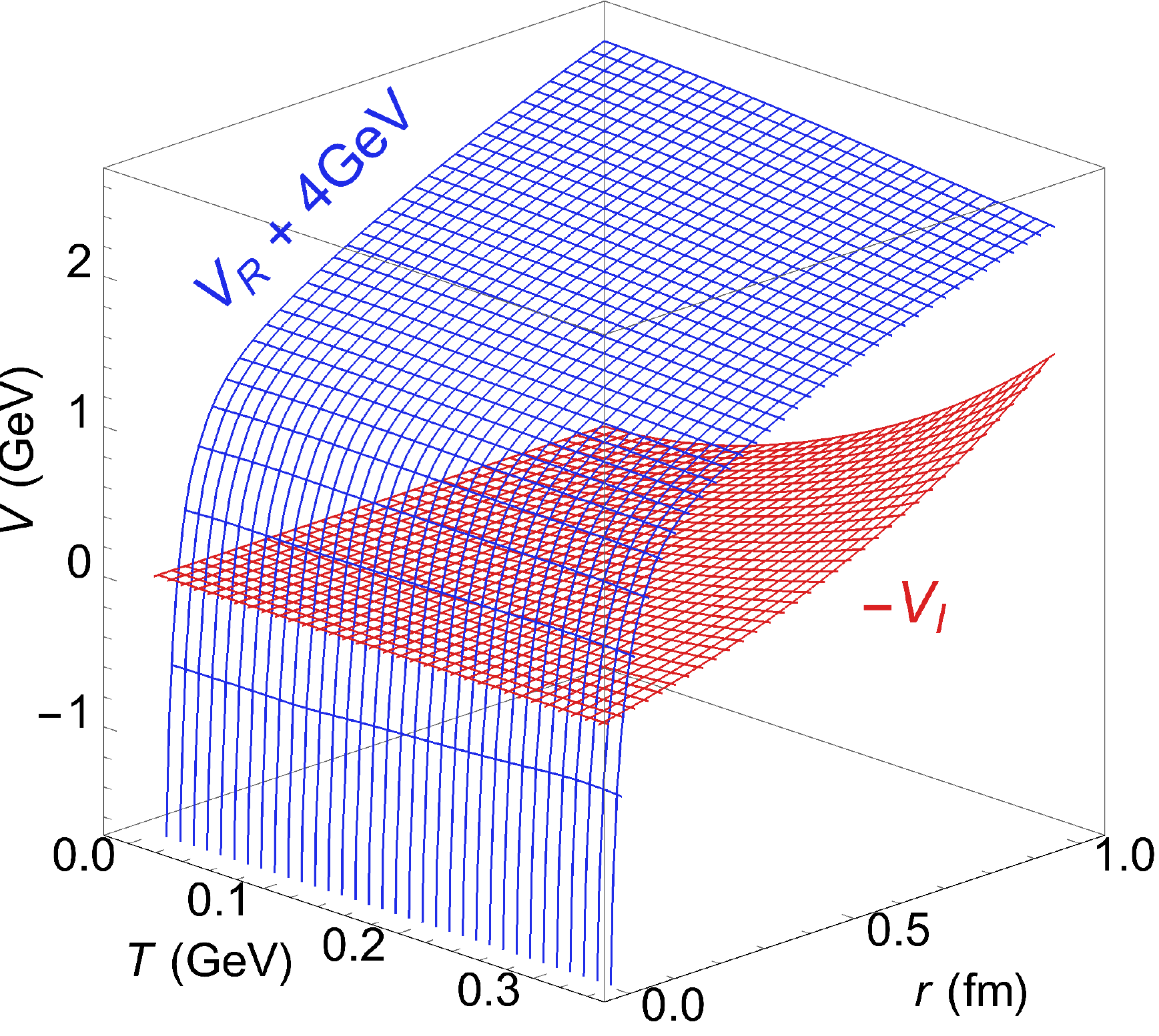}
\caption{Real (blue) and imaginary (red) part of interaction potentials versus temperature $T$ and quark-antiquark distance $r$ extracted via DNNs.
\label{fig.potential_DNN_3D}}
\end{figure}

The failure of the only known analytic form to describe the LQCD results necessitates a model-independent extraction of $V(T,r)$ using an adequate unbiased parameterization. To achieve this, we devised the above outlined method by coupling Schr\"odinger equation with DNNs. Using this set-up, we optimized the DNNs' parameters and achieved good agreement with the LQCD results~\cite{Larsen:2019zqv}. The optimized fitting for the mass shifts and thermal widths are shown by the solid symbols in Fig.~\ref{fig.mass} (left and middle), with the corresponding $\chi^2$-per-data-point to be $16.5/30$. The $T$- and $r$-dependence of the real (top) and imaginary (bottom) are shown in Fig.~\ref{fig.mass} (right). We see signs that with increasing temperature $V_R(T,r)$ becomes flatter at large $r$, as expected from color screening effect.  However, the temperature dependence of $V_R(T,r)$ is very mild between $T\approx151-334$~MeV, and closely approximates its vacuum counterpart. In the same temperature range, $V_I(T,r)$ show significant monotonic increase both with temperature and distance. In the succeeding section, we also performed similar analyses using temperature-independent DNNs and polynomials in $r$ to represent the functional form of the potential. We obtained consistent results with our original implementation. While the $r$-dependence could be retrieved by simpler parameterizations, the $T$-dependence is non-trivial to be captured in an unbiased manner. Using DNN to represent the potentially non-linear temperature and distance dependence we reconstructed the potential with reliable uncertainty for the temperatures in the region $T \in [0, 334]$~MeV. These results are shown in Fig.~\ref{fig.potential_DNN_3D}.

\begin{figure}[!hbtp]\centering
\includegraphics[width=0.45\textwidth]{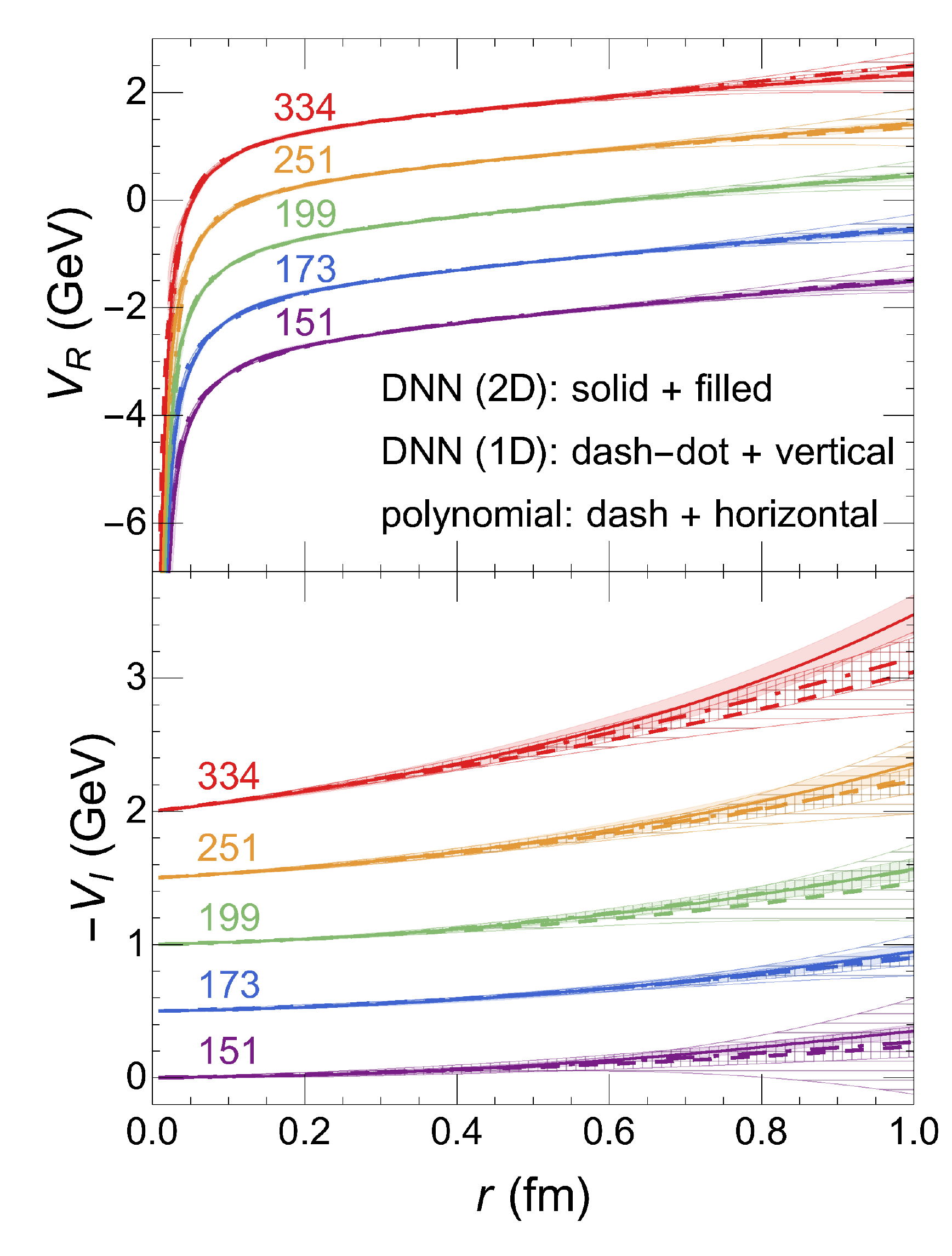}
\caption{Real (top) and imaginary (bottom) interaction potentials versus quark distance $r$ extracted by $T$-dependent DNNs (as known as DNN(2D), solid line with filled uncertainty bands), $T$-independent DNNs (as known as DNN(1D), dash-dotted line with vertical hashed uncertainty bands) and polynomial parameterizations (dashed line with horizontal hashed uncertainty bands);
Different colors respectively represent temperature $T = 151$ (purple), $173$ (blue), $199$ (green), $251 $(orange), and $334$ (red)~MeV which are also ordered from bottom to top. For better visualization, the curves are shifted vertically. The error bands represent the $68\%$($1\sigma$) confidence interval.
\label{fig.potential_compare}}
\end{figure}
\begin{figure*}[!hbtp]\centering
\includegraphics[width=0.8\textwidth]{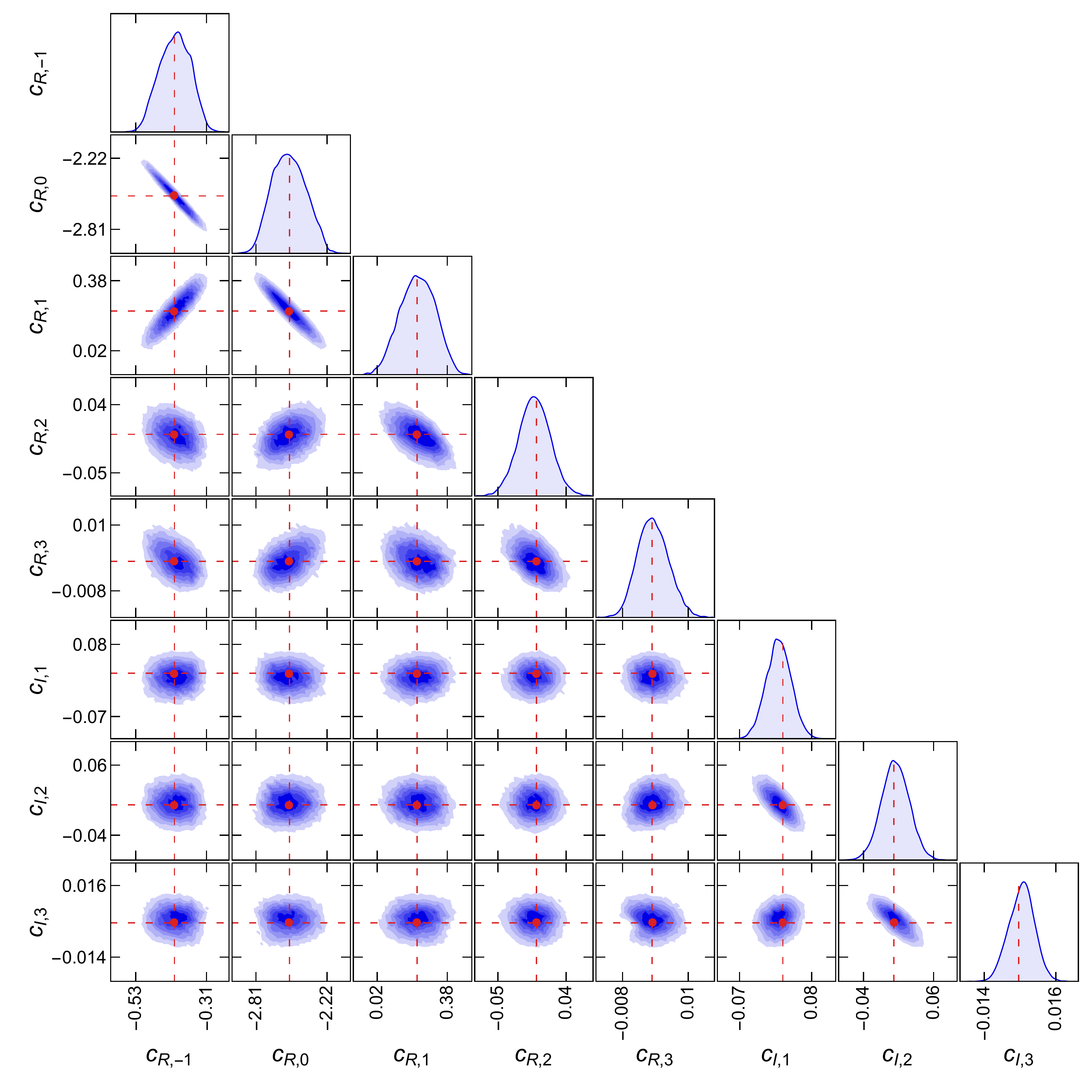}
\caption{Marginal likelihood distribution for the polynomial coefficients at $T=151$~MeV. Red dots and dashed lines indicate the most optimal parameter set. The unit of the coefficients are listed in \protect{Table~\ref{tab.poly_parameter}}.
\label{fig.marginal}}
\end{figure*}

\section{Consistency Tests}\label{sec.consistency}

\subsection{Temperature-Independent Parameterizations with DNN or Polynomials}\label{sec.consistency_potential}

In order to examine the consistency of potentials obtained in Sec.~\ref{sec.potential_dnn} from aspects of parameterizations, we performed two independent tests with two different parameterization schemes both being temperature-independent: a) the DNNs with only distance $r$ to be the input argument, and b) polynomial parameterization, of the real and imaginary potentials.
The flow chart of such model training is similar to what has been discussed in Sec.~\ref{sec.methodology} and shown in Fig.~\ref{fig.flow_chart}. 
The only modification one needs to apply is to replace the $(T,r) \to V_{R/I}(T,r)$ DNNs (the left upper corner of Fig.~\ref{fig.flow_chart}) by the $(r) \to V_{R/I}(r)$ DNNs for the parameterization scheme (a), while for scheme (b), by the polynomial parameterization as follows,
\begin{align}
\begin{split}
V_R(r) = \sum_{i=-1}^{3} c_{R,i} \, r^i  ,\qquad
V_I(r) = - \sum_{i=1}^{3} c_{I,i} \, r^i .
\end{split}\label{eq.V_polynomial}
\end{align}
For the polynomial parameterization, we have taken into account a physical conjecture that $V_I(r)$ vanishes when $r\to0$. Also, we applied our prior belief that higher-order coefficients shall not be large, hence we employed the regularizer $J_\text{reg} = \sum_i \lambda_i (c_{R,i}^2 + c_{I,i}^2)$, with $\lambda_{i\leq1} = 0$, $\lambda_{2} = 1000/\text{GeV}^6$ and $\lambda_{3} = 5000/\text{GeV}^8$.
In the $T$-independent DNNs, we used simpler network structures, i.e. $1\times16\times16\times1$ for the real potential, and $1\times4\times4\times1$ for the imaginary part, with regularizer $\lambda_R=0.1$ for the former while $\lambda_I=0.001$ for the latter which are consistent with the complexity of the corresponding network structure as well. Similar to the preceding section, we used \texttt{elu}(\texttt{linear}) activation functions for the hidden(output) layers in the DNNs. We have also applied the $\lim_{r\to0}V_I=0$ conjecture by letting $V_I(r) = r\; V_I^\text{DNN}(r)$.
Noting that lattice QCD simulation~\cite{Larsen:2019zqv} provides independent sets of bottomonia mass and width at different temperatures, we perform the Bayesian analysis at each temperature point separately, and extract the optimal parameter set together with the corresponding likelihood functions.
Again, the prior distribution is defined according to the regularizer.
In Fig.~\ref{fig.potential_compare}, we compare the complex potentials obtained in such three schemes, and find nice agreement between them.

\begin{table}[!hbtp]
\begin{tabular}{l|ccccc}
\hline\hline
T (MeV) & 151 & 173 & 199 & 251 & 334 \\
\hline
$c_{R,-1}$
 &	-0.41 &	-0.41 &	-0.41 &	-0.40 &	-0.39	\\
$c_{R,0}~(\text{GeV})$
 &	-2.53 &	-2.53 &	-2.53 &	-2.54 &	-2.55	\\
$c_{R,1}~(\text{GeV}^2)$
 &	0.22 &	0.22 &	0.22 &	0.21 &	0.20	\\
$10^3\times c_{R,2}~(\text{GeV}^3)$
 &	0.84 &	0.47 &	-0.31 &	-2.26 &	-0.14	\\
$10^4\times c_{R,3}~(\text{GeV}^4)$
 &	0.82 &	0.43 &	-0.24 &	-2.82 &	1.38	\\
$10\times c_{I,1}~(\text{GeV}^2)$
 &	0.20 &	0.31 &	0.34 &	0.65 &	1.46	\\
$10^2\times c_{I,2}~(\text{GeV}^3)$
 &	0.34 &	0.58 &	0.68 &	1.08 &	0.76	\\
$10^3\times c_{I,3}~(\text{GeV}^4)$
 &	0.40 &	0.96 &	0.98 &	1.28 &	1.00	\\
\hline\hline
\end{tabular}
\caption{Optimal values of the polynomial coefficients at different temperatures.
\label{tab.poly_parameter}}
\end{table}

The polynomial parameterization scheme has in total eight coefficients. Hence, it is possible to list the optimal values of the polynomial coefficients at different temperatures as well as their marginal likelihood distribution, for better visualizing the fitting quality and the correlation between different parameters.
We list the optimal coefficient set in Table~\ref{tab.poly_parameter}, and show the marginal likelihood distribution for $T = 151$~MeV in Fig.~\ref{fig.marginal}. One can see the strong correlation between ``neighboring'' coefficients.

In addition, we note that while the Schr\"odinger equation is solved within the range $r \in [0,2]$~fm, the potentials can be well-constrained only within the range $r\leq1$~fm.
Such a limitation can be well understood:  the wave-functions of such bound states concentrate in the $r\leq1$~fm region. According to the Hellmann--Feynman theorem, the mass spectrum is not sensitive to the potential in the $1< r \leq 2$~fm region, and it can hardly constrain the potential for that region.

To conclude, we emphasize that although the polynomial parameterization provides a relatively simple picture of the distance dependence of potential, it has difficulties in describing the temperature dependence without enough priors. Thus, one needs to generically employ an unbiased but robust parameterization scheme to obtain the 2D potential depending on both distance and temperature, for which DNNs provide the proper parameterization with moreover the well-developed optimization approach in practice.

\begin{figure*}[!hbtp]\centering
\includegraphics[width=0.35\textwidth]{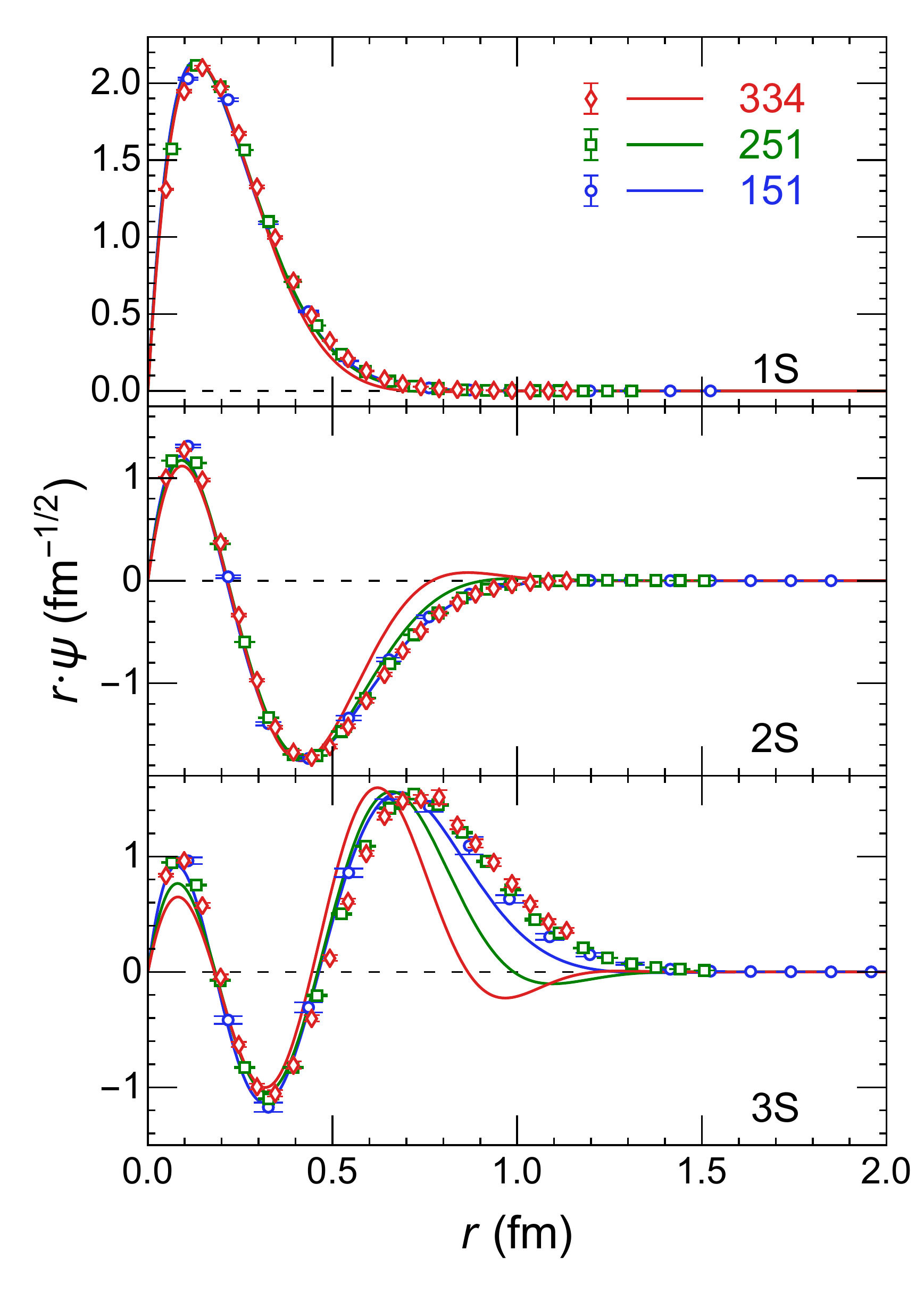}\qquad\qquad
\includegraphics[width=0.35\textwidth]{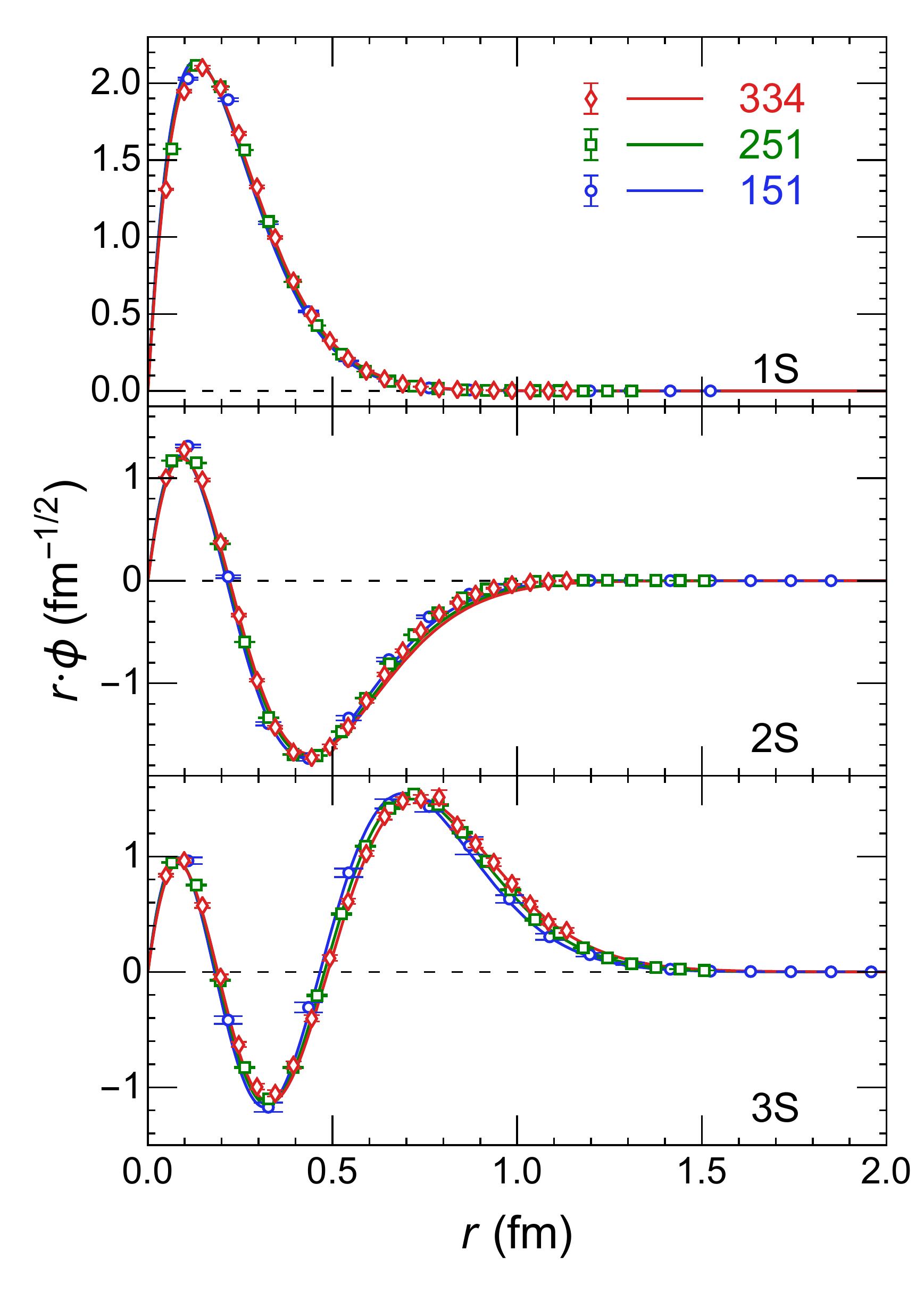}
\caption{(Left) Comparison of the real part of finite temperature wave-functions(curves) and Bethe--Salpeter amplitudes(symbol).
Results at $T$=151, 251, and 334 MeV are respectively colored in blue, green, and red. (Right) Same as Left but for ``pseudo-wave-function'' obtained only from the real potential. See text for explanation.
\label{fig.wavfunc_complete}}
\end{figure*}
\subsection{Comparing the Wave-Functions with the Bethe--Salpeter Amplitude at Finite Temperature}\label{sec.consistency_wavfunc}

In this subsection, we compare the finite temperature wave-functions with the corresponding Bethe--Salpeter(BS) amplitudes from the lattice QCD calculation~\cite{Larsen:2020rjk}, which is obtained consistently with the masses and widths~\cite{Larsen:2019zqv}. With such complementary information, the comparison serve as an independent test of the finite temperature potential. We compare the real part of wave-functions at different temperatures in Fig.~\ref{fig.wavfunc_complete} (left). We observe mild temperature dependence of the BS amplitudes, while the wave-functions are obviously different at higher temperature. We note that the real part of the interaction potentials show weak dependence on temperature, and the change of wave-function is dominated by the imaginary potential. 

\begin{figure}[!hbtp]\centering
\includegraphics[width=0.35\textwidth]{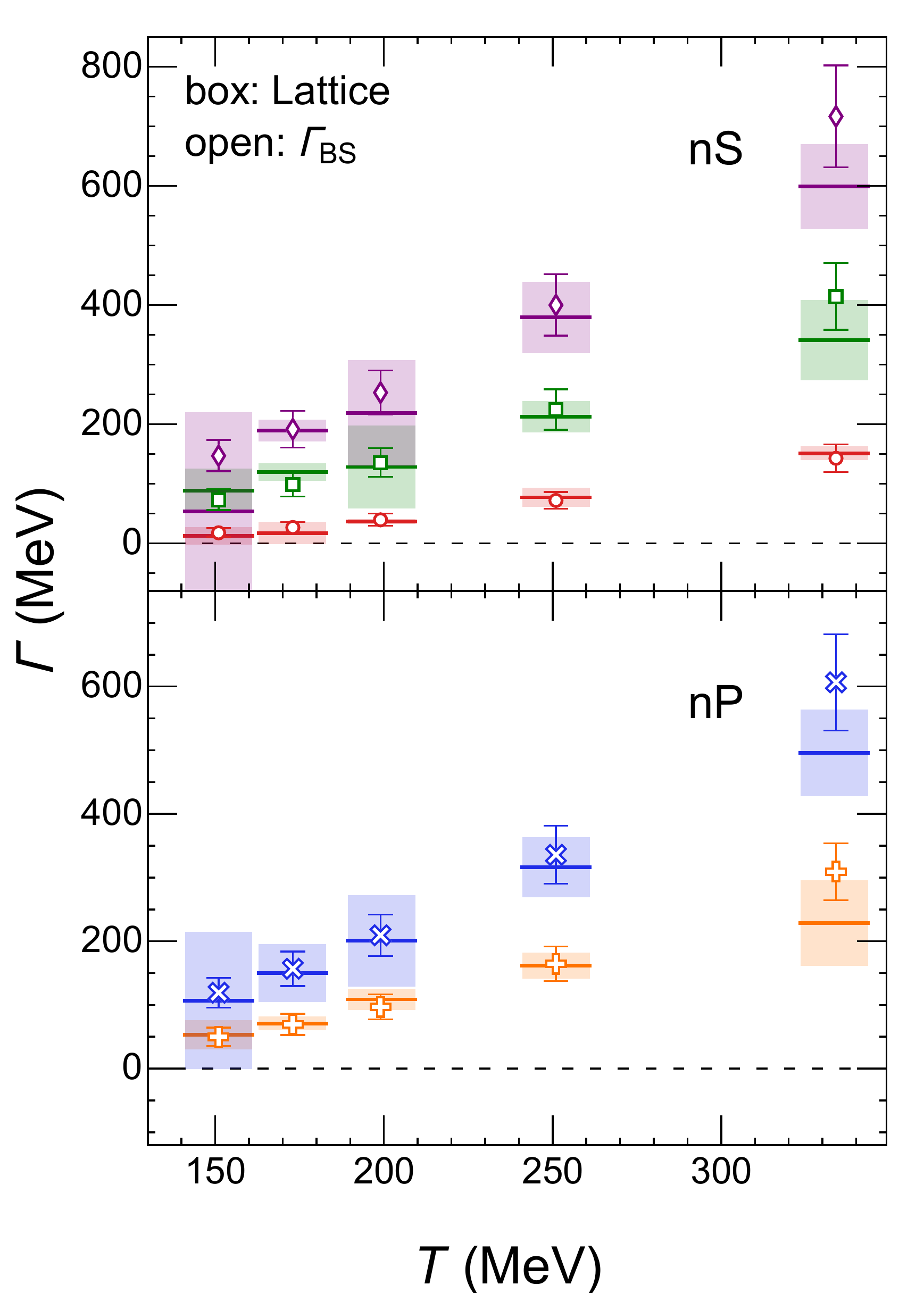}
\caption{Expectation of the thermal width based on the Bethe--Salpeter amplitudes.
\label{fig.Width_BS}}
\end{figure}

As noted in Ref.~\cite{Larsen:2020rjk}, due their non-trivial Euclidean-time dependence, the BSAs at $T>0$ fail to capture the thermal broadening of the states, rather resemble the vacuum wavefunctions. Consequently, we solve the ``pseudo-wave-functions'', denoted as $\phi$, according to the real potential in Fig.~\ref{fig.potential_DNN_3D},
\begin{equation}
-\frac{\nabla^2}{m_b} \phi_n + V_R(T,r) \phi_n = \widetilde{E}_{n} \phi_n,
\label{eq.bs}
\end{equation}
and compare them with the BS amplitude in Fig.~\ref{fig.wavfunc_complete} (right), and find excellent agreement especially regarding the large-$r$ tail at different temperatures.
Such comparison serve as an independent test of the real part of the interaction potential at finite temperature.
In particular, the tail behavior of the wave-functions is sensitive to the flatness of the potential at $r \gtrsim 0.5~\text{fm}$. 
The excellent agreement shown in Fig.~\ref{fig.wavfunc_complete} (right), especially for the 3S state at all temperatures, confirms the weak screening effect observed in the real part of the potential.

It would be interesting to check the role of the complex wave-functions. In the main content, we solve the complex wave-functions according to the complex potential and obtain the mass and width from the complex energy eigenvalue. An alternative method is to treat the imaginary potential as a perturbation, and extract the wave-function ($\phi_n$) according to the real part of the potential, as in Eq.~\eqref{eq.bs}. 
As has been discussed above, $\phi_n$'s are equivalent to the BS amplitudes in the lattice calculations, and we refer to it as the BS amplitudes. Then, we compute the thermal width as the BS-expectation of the imaginary potential,
\begin{align}
    \Gamma_\text{BS} \equiv 
    - \int |\phi_\text{BS}(r)|^2 V_I(r) r^2 \mathrm{d}r.
\end{align}
In Fig.~\ref{fig.Width_BS}, we compare the $\Gamma_\text{BS}$, from the perturbative treatment, with the complete thermal width from the lattice result. We find that at lower temperatures, at which $\Gamma$'s are small, the perturbation results agree well with the complete ones, whereas, at high temperature, e.g., $334$~MeV, $\Gamma_\text{BS}$ are slightly, but systematically, higher than the lattice results.

\section{Conclusion and Discussion} \label{sec.conclusion}

In this work, we report a surprising empirical finding: LQCD results~\cite{Larsen:2019zqv} for the masses and thermal widths of up to $3S$ and $2P$ bottomonium states in QGP admits a consistent quantum mechanical description based on an complex-valued potential and non-relativistic Schr\"odinger equation.
By coupling the Schr\"odinger equation to a DNN, we introduced a novel method for unbiased extractions of the real and imaginary parts of the heavy quark potential, and invoked Bayesian inference to quantify the potential uncertainties in a non-local fashion. With such a model-independent method, we obtained the empirical $V_R(T,r)$ and $V_I(T,r)$ for $r\lesssim1$~fm and $T\lesssim334$~MeV, which can successfully map the QCD spectrum of the lowest-lying bottomonium states in QGP to a quantum mechanical system.

The heavy quark potential obtained here renders an empirical mapping from the masses and thermal widths of bottomonium states at finite temperature to an effective quantum mechanics framework, based on the LQCD calculations of bottomonium state using a 2+1 flavor dynamical gauge field background with nearly physical values of up, down, and strange quark masses.  
Direct quantitative comparison with extant LQCD calculations of static quark  potentials~\cite{Rothkopf:2011db,Burnier:2014ssa,Burnier:2015tda,Bala:2019cqu} is difficult. However, our result for the heavy quark potential is qualitatively different from those potentials. Unlike the previous studies, the $V_R$ obtained in this work show very little signs of color-electric Debye screening for $r\lesssim 1$~fm for the entire temperature range $T \in [0, 334]$~MeV. The $V_I$ here is much larger in magnitude and increases more rapidly, both with $T$ and $r$, than the one-loop HTL-motivated extractions. On the other hand, it is reassuring that the potential obtained here is quantitatively consistent with the very recent LQCD calculations~\cite{Bala:2021fkm} on the peak position and the width of the Gaussian-form spectral function, as functions of the separation $r$ and temperature $T$. Agreement -- in the sense of the strongly-coupled behavior as well as large magnitude of imaginary potential -- is observed in the comparison with heavy-quark potentials computed in the $T$-matrix approach~\cite{Liu:2016ysz,Liu:2017qah} and phenomenologically extracted from bottomonium data~\cite{Du:2017qkv,Du:2019tjf,Strickland:2011mw,Islam:2020bnp}, despite of some difference in the exact value. Meanwhile, it might be worth noting that large imaginary part, are also seen in Ref.~\cite{Bala:2019cqu}, and is not ruled out by Refs.~\cite{Burnier:2014ssa,Burnier:2015tda}, since the large errors for their results.
It would be very interesting to see the phenomenological consequences~\cite{Islam:2020gdv} of this heavy quark potential, model-independently extracted from the non-perturbative LQCD calculations. 

Further, we carried out detailed comparisons of the real parts of the wavefunctions with the BSAs obtained from LQCD calculations~\cite{Larsen:2020rjk}. As noted in Ref.~\cite{Larsen:2020rjk}, due to their non-trivial Euclidean-time dependence, the BSAs at $T>0$ fail to capture the thermal broadening of the states and resemble the vacuum wavefunctions. Our comparisons seem to support this picture. While the real parts of the actual wavefunctions show deviations from the BSAs at large $r$, the ``pseudo'' wavefunctions obtained using only $V_R$ (with $V_I=0$) reproduce the BSAs. Furthermore, we also find that the ``pseudo'' thermal widths $\Gamma_\text{BS}(T)= -\int |\psi_\text{BS}(T,r)|^2 V_I(T,r) r^2 \mathrm{d}r \approx \Gamma(T)$, suggest that $V_I(T,r)$ might be considered as a perturbation on top of an approximately vacuum-like excitation. Based on our results, one might speculate that, for phenomenologically relevant temperatures $T\lesssim334$~MeV, bottomonia are approximately vacuum-like excitation but of very short lifetimes that are inversely proportional to their large thermal widths. At high enough temperatures, we anticipate that this speculative picture would smoothly turn over to the more conventional picture based on quarkonia melting due to color-electric Debye screening~\cite{Satz:2005hx,Zhao:2020jqu,Mocsy:2007jz} and perturbative Landau damping~\cite{Laine:2006ns,Beraudo:2007ky}.

\vspace{1cm}
\emph{Acknowledgements.---} 
The authors thank Min He, Matthew Heffernan, Rasmus Larsen, Simon Mak, Peter Petreczky, Ralf Rapp, Alexander Rothkopf, Michael Strickland, and Nan Su for many insightful discussions.

This material is based upon work supported by: (i) The NSFC under grant Nos. 11890712 and 12075129 and Guangdong Major Project of Basic and Applied Basic Research No. 2020B0301030008 (J.Z. and P.Z.);  (ii) The Natural Sciences and Engineering Research Council of Canada (S.S.); (iii) The Fonds de recherche du Qu\'ebec - Nature et technologies (FRQNT) through the Programmede Bourses d'Excellencepour \'Etudiants \'Etrangers (PBEEE) scholarship (S.S.); (iv) The BMBF funding under the ErUM-Data project and the AI grant at FIAS of SAMSON AG, Frankfurt (K.Z.); (v) The GPU Grant of the NVIDIA Corporation (K.Z.); (vi) The U.S. Department of Energy, Office of Science, Office of Nuclear Physics through the Contract No.~DE-SC0012704 (S.M.); (vii) The U.S. Department of Energy, Office of Science, Office of Nuclear Physics and Office of Advanced Scientific Computing Research, within the framework of Scientific Discovery through Advance Computing (SciDAC) award Computing the Properties of Matter with Leadership Computing Resources (S.M.).

\begin{appendix}
\section{About the Imaginary Energy and Width of the Spectral Function}\label{sec.spectral_function}
In this work, we have assumed that the width obtained in Ref.~\cite{Larsen:2019zqv} is the imaginary part of the energy eigenvalue, $\Gamma_n = |\mathrm{Im}[E_n]|$. Such relation is not obvious as the lattice QCD results assume a Gaussian form for the spectral function. In this section, we investigate the relation between the imaginary part of the energy and the Gaussian width.

Following the procedure in Ref.~\cite{Burnier:2007qm}, we start the correlation as a spatial Dirac-$\delta$ function,
\begin{align}
C^{>}(0,\mathbf r) = \delta^{(3)} (\mathbf r)
\end{align}
and evolve the forward and backward propagator according to the Hamiltonian
\begin{align}\left\{\begin{array}{cc}
\widehat{H} \; C^{>}(t,\mathbf r) = i \partial_t C^{>}(t,\mathbf r), &\qquad t>0,\\
\widehat{H}^\dagger C^{>}(t,\mathbf r) = i \partial_t C^{>}(t,\mathbf r), &\qquad t<0.
\end{array}\right.
\end{align}
We note that the eigenfunctions of the Hamiltonian, $\{\psi_n\}$, form a complete set of the Hilbert space.  
Hence, we expand the Dirac-$\delta$ function in series of the wave-functions,
\begin{align}
c_n \equiv&\; \int \mathrm{d}^3\mathbf r C^{>}(0,\mathbf r) \psi_n^* (\mathbf r)
 = \psi^*_n(0),
\end{align}
and the time dependence of the correlation function can be express as the super position of different eigenmodes:
\begin{align}
C^{>}(t,\mathbf r) = \left\{\begin{array}{cc}
\sum_n c_n e^{-i E_n t} \times \psi_n(\mathbf r), &\qquad t>0,\\
\sum_n c_n^* e^{-i E_n^* t} \times \psi_n^*(\mathbf r), &\qquad t<0,
\end{array}\right.
\end{align}
where $\sum_n$ denotes summation over all bound-states, as well as the integral over scattering continuum when applicable.

With these, we find that the spectral function
\begin{align}
\rho(\omega) \equiv&\; \int_{-\infty}^{+\infty} \mathrm{d}t \; e^{i \omega t} C^{>}(t, 0) \\
=&\;  \sum_n
	 \frac{ -2 |\psi_n(0)|^2 \, \mathrm{Im}[E_{n}]}
	 {(\omega - \mathrm{Re}[E_{n}])^2 + (\mathrm{Im}[E_{n}])^2}
\end{align}
takes the Lorentzian form, with the Lorentzian width being the imaginary energy eigenvalue.
To guarantee that the amplitude $|\psi(t)\rangle$ decays rather than explodes, the imaginary energy should always be non-positive, and 
\begin{equation}
\Gamma_n^\mathrm{Lor} = -\mathrm{Im}[E_n].
\end{equation}

On the other hand, in Refs.~\cite{Larsen:2019bwy,Larsen:2019zqv,Larsen:2020rjk}, the masses and widths are extracted under the assumption of a Gaussian  spectral function
\begin{equation}
\rho (\omega) \propto \sum_n \exp[-\frac{(\omega - M_n)^2}{2\Gamma_n^2}]\,.
\end{equation}
While the mass can be uniquely defined as the peak position, there is no obvious way to map the Lorentzian width with the Gaussian one. In this work, we take $\Gamma_n^\mathrm{Lor} = \Gamma_n^\mathrm{Gau}$, as they both represent the characteristic width. In a different point of view, otherwise, one might match them according to the half-maximum of the spectral function. In the latter case, one would find $\Gamma_n^\mathrm{Lor} = \sqrt{2\ln2}\, \Gamma_n^\mathrm{Gau}$, hence $\mathrm{Im}[E_n] = -1.18 \Gamma_n^\mathrm{Gau}$. If taking the latter mapping, the decay width of the bottomonium states shall be multiplied by a factor of $1.18$, and the extracted $V_I$ shall increase by $\sim18\%$.

\begin{figure}[!hbtp]\centering
\includegraphics[width=0.35\textwidth]{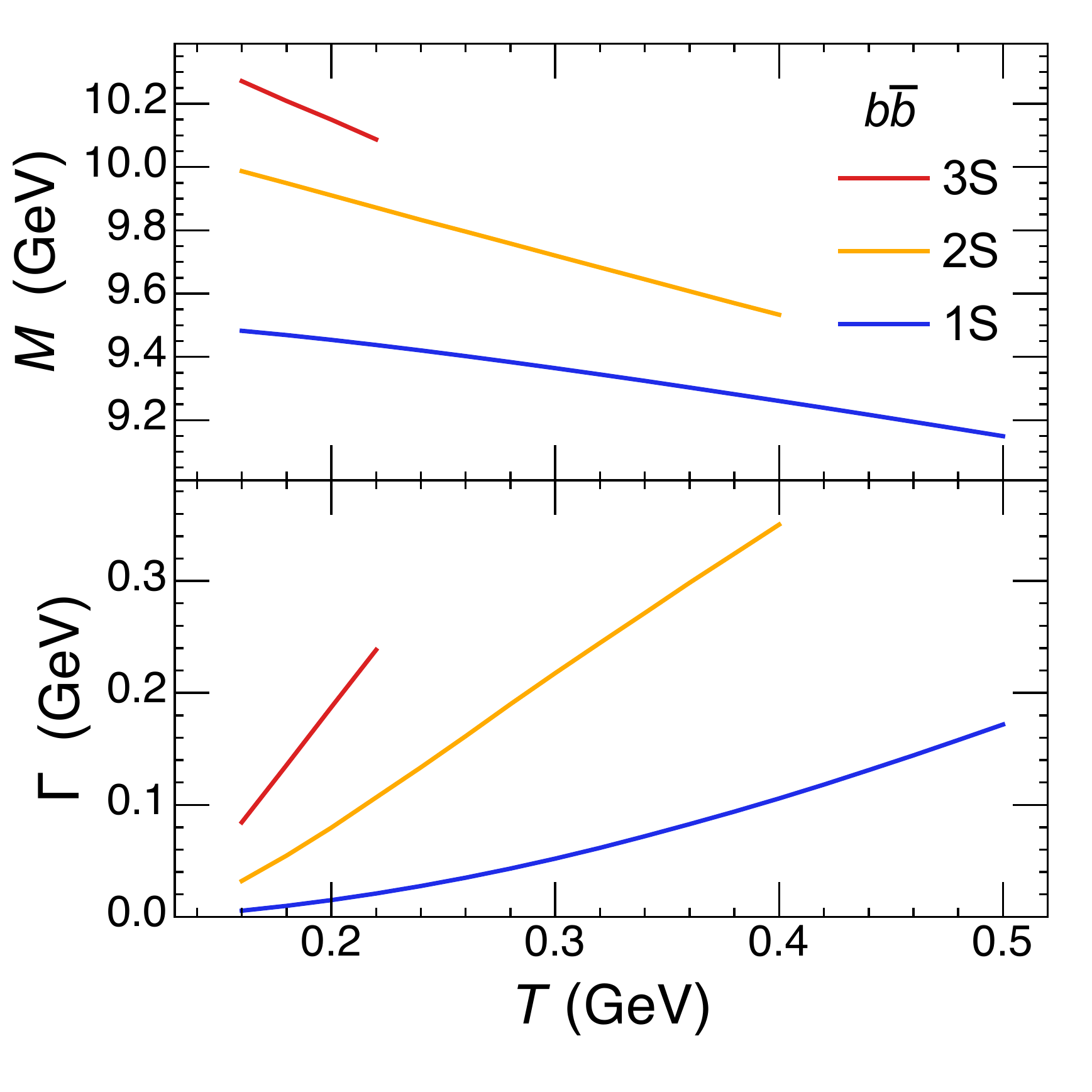}
\caption{Masses and thermal widths of bottomonium states by solving bound-state problem of the Schr\"odinger Hamiltonian with complex-valued potential listed in~\cite{Lafferty:2019jpr}.
\label{fig.schroedinger_test}}
\end{figure}
In this Appendix, we have shown, analytically, that mass and thermal widths obtained by solving the evolution of correlators shall be equivalent to those from solving the energy bound states. One may wonder if such an equivalence would still hold in numerical procedures, especially given the possible uncertainty of fitting the peaks of the spectral function. To answer this question, we perform a numerical verification as follows.
We start from the complex-valued potential listed in~\cite{Lafferty:2019jpr}, solve the bound-state problem for the Schr\"odinger Hamiltonian according to the inverse power method~\cite{1994JCoPh.115..470C}, and obtain the complex-valued energy eigenvalues for various bottomonium states. Their masses and thermal widths --- respectively being the real and imaginary part of the energy eigenvalues --- are shown in Fig.~\ref{fig.schroedinger_test}. We find our results are consistent with those of~\cite{Lafferty:2019jpr}, which computes the masses and thermal widths from the correlator evolution as outlined in~\cite{Burnier:2007qm}.
With both the analytical derivation and numerical verifications, we conclude that these two procedures are equivalent in computing the mass and thermal widths from Schr\"odinger Hamiltonian with time-independent potentials.

\end{appendix}
\bibliographystyle{apsrev4-1.bst}
\bibliography{Ref}

\begin{thebibliography}{75}%
\makeatletter
\providecommand \@ifxundefined [1]{%
 \@ifx{#1\undefined}
}%
\providecommand \@ifnum [1]{%
 \ifnum #1\expandafter \@firstoftwo
 \else \expandafter \@secondoftwo
 \fi
}%
\providecommand \@ifx [1]{%
 \ifx #1\expandafter \@firstoftwo
 \else \expandafter \@secondoftwo
 \fi
}%
\providecommand \natexlab [1]{#1}%
\providecommand \enquote  [1]{``#1''}%
\providecommand \bibnamefont  [1]{#1}%
\providecommand \bibfnamefont [1]{#1}%
\providecommand \citenamefont [1]{#1}%
\providecommand \href@noop [0]{\@secondoftwo}%
\providecommand \href [0]{\begingroup \@sanitize@url \@href}%
\providecommand \@href[1]{\@@startlink{#1}\@@href}%
\providecommand \@@href[1]{\endgroup#1\@@endlink}%
\providecommand \@sanitize@url [0]{\catcode `\\12\catcode `\$12\catcode
  `\&12\catcode `\#12\catcode `\^12\catcode `\_12\catcode `\%12\relax}%
\providecommand \@@startlink[1]{}%
\providecommand \@@endlink[0]{}%
\providecommand \url  [0]{\begingroup\@sanitize@url \@url }%
\providecommand \@url [1]{\endgroup\@href {#1}{\urlprefix }}%
\providecommand \urlprefix  [0]{URL }%
\providecommand \Eprint [0]{\href }%
\providecommand \doibase [0]{http://dx.doi.org/}%
\providecommand \selectlanguage [0]{\@gobble}%
\providecommand \bibinfo  [0]{\@secondoftwo}%
\providecommand \bibfield  [0]{\@secondoftwo}%
\providecommand \translation [1]{[#1]}%
\providecommand \BibitemOpen [0]{}%
\providecommand \bibitemStop [0]{}%
\providecommand \bibitemNoStop [0]{.\EOS\space}%
\providecommand \EOS [0]{\spacefactor3000\relax}%
\providecommand \BibitemShut  [1]{\csname bibitem#1\endcsname}%
\let\auto@bib@innerbib\@empty
\bibitem [{\citenamefont {Matsui}\ and\ \citenamefont
  {Satz}(1986)}]{Matsui:1986dk}%
  \BibitemOpen
  \bibfield  {author} {\bibinfo {author} {\bibfnamefont {T.}~\bibnamefont
  {Matsui}}\ and\ \bibinfo {author} {\bibfnamefont {H.}~\bibnamefont {Satz}},\
  }\href {\doibase 10.1016/0370-2693(86)91404-8} {\bibfield  {journal}
  {\bibinfo  {journal} {Phys. Lett. B}\ }\textbf {\bibinfo {volume} {178}},\
  \bibinfo {pages} {416} (\bibinfo {year} {1986})}\BibitemShut {NoStop}%
\bibitem [{\citenamefont {Karsch}\ \emph {et~al.}(1988)\citenamefont {Karsch},
  \citenamefont {Mehr},\ and\ \citenamefont {Satz}}]{Karsch:1987pv}%
  \BibitemOpen
  \bibfield  {author} {\bibinfo {author} {\bibfnamefont {F.}~\bibnamefont
  {Karsch}}, \bibinfo {author} {\bibfnamefont {M.~T.}\ \bibnamefont {Mehr}}, \
  and\ \bibinfo {author} {\bibfnamefont {H.}~\bibnamefont {Satz}},\ }\href
  {\doibase 10.1007/BF01549722} {\bibfield  {journal} {\bibinfo  {journal} {Z.
  Phys. C}\ }\textbf {\bibinfo {volume} {37}},\ \bibinfo {pages} {617}
  (\bibinfo {year} {1988})}\BibitemShut {NoStop}%
\bibitem [{\citenamefont {Blaizot}\ and\ \citenamefont
  {Ollitrault}(1996)}]{Blaizot:1996nq}%
  \BibitemOpen
  \bibfield  {author} {\bibinfo {author} {\bibfnamefont {J.-P.}\ \bibnamefont
  {Blaizot}}\ and\ \bibinfo {author} {\bibfnamefont {J.-Y.}\ \bibnamefont
  {Ollitrault}},\ }\href {\doibase 10.1103/PhysRevLett.77.1703} {\bibfield
  {journal} {\bibinfo  {journal} {Phys. Rev. Lett.}\ }\textbf {\bibinfo
  {volume} {77}},\ \bibinfo {pages} {1703} (\bibinfo {year} {1996})},\ \Eprint
  {http://arxiv.org/abs/hep-ph/9606289} {arXiv:hep-ph/9606289} \BibitemShut
  {NoStop}%
\bibitem [{\citenamefont {Braun-Munzinger}\ and\ \citenamefont
  {Stachel}(2000)}]{BraunMunzinger:2000px}%
  \BibitemOpen
  \bibfield  {author} {\bibinfo {author} {\bibfnamefont {P.}~\bibnamefont
  {Braun-Munzinger}}\ and\ \bibinfo {author} {\bibfnamefont {J.}~\bibnamefont
  {Stachel}},\ }\href {\doibase 10.1016/S0370-2693(00)00991-6} {\bibfield
  {journal} {\bibinfo  {journal} {Phys. Lett. B}\ }\textbf {\bibinfo {volume}
  {490}},\ \bibinfo {pages} {196} (\bibinfo {year} {2000})},\ \Eprint
  {http://arxiv.org/abs/nucl-th/0007059} {arXiv:nucl-th/0007059} \BibitemShut
  {NoStop}%
\bibitem [{\citenamefont {Digal}\ \emph {et~al.}(2001)\citenamefont {Digal},
  \citenamefont {Petreczky},\ and\ \citenamefont {Satz}}]{Digal:2001ue}%
  \BibitemOpen
  \bibfield  {author} {\bibinfo {author} {\bibfnamefont {S.}~\bibnamefont
  {Digal}}, \bibinfo {author} {\bibfnamefont {P.}~\bibnamefont {Petreczky}}, \
  and\ \bibinfo {author} {\bibfnamefont {H.}~\bibnamefont {Satz}},\ }\href
  {\doibase 10.1103/PhysRevD.64.094015} {\bibfield  {journal} {\bibinfo
  {journal} {Phys. Rev. D}\ }\textbf {\bibinfo {volume} {64}},\ \bibinfo
  {pages} {094015} (\bibinfo {year} {2001})},\ \Eprint
  {http://arxiv.org/abs/hep-ph/0106017} {arXiv:hep-ph/0106017} \BibitemShut
  {NoStop}%
\bibitem [{\citenamefont {Grandchamp}\ \emph {et~al.}(2004)\citenamefont
  {Grandchamp}, \citenamefont {Rapp},\ and\ \citenamefont
  {Brown}}]{Grandchamp:2003uw}%
  \BibitemOpen
  \bibfield  {author} {\bibinfo {author} {\bibfnamefont {L.}~\bibnamefont
  {Grandchamp}}, \bibinfo {author} {\bibfnamefont {R.}~\bibnamefont {Rapp}}, \
  and\ \bibinfo {author} {\bibfnamefont {G.~E.}\ \bibnamefont {Brown}},\ }\href
  {\doibase 10.1103/PhysRevLett.92.212301} {\bibfield  {journal} {\bibinfo
  {journal} {Phys. Rev. Lett.}\ }\textbf {\bibinfo {volume} {92}},\ \bibinfo
  {pages} {212301} (\bibinfo {year} {2004})},\ \Eprint
  {http://arxiv.org/abs/hep-ph/0306077} {arXiv:hep-ph/0306077} \BibitemShut
  {NoStop}%
\bibitem [{\citenamefont {Song}\ \emph {et~al.}(2011)\citenamefont {Song},
  \citenamefont {Han},\ and\ \citenamefont {Ko}}]{Song:2011xi}%
  \BibitemOpen
  \bibfield  {author} {\bibinfo {author} {\bibfnamefont {T.}~\bibnamefont
  {Song}}, \bibinfo {author} {\bibfnamefont {K.~C.}\ \bibnamefont {Han}}, \
  and\ \bibinfo {author} {\bibfnamefont {C.~M.}\ \bibnamefont {Ko}},\ }\href
  {\doibase 10.1103/PhysRevC.84.034907} {\bibfield  {journal} {\bibinfo
  {journal} {Phys. Rev. C}\ }\textbf {\bibinfo {volume} {84}},\ \bibinfo
  {pages} {034907} (\bibinfo {year} {2011})},\ \Eprint
  {http://arxiv.org/abs/1103.6197} {arXiv:1103.6197 [nucl-th]} \BibitemShut
  {NoStop}%
\bibitem [{\citenamefont {Du}\ and\ \citenamefont {Rapp}(2015)}]{Du:2015wha}%
  \BibitemOpen
  \bibfield  {author} {\bibinfo {author} {\bibfnamefont {X.}~\bibnamefont
  {Du}}\ and\ \bibinfo {author} {\bibfnamefont {R.}~\bibnamefont {Rapp}},\
  }\href {\doibase 10.1016/j.nuclphysa.2015.09.006} {\bibfield  {journal}
  {\bibinfo  {journal} {Nucl. Phys. A}\ }\textbf {\bibinfo {volume} {943}},\
  \bibinfo {pages} {147} (\bibinfo {year} {2015})},\ \Eprint
  {http://arxiv.org/abs/1504.00670} {arXiv:1504.00670 [hep-ph]} \BibitemShut
  {NoStop}%
\bibitem [{\citenamefont {Liu}\ \emph {et~al.}(2011)\citenamefont {Liu},
  \citenamefont {Chen}, \citenamefont {Xu},\ and\ \citenamefont
  {Zhuang}}]{Liu:2010ej}%
  \BibitemOpen
  \bibfield  {author} {\bibinfo {author} {\bibfnamefont {Y.}~\bibnamefont
  {Liu}}, \bibinfo {author} {\bibfnamefont {B.}~\bibnamefont {Chen}}, \bibinfo
  {author} {\bibfnamefont {N.}~\bibnamefont {Xu}}, \ and\ \bibinfo {author}
  {\bibfnamefont {P.}~\bibnamefont {Zhuang}},\ }\href {\doibase
  10.1016/j.physletb.2011.01.026} {\bibfield  {journal} {\bibinfo  {journal}
  {Phys. Lett. B}\ }\textbf {\bibinfo {volume} {697}},\ \bibinfo {pages} {32}
  (\bibinfo {year} {2011})},\ \Eprint {http://arxiv.org/abs/1009.2585}
  {arXiv:1009.2585 [nucl-th]} \BibitemShut {NoStop}%
\bibitem [{\citenamefont {Zhou}\ \emph {et~al.}(2014)\citenamefont {Zhou},
  \citenamefont {Xu}, \citenamefont {Xu},\ and\ \citenamefont
  {Zhuang}}]{Zhou:2014kka}%
  \BibitemOpen
  \bibfield  {author} {\bibinfo {author} {\bibfnamefont {K.}~\bibnamefont
  {Zhou}}, \bibinfo {author} {\bibfnamefont {N.}~\bibnamefont {Xu}}, \bibinfo
  {author} {\bibfnamefont {Z.}~\bibnamefont {Xu}}, \ and\ \bibinfo {author}
  {\bibfnamefont {P.}~\bibnamefont {Zhuang}},\ }\href {\doibase
  10.1103/PhysRevC.89.054911} {\bibfield  {journal} {\bibinfo  {journal} {Phys.
  Rev. C}\ }\textbf {\bibinfo {volume} {89}},\ \bibinfo {pages} {054911}
  (\bibinfo {year} {2014})},\ \Eprint {http://arxiv.org/abs/1401.5845}
  {arXiv:1401.5845 [nucl-th]} \BibitemShut {NoStop}%
\bibitem [{\citenamefont {Katz}\ and\ \citenamefont
  {Gossiaux}(2016)}]{Katz:2015qja}%
  \BibitemOpen
  \bibfield  {author} {\bibinfo {author} {\bibfnamefont {R.}~\bibnamefont
  {Katz}}\ and\ \bibinfo {author} {\bibfnamefont {P.~B.}\ \bibnamefont
  {Gossiaux}},\ }\href {\doibase 10.1016/j.aop.2016.02.005} {\bibfield
  {journal} {\bibinfo  {journal} {Annals Phys.}\ }\textbf {\bibinfo {volume}
  {368}},\ \bibinfo {pages} {267} (\bibinfo {year} {2016})},\ \Eprint
  {http://arxiv.org/abs/1504.08087} {arXiv:1504.08087 [quant-ph]} \BibitemShut
  {NoStop}%
\bibitem [{\citenamefont {Yao}\ \emph {et~al.}(2020)\citenamefont {Yao},
  \citenamefont {Ke}, \citenamefont {Xu}, \citenamefont {Bass},\ and\
  \citenamefont {M\"uller}}]{Yao:2020xzw}%
  \BibitemOpen
  \bibfield  {author} {\bibinfo {author} {\bibfnamefont {X.}~\bibnamefont
  {Yao}}, \bibinfo {author} {\bibfnamefont {W.}~\bibnamefont {Ke}}, \bibinfo
  {author} {\bibfnamefont {Y.}~\bibnamefont {Xu}}, \bibinfo {author}
  {\bibfnamefont {S.~A.}\ \bibnamefont {Bass}}, \ and\ \bibinfo {author}
  {\bibfnamefont {B.}~\bibnamefont {M\"uller}},\ }\href {\doibase
  10.1007/JHEP01(2021)046} {\bibfield  {journal} {\bibinfo  {journal} {JHEP}\
  }\textbf {\bibinfo {volume} {21}},\ \bibinfo {pages} {046} (\bibinfo {year}
  {2020})},\ \Eprint {http://arxiv.org/abs/2004.06746} {arXiv:2004.06746
  [hep-ph]} \BibitemShut {NoStop}%
\bibitem [{\citenamefont {Islam}\ and\ \citenamefont
  {Strickland}(2020{\natexlab{a}})}]{Islam:2020gdv}%
  \BibitemOpen
  \bibfield  {author} {\bibinfo {author} {\bibfnamefont {A.}~\bibnamefont
  {Islam}}\ and\ \bibinfo {author} {\bibfnamefont {M.}~\bibnamefont
  {Strickland}},\ }\href {\doibase 10.1016/j.physletb.2020.135949} {\bibfield
  {journal} {\bibinfo  {journal} {Phys. Lett. B}\ }\textbf {\bibinfo {volume}
  {811}},\ \bibinfo {pages} {135949} (\bibinfo {year} {2020}{\natexlab{a}})},\
  \Eprint {http://arxiv.org/abs/2007.10211} {arXiv:2007.10211 [hep-ph]}
  \BibitemShut {NoStop}%
\bibitem [{\citenamefont {Chatrchyan}\ \emph {et~al.}(2011)\citenamefont
  {Chatrchyan} \emph {et~al.}}]{Chatrchyan:2011pe}%
  \BibitemOpen
  \bibfield  {author} {\bibinfo {author} {\bibfnamefont {S.}~\bibnamefont
  {Chatrchyan}} \emph {et~al.} (\bibinfo {collaboration} {CMS}),\ }\href
  {\doibase 10.1103/PhysRevLett.107.052302} {\bibfield  {journal} {\bibinfo
  {journal} {Phys. Rev. Lett.}\ }\textbf {\bibinfo {volume} {107}},\ \bibinfo
  {pages} {052302} (\bibinfo {year} {2011})},\ \Eprint
  {http://arxiv.org/abs/1105.4894} {arXiv:1105.4894 [nucl-ex]} \BibitemShut
  {NoStop}%
\bibitem [{\citenamefont {Chatrchyan}\ \emph {et~al.}(2012)\citenamefont
  {Chatrchyan} \emph {et~al.}}]{Chatrchyan:2012lxa}%
  \BibitemOpen
  \bibfield  {author} {\bibinfo {author} {\bibfnamefont {S.}~\bibnamefont
  {Chatrchyan}} \emph {et~al.} (\bibinfo {collaboration} {CMS}),\ }\href
  {\doibase 10.1103/PhysRevLett.109.222301} {\bibfield  {journal} {\bibinfo
  {journal} {Phys. Rev. Lett.}\ }\textbf {\bibinfo {volume} {109}},\ \bibinfo
  {pages} {222301} (\bibinfo {year} {2012})},\ \bibinfo {note} {[Erratum:
  Phys.Rev.Lett. 120, 199903 (2018)]},\ \Eprint
  {http://arxiv.org/abs/1208.2826} {arXiv:1208.2826 [nucl-ex]} \BibitemShut
  {NoStop}%
\bibitem [{\citenamefont {Khachatryan}\ \emph {et~al.}(2017)\citenamefont
  {Khachatryan} \emph {et~al.}}]{Khachatryan:2016xxp}%
  \BibitemOpen
  \bibfield  {author} {\bibinfo {author} {\bibfnamefont {V.}~\bibnamefont
  {Khachatryan}} \emph {et~al.} (\bibinfo {collaboration} {CMS}),\ }\href
  {\doibase 10.1016/j.physletb.2017.04.031} {\bibfield  {journal} {\bibinfo
  {journal} {Phys. Lett. B}\ }\textbf {\bibinfo {volume} {770}},\ \bibinfo
  {pages} {357} (\bibinfo {year} {2017})},\ \Eprint
  {http://arxiv.org/abs/1611.01510} {arXiv:1611.01510 [nucl-ex]} \BibitemShut
  {NoStop}%
\bibitem [{\citenamefont {Sirunyan}\ \emph {et~al.}(2018)\citenamefont
  {Sirunyan} \emph {et~al.}}]{Sirunyan:2017lzi}%
  \BibitemOpen
  \bibfield  {author} {\bibinfo {author} {\bibfnamefont {A.~M.}\ \bibnamefont
  {Sirunyan}} \emph {et~al.} (\bibinfo {collaboration} {CMS}),\ }\href
  {\doibase 10.1103/PhysRevLett.120.142301} {\bibfield  {journal} {\bibinfo
  {journal} {Phys. Rev. Lett.}\ }\textbf {\bibinfo {volume} {120}},\ \bibinfo
  {pages} {142301} (\bibinfo {year} {2018})},\ \Eprint
  {http://arxiv.org/abs/1706.05984} {arXiv:1706.05984 [hep-ex]} \BibitemShut
  {NoStop}%
\bibitem [{\citenamefont {Caswell}\ and\ \citenamefont
  {Lepage}(1986)}]{Caswell:1985ui}%
  \BibitemOpen
  \bibfield  {author} {\bibinfo {author} {\bibfnamefont {W.~E.}\ \bibnamefont
  {Caswell}}\ and\ \bibinfo {author} {\bibfnamefont {G.~P.}\ \bibnamefont
  {Lepage}},\ }\href {\doibase 10.1016/0370-2693(86)91297-9} {\bibfield
  {journal} {\bibinfo  {journal} {Phys. Lett. B}\ }\textbf {\bibinfo {volume}
  {167}},\ \bibinfo {pages} {437} (\bibinfo {year} {1986})}\BibitemShut
  {NoStop}%
\bibitem [{\citenamefont {Brambilla}\ \emph {et~al.}(2000)\citenamefont
  {Brambilla}, \citenamefont {Pineda}, \citenamefont {Soto},\ and\
  \citenamefont {Vairo}}]{Brambilla:1999xf}%
  \BibitemOpen
  \bibfield  {author} {\bibinfo {author} {\bibfnamefont {N.}~\bibnamefont
  {Brambilla}}, \bibinfo {author} {\bibfnamefont {A.}~\bibnamefont {Pineda}},
  \bibinfo {author} {\bibfnamefont {J.}~\bibnamefont {Soto}}, \ and\ \bibinfo
  {author} {\bibfnamefont {A.}~\bibnamefont {Vairo}},\ }\href {\doibase
  10.1016/S0550-3213(99)00693-8} {\bibfield  {journal} {\bibinfo  {journal}
  {Nucl. Phys. B}\ }\textbf {\bibinfo {volume} {566}},\ \bibinfo {pages} {275}
  (\bibinfo {year} {2000})},\ \Eprint {http://arxiv.org/abs/hep-ph/9907240}
  {arXiv:hep-ph/9907240} \BibitemShut {NoStop}%
\bibitem [{\citenamefont {Satz}(2006)}]{Satz:2005hx}%
  \BibitemOpen
  \bibfield  {author} {\bibinfo {author} {\bibfnamefont {H.}~\bibnamefont
  {Satz}},\ }\href {\doibase 10.1088/0954-3899/32/3/R01} {\bibfield  {journal}
  {\bibinfo  {journal} {J. Phys. G}\ }\textbf {\bibinfo {volume} {32}},\
  \bibinfo {pages} {R25} (\bibinfo {year} {2006})},\ \Eprint
  {http://arxiv.org/abs/hep-ph/0512217} {arXiv:hep-ph/0512217} \BibitemShut
  {NoStop}%
\bibitem [{\citenamefont {Zhao}\ \emph {et~al.}(2020)\citenamefont {Zhao},
  \citenamefont {Zhou}, \citenamefont {Chen},\ and\ \citenamefont
  {Zhuang}}]{Zhao:2020jqu}%
  \BibitemOpen
  \bibfield  {author} {\bibinfo {author} {\bibfnamefont {J.}~\bibnamefont
  {Zhao}}, \bibinfo {author} {\bibfnamefont {K.}~\bibnamefont {Zhou}}, \bibinfo
  {author} {\bibfnamefont {S.}~\bibnamefont {Chen}}, \ and\ \bibinfo {author}
  {\bibfnamefont {P.}~\bibnamefont {Zhuang}},\ }\href {\doibase
  10.1016/j.ppnp.2020.103801} {\bibfield  {journal} {\bibinfo  {journal} {Prog.
  Part. Nucl. Phys.}\ }\textbf {\bibinfo {volume} {114}},\ \bibinfo {pages}
  {103801} (\bibinfo {year} {2020})},\ \Eprint
  {http://arxiv.org/abs/2005.08277} {arXiv:2005.08277 [nucl-th]} \BibitemShut
  {NoStop}%
\bibitem [{\citenamefont {Crater}\ \emph {et~al.}(2009)\citenamefont {Crater},
  \citenamefont {Yoon},\ and\ \citenamefont {Wong}}]{Crater:2008rt}%
  \BibitemOpen
  \bibfield  {author} {\bibinfo {author} {\bibfnamefont {H.~W.}\ \bibnamefont
  {Crater}}, \bibinfo {author} {\bibfnamefont {J.-H.}\ \bibnamefont {Yoon}}, \
  and\ \bibinfo {author} {\bibfnamefont {C.-Y.}\ \bibnamefont {Wong}},\ }\href
  {\doibase 10.1103/PhysRevD.79.034011} {\bibfield  {journal} {\bibinfo
  {journal} {Phys. Rev. D}\ }\textbf {\bibinfo {volume} {79}},\ \bibinfo
  {pages} {034011} (\bibinfo {year} {2009})},\ \Eprint
  {http://arxiv.org/abs/0811.0732} {arXiv:0811.0732 [hep-ph]} \BibitemShut
  {NoStop}%
\bibitem [{\citenamefont {Guo}\ \emph {et~al.}(2012)\citenamefont {Guo},
  \citenamefont {Shi},\ and\ \citenamefont {Zhuang}}]{Guo:2012hx}%
  \BibitemOpen
  \bibfield  {author} {\bibinfo {author} {\bibfnamefont {X.}~\bibnamefont
  {Guo}}, \bibinfo {author} {\bibfnamefont {S.}~\bibnamefont {Shi}}, \ and\
  \bibinfo {author} {\bibfnamefont {P.}~\bibnamefont {Zhuang}},\ }\href
  {\doibase 10.1016/j.physletb.2012.10.032} {\bibfield  {journal} {\bibinfo
  {journal} {Phys. Lett. B}\ }\textbf {\bibinfo {volume} {718}},\ \bibinfo
  {pages} {143} (\bibinfo {year} {2012})},\ \Eprint
  {http://arxiv.org/abs/1209.5873} {arXiv:1209.5873 [hep-ph]} \BibitemShut
  {NoStop}%
\bibitem [{\citenamefont {Laine}\ \emph {et~al.}(2007)\citenamefont {Laine},
  \citenamefont {Philipsen}, \citenamefont {Romatschke},\ and\ \citenamefont
  {Tassler}}]{Laine:2006ns}%
  \BibitemOpen
  \bibfield  {author} {\bibinfo {author} {\bibfnamefont {M.}~\bibnamefont
  {Laine}}, \bibinfo {author} {\bibfnamefont {O.}~\bibnamefont {Philipsen}},
  \bibinfo {author} {\bibfnamefont {P.}~\bibnamefont {Romatschke}}, \ and\
  \bibinfo {author} {\bibfnamefont {M.}~\bibnamefont {Tassler}},\ }\href
  {\doibase 10.1088/1126-6708/2007/03/054} {\bibfield  {journal} {\bibinfo
  {journal} {JHEP}\ }\textbf {\bibinfo {volume} {03}},\ \bibinfo {pages} {054}
  (\bibinfo {year} {2007})},\ \Eprint {http://arxiv.org/abs/hep-ph/0611300}
  {arXiv:hep-ph/0611300} \BibitemShut {NoStop}%
\bibitem [{\citenamefont {Beraudo}\ \emph {et~al.}(2008)\citenamefont
  {Beraudo}, \citenamefont {Blaizot},\ and\ \citenamefont
  {Ratti}}]{Beraudo:2007ky}%
  \BibitemOpen
  \bibfield  {author} {\bibinfo {author} {\bibfnamefont {A.}~\bibnamefont
  {Beraudo}}, \bibinfo {author} {\bibfnamefont {J.~P.}\ \bibnamefont
  {Blaizot}}, \ and\ \bibinfo {author} {\bibfnamefont {C.}~\bibnamefont
  {Ratti}},\ }\href {\doibase 10.1016/j.nuclphysa.2008.03.001} {\bibfield
  {journal} {\bibinfo  {journal} {Nucl. Phys. A}\ }\textbf {\bibinfo {volume}
  {806}},\ \bibinfo {pages} {312} (\bibinfo {year} {2008})},\ \Eprint
  {http://arxiv.org/abs/0712.4394} {arXiv:0712.4394 [nucl-th]} \BibitemShut
  {NoStop}%
\bibitem [{\citenamefont {Brambilla}\ \emph {et~al.}(2008)\citenamefont
  {Brambilla}, \citenamefont {Ghiglieri}, \citenamefont {Vairo},\ and\
  \citenamefont {Petreczky}}]{Brambilla:2008cx}%
  \BibitemOpen
  \bibfield  {author} {\bibinfo {author} {\bibfnamefont {N.}~\bibnamefont
  {Brambilla}}, \bibinfo {author} {\bibfnamefont {J.}~\bibnamefont
  {Ghiglieri}}, \bibinfo {author} {\bibfnamefont {A.}~\bibnamefont {Vairo}}, \
  and\ \bibinfo {author} {\bibfnamefont {P.}~\bibnamefont {Petreczky}},\ }\href
  {\doibase 10.1103/PhysRevD.78.014017} {\bibfield  {journal} {\bibinfo
  {journal} {Phys. Rev. D}\ }\textbf {\bibinfo {volume} {78}},\ \bibinfo
  {pages} {014017} (\bibinfo {year} {2008})},\ \Eprint
  {http://arxiv.org/abs/0804.0993} {arXiv:0804.0993 [hep-ph]} \BibitemShut
  {NoStop}%
\bibitem [{\citenamefont {Brambilla}\ \emph {et~al.}(2010)\citenamefont
  {Brambilla}, \citenamefont {Escobedo}, \citenamefont {Ghiglieri},
  \citenamefont {Soto},\ and\ \citenamefont {Vairo}}]{Brambilla:2010vq}%
  \BibitemOpen
  \bibfield  {author} {\bibinfo {author} {\bibfnamefont {N.}~\bibnamefont
  {Brambilla}}, \bibinfo {author} {\bibfnamefont {M.~A.}\ \bibnamefont
  {Escobedo}}, \bibinfo {author} {\bibfnamefont {J.}~\bibnamefont {Ghiglieri}},
  \bibinfo {author} {\bibfnamefont {J.}~\bibnamefont {Soto}}, \ and\ \bibinfo
  {author} {\bibfnamefont {A.}~\bibnamefont {Vairo}},\ }\href {\doibase
  10.1007/JHEP09(2010)038} {\bibfield  {journal} {\bibinfo  {journal} {JHEP}\
  }\textbf {\bibinfo {volume} {09}},\ \bibinfo {pages} {038} (\bibinfo {year}
  {2010})},\ \Eprint {http://arxiv.org/abs/1007.4156} {arXiv:1007.4156
  [hep-ph]} \BibitemShut {NoStop}%
\bibitem [{\citenamefont {Rothkopf}\ \emph {et~al.}(2012)\citenamefont
  {Rothkopf}, \citenamefont {Hatsuda},\ and\ \citenamefont
  {Sasaki}}]{Rothkopf:2011db}%
  \BibitemOpen
  \bibfield  {author} {\bibinfo {author} {\bibfnamefont {A.}~\bibnamefont
  {Rothkopf}}, \bibinfo {author} {\bibfnamefont {T.}~\bibnamefont {Hatsuda}}, \
  and\ \bibinfo {author} {\bibfnamefont {S.}~\bibnamefont {Sasaki}},\ }\href
  {\doibase 10.1103/PhysRevLett.108.162001} {\bibfield  {journal} {\bibinfo
  {journal} {Phys. Rev. Lett.}\ }\textbf {\bibinfo {volume} {108}},\ \bibinfo
  {pages} {162001} (\bibinfo {year} {2012})},\ \Eprint
  {http://arxiv.org/abs/1108.1579} {arXiv:1108.1579 [hep-lat]} \BibitemShut
  {NoStop}%
\bibitem [{\citenamefont {Burnier}\ \emph
  {et~al.}(2015{\natexlab{a}})\citenamefont {Burnier}, \citenamefont
  {Kaczmarek},\ and\ \citenamefont {Rothkopf}}]{Burnier:2014ssa}%
  \BibitemOpen
  \bibfield  {author} {\bibinfo {author} {\bibfnamefont {Y.}~\bibnamefont
  {Burnier}}, \bibinfo {author} {\bibfnamefont {O.}~\bibnamefont {Kaczmarek}},
  \ and\ \bibinfo {author} {\bibfnamefont {A.}~\bibnamefont {Rothkopf}},\
  }\href {\doibase 10.1103/PhysRevLett.114.082001} {\bibfield  {journal}
  {\bibinfo  {journal} {Phys. Rev. Lett.}\ }\textbf {\bibinfo {volume} {114}},\
  \bibinfo {pages} {082001} (\bibinfo {year} {2015}{\natexlab{a}})},\ \Eprint
  {http://arxiv.org/abs/1410.2546} {arXiv:1410.2546 [hep-lat]} \BibitemShut
  {NoStop}%
\bibitem [{\citenamefont {Burnier}\ \emph
  {et~al.}(2015{\natexlab{b}})\citenamefont {Burnier}, \citenamefont
  {Kaczmarek},\ and\ \citenamefont {Rothkopf}}]{Burnier:2015tda}%
  \BibitemOpen
  \bibfield  {author} {\bibinfo {author} {\bibfnamefont {Y.}~\bibnamefont
  {Burnier}}, \bibinfo {author} {\bibfnamefont {O.}~\bibnamefont {Kaczmarek}},
  \ and\ \bibinfo {author} {\bibfnamefont {A.}~\bibnamefont {Rothkopf}},\
  }\href {\doibase 10.1007/JHEP12(2015)101} {\bibfield  {journal} {\bibinfo
  {journal} {JHEP}\ }\textbf {\bibinfo {volume} {12}},\ \bibinfo {pages} {101}
  (\bibinfo {year} {2015}{\natexlab{b}})},\ \Eprint
  {http://arxiv.org/abs/1509.07366} {arXiv:1509.07366 [hep-ph]} \BibitemShut
  {NoStop}%
\bibitem [{\citenamefont {Bala}\ and\ \citenamefont
  {Datta}(2020)}]{Bala:2019cqu}%
  \BibitemOpen
  \bibfield  {author} {\bibinfo {author} {\bibfnamefont {D.}~\bibnamefont
  {Bala}}\ and\ \bibinfo {author} {\bibfnamefont {S.}~\bibnamefont {Datta}},\
  }\href {\doibase 10.1103/PhysRevD.101.034507} {\bibfield  {journal} {\bibinfo
   {journal} {Phys. Rev. D}\ }\textbf {\bibinfo {volume} {101}},\ \bibinfo
  {pages} {034507} (\bibinfo {year} {2020})},\ \Eprint
  {http://arxiv.org/abs/1909.10548} {arXiv:1909.10548 [hep-lat]} \BibitemShut
  {NoStop}%
\bibitem [{\citenamefont {Larsen}\ \emph {et~al.}(2019)\citenamefont {Larsen},
  \citenamefont {Meinel}, \citenamefont {Mukherjee},\ and\ \citenamefont
  {Petreczky}}]{Larsen:2019bwy}%
  \BibitemOpen
  \bibfield  {author} {\bibinfo {author} {\bibfnamefont {R.}~\bibnamefont
  {Larsen}}, \bibinfo {author} {\bibfnamefont {S.}~\bibnamefont {Meinel}},
  \bibinfo {author} {\bibfnamefont {S.}~\bibnamefont {Mukherjee}}, \ and\
  \bibinfo {author} {\bibfnamefont {P.}~\bibnamefont {Petreczky}},\ }\href
  {\doibase 10.1103/PhysRevD.100.074506} {\bibfield  {journal} {\bibinfo
  {journal} {Phys. Rev. D}\ }\textbf {\bibinfo {volume} {100}},\ \bibinfo
  {pages} {074506} (\bibinfo {year} {2019})},\ \Eprint
  {http://arxiv.org/abs/1908.08437} {arXiv:1908.08437 [hep-lat]} \BibitemShut
  {NoStop}%
\bibitem [{\citenamefont {Larsen}\ \emph
  {et~al.}(2020{\natexlab{a}})\citenamefont {Larsen}, \citenamefont {Meinel},
  \citenamefont {Mukherjee},\ and\ \citenamefont {Petreczky}}]{Larsen:2019zqv}%
  \BibitemOpen
  \bibfield  {author} {\bibinfo {author} {\bibfnamefont {R.}~\bibnamefont
  {Larsen}}, \bibinfo {author} {\bibfnamefont {S.}~\bibnamefont {Meinel}},
  \bibinfo {author} {\bibfnamefont {S.}~\bibnamefont {Mukherjee}}, \ and\
  \bibinfo {author} {\bibfnamefont {P.}~\bibnamefont {Petreczky}},\ }\href
  {\doibase 10.1016/j.physletb.2019.135119} {\bibfield  {journal} {\bibinfo
  {journal} {Phys. Lett. B}\ }\textbf {\bibinfo {volume} {800}},\ \bibinfo
  {pages} {135119} (\bibinfo {year} {2020}{\natexlab{a}})},\ \Eprint
  {http://arxiv.org/abs/1910.07374} {arXiv:1910.07374 [hep-lat]} \BibitemShut
  {NoStop}%
\bibitem [{\citenamefont {Larsen}\ \emph
  {et~al.}(2020{\natexlab{b}})\citenamefont {Larsen}, \citenamefont {Meinel},
  \citenamefont {Mukherjee},\ and\ \citenamefont {Petreczky}}]{Larsen:2020rjk}%
  \BibitemOpen
  \bibfield  {author} {\bibinfo {author} {\bibfnamefont {R.}~\bibnamefont
  {Larsen}}, \bibinfo {author} {\bibfnamefont {S.}~\bibnamefont {Meinel}},
  \bibinfo {author} {\bibfnamefont {S.}~\bibnamefont {Mukherjee}}, \ and\
  \bibinfo {author} {\bibfnamefont {P.}~\bibnamefont {Petreczky}},\ }\href
  {\doibase 10.1103/PhysRevD.102.114508} {\bibfield  {journal} {\bibinfo
  {journal} {Phys. Rev. D}\ }\textbf {\bibinfo {volume} {102}},\ \bibinfo
  {pages} {114508} (\bibinfo {year} {2020}{\natexlab{b}})},\ \Eprint
  {http://arxiv.org/abs/2008.00100} {arXiv:2008.00100 [hep-lat]} \BibitemShut
  {NoStop}%
\bibitem [{Note1()}]{Note1}%
  \BibitemOpen
  \bibinfo {note} {It might be worth noting that the potential here is the
  effective potential for a Schr{\"o}dinger equation.}\BibitemShut {Stop}%
\bibitem [{\citenamefont {Tanabashi}\ \emph {et~al.}(2018)\citenamefont
  {Tanabashi} \emph {et~al.}}]{Tanabashi:2018oca}%
  \BibitemOpen
  \bibfield  {author} {\bibinfo {author} {\bibfnamefont {M.}~\bibnamefont
  {Tanabashi}} \emph {et~al.} (\bibinfo {collaboration} {Particle Data
  Group}),\ }\href {\doibase 10.1103/PhysRevD.98.030001} {\bibfield  {journal}
  {\bibinfo  {journal} {Phys. Rev. D}\ }\textbf {\bibinfo {volume} {98}},\
  \bibinfo {pages} {030001} (\bibinfo {year} {2018})}\BibitemShut {NoStop}%
\bibitem [{\citenamefont {Leshno}\ and\ \citenamefont
  {Schocken}(1993)}]{Leshno93multilayerfeedforward}%
  \BibitemOpen
  \bibfield  {author} {\bibinfo {author} {\bibfnamefont {M.}~\bibnamefont
  {Leshno}}\ and\ \bibinfo {author} {\bibfnamefont {S.}~\bibnamefont
  {Schocken}},\ }\href@noop {} {\bibfield  {journal} {\bibinfo  {journal}
  {Neural Networks}\ }\textbf {\bibinfo {volume} {6}},\ \bibinfo {pages} {861}
  (\bibinfo {year} {1993})}\BibitemShut {NoStop}%
\bibitem [{\citenamefont {Kratsios}(2021)}]{Kratsios_2021}%
  \BibitemOpen
  \bibfield  {author} {\bibinfo {author} {\bibfnamefont {A.}~\bibnamefont
  {Kratsios}},\ }\href {\doibase 10.1007/s10472-020-09723-1} {\bibfield
  {journal} {\bibinfo  {journal} {Annals of Mathematics and Artificial
  Intelligence}\ } (\bibinfo {year} {2021}),\
  10.1007/s10472-020-09723-1}\BibitemShut {NoStop}%
\bibitem [{\citenamefont {Forte}\ \emph {et~al.}(2002)\citenamefont {Forte},
  \citenamefont {Garrido}, \citenamefont {Latorre},\ and\ \citenamefont
  {Piccione}}]{Forte_2002}%
  \BibitemOpen
  \bibfield  {author} {\bibinfo {author} {\bibfnamefont {S.}~\bibnamefont
  {Forte}}, \bibinfo {author} {\bibfnamefont {L.~s.}\ \bibnamefont {Garrido}},
  \bibinfo {author} {\bibfnamefont {J.~I.}\ \bibnamefont {Latorre}}, \ and\
  \bibinfo {author} {\bibfnamefont {A.}~\bibnamefont {Piccione}},\ }\href
  {\doibase 10.1088/1126-6708/2002/05/062} {\bibfield  {journal} {\bibinfo
  {journal} {Journal of High Energy Physics}\ }\textbf {\bibinfo {volume}
  {2002}},\ \bibinfo {pages} {062–062} (\bibinfo {year} {2002})}\BibitemShut
  {NoStop}%
\bibitem [{\citenamefont {Collaboration}\ \emph {et~al.}(2007)\citenamefont
  {Collaboration}, \citenamefont {Debbio}, \citenamefont {Forte}, \citenamefont
  {Latorre}, \citenamefont {Piccione},\ and\ \citenamefont
  {Rojo}}]{Collaboration_2007}%
  \BibitemOpen
  \bibfield  {author} {\bibinfo {author} {\bibfnamefont {T.~N.}\ \bibnamefont
  {Collaboration}}, \bibinfo {author} {\bibfnamefont {L.~D.}\ \bibnamefont
  {Debbio}}, \bibinfo {author} {\bibfnamefont {S.}~\bibnamefont {Forte}},
  \bibinfo {author} {\bibfnamefont {J.~I.}\ \bibnamefont {Latorre}}, \bibinfo
  {author} {\bibfnamefont {A.}~\bibnamefont {Piccione}}, \ and\ \bibinfo
  {author} {\bibfnamefont {J.}~\bibnamefont {Rojo}},\ }\href {\doibase
  10.1088/1126-6708/2007/03/039} {\bibfield  {journal} {\bibinfo  {journal}
  {Journal of High Energy Physics}\ }\textbf {\bibinfo {volume} {2007}},\
  \bibinfo {pages} {039–039} (\bibinfo {year} {2007})}\BibitemShut {NoStop}%
\bibitem [{\citenamefont {Kades}\ \emph {et~al.}(2020)\citenamefont {Kades},
  \citenamefont {Pawlowski}, \citenamefont {Rothkopf}, \citenamefont
  {Scherzer}, \citenamefont {Urban}, \citenamefont {Wetzel}, \citenamefont
  {Wink},\ and\ \citenamefont {Ziegler}}]{Kades:2019wtd}%
  \BibitemOpen
  \bibfield  {author} {\bibinfo {author} {\bibfnamefont {L.}~\bibnamefont
  {Kades}}, \bibinfo {author} {\bibfnamefont {J.~M.}\ \bibnamefont
  {Pawlowski}}, \bibinfo {author} {\bibfnamefont {A.}~\bibnamefont {Rothkopf}},
  \bibinfo {author} {\bibfnamefont {M.}~\bibnamefont {Scherzer}}, \bibinfo
  {author} {\bibfnamefont {J.~M.}\ \bibnamefont {Urban}}, \bibinfo {author}
  {\bibfnamefont {S.~J.}\ \bibnamefont {Wetzel}}, \bibinfo {author}
  {\bibfnamefont {N.}~\bibnamefont {Wink}}, \ and\ \bibinfo {author}
  {\bibfnamefont {F.~P.~G.}\ \bibnamefont {Ziegler}},\ }\href {\doibase
  10.1103/PhysRevD.102.096001} {\bibfield  {journal} {\bibinfo  {journal}
  {Phys. Rev. D}\ }\textbf {\bibinfo {volume} {102}},\ \bibinfo {pages}
  {096001} (\bibinfo {year} {2020})},\ \Eprint
  {http://arxiv.org/abs/1905.04305} {arXiv:1905.04305 [physics.comp-ph]}
  \BibitemShut {NoStop}%
\bibitem [{\citenamefont {Zhou}\ \emph {et~al.}(2021)\citenamefont {Zhou},
  \citenamefont {Gao}, \citenamefont {Chao}, \citenamefont {Liu},\ and\
  \citenamefont {Song}}]{Zhou:2021bvw}%
  \BibitemOpen
  \bibfield  {author} {\bibinfo {author} {\bibfnamefont {M.}~\bibnamefont
  {Zhou}}, \bibinfo {author} {\bibfnamefont {F.}~\bibnamefont {Gao}}, \bibinfo
  {author} {\bibfnamefont {J.}~\bibnamefont {Chao}}, \bibinfo {author}
  {\bibfnamefont {Y.-X.}\ \bibnamefont {Liu}}, \ and\ \bibinfo {author}
  {\bibfnamefont {H.}~\bibnamefont {Song}},\ }\href {\doibase
  10.1103/PhysRevD.104.076011} {\bibfield  {journal} {\bibinfo  {journal}
  {Phys. Rev. D}\ }\textbf {\bibinfo {volume} {104}},\ \bibinfo {pages}
  {076011} (\bibinfo {year} {2021})},\ \Eprint
  {http://arxiv.org/abs/2106.08168} {arXiv:2106.08168 [hep-ph]} \BibitemShut
  {NoStop}%
\bibitem [{\citenamefont {Chen}\ \emph {et~al.}(2021)\citenamefont {Chen},
  \citenamefont {Ding}, \citenamefont {Liu}, \citenamefont {Papp},\ and\
  \citenamefont {Yang}}]{Chen:2021giw}%
  \BibitemOpen
  \bibfield  {author} {\bibinfo {author} {\bibfnamefont {S.~Y.}\ \bibnamefont
  {Chen}}, \bibinfo {author} {\bibfnamefont {H.~T.}\ \bibnamefont {Ding}},
  \bibinfo {author} {\bibfnamefont {F.~Y.}\ \bibnamefont {Liu}}, \bibinfo
  {author} {\bibfnamefont {G.}~\bibnamefont {Papp}}, \ and\ \bibinfo {author}
  {\bibfnamefont {C.~B.}\ \bibnamefont {Yang}},\ }\href@noop {} {\  (\bibinfo
  {year} {2021})},\ \Eprint {http://arxiv.org/abs/2110.13521} {arXiv:2110.13521
  [hep-lat]} \BibitemShut {NoStop}%
\bibitem [{\citenamefont {Carrasquilla}\ and\ \citenamefont
  {Melko}(2017)}]{Carrasquilla_2017}%
  \BibitemOpen
  \bibfield  {author} {\bibinfo {author} {\bibfnamefont {J.}~\bibnamefont
  {Carrasquilla}}\ and\ \bibinfo {author} {\bibfnamefont {R.~G.}\ \bibnamefont
  {Melko}},\ }\href {\doibase 10.1038/nphys4035} {\bibfield  {journal}
  {\bibinfo  {journal} {Nature Physics}\ }\textbf {\bibinfo {volume} {13}},\
  \bibinfo {pages} {431–434} (\bibinfo {year} {2017})}\BibitemShut {NoStop}%
\bibitem [{\citenamefont {Pang}\ \emph {et~al.}(2018)\citenamefont {Pang},
  \citenamefont {Zhou}, \citenamefont {Su}, \citenamefont {Petersen},
  \citenamefont {Stöcker},\ and\ \citenamefont {Wang}}]{Pang:2016vdc}%
  \BibitemOpen
  \bibfield  {author} {\bibinfo {author} {\bibfnamefont {L.-G.}\ \bibnamefont
  {Pang}}, \bibinfo {author} {\bibfnamefont {K.}~\bibnamefont {Zhou}}, \bibinfo
  {author} {\bibfnamefont {N.}~\bibnamefont {Su}}, \bibinfo {author}
  {\bibfnamefont {H.}~\bibnamefont {Petersen}}, \bibinfo {author}
  {\bibfnamefont {H.}~\bibnamefont {Stöcker}}, \ and\ \bibinfo {author}
  {\bibfnamefont {X.-N.}\ \bibnamefont {Wang}},\ }\href {\doibase
  10.1038/s41467-017-02726-3} {\bibfield  {journal} {\bibinfo  {journal}
  {Nature Commun.}\ }\textbf {\bibinfo {volume} {9}},\ \bibinfo {pages} {210}
  (\bibinfo {year} {2018})},\ \Eprint {http://arxiv.org/abs/1612.04262}
  {arXiv:1612.04262 [hep-ph]} \BibitemShut {NoStop}%
\bibitem [{\citenamefont {Wang}\ \emph {et~al.}(2020)\citenamefont {Wang},
  \citenamefont {Ma}, \citenamefont {Wada}, \citenamefont {Chen}, \citenamefont
  {He}, \citenamefont {Liu},\ and\ \citenamefont {Sun}}]{Wang:2020tgb}%
  \BibitemOpen
  \bibfield  {author} {\bibinfo {author} {\bibfnamefont {R.}~\bibnamefont
  {Wang}}, \bibinfo {author} {\bibfnamefont {Y.-G.}\ \bibnamefont {Ma}},
  \bibinfo {author} {\bibfnamefont {R.}~\bibnamefont {Wada}}, \bibinfo {author}
  {\bibfnamefont {L.-W.}\ \bibnamefont {Chen}}, \bibinfo {author}
  {\bibfnamefont {W.-B.}\ \bibnamefont {He}}, \bibinfo {author} {\bibfnamefont
  {H.-L.}\ \bibnamefont {Liu}}, \ and\ \bibinfo {author} {\bibfnamefont
  {K.-J.}\ \bibnamefont {Sun}},\ }\href {\doibase
  10.1103/PhysRevResearch.2.043202} {\bibfield  {journal} {\bibinfo  {journal}
  {Phys. Rev. Res.}\ }\textbf {\bibinfo {volume} {2}},\ \bibinfo {pages}
  {043202} (\bibinfo {year} {2020})},\ \Eprint
  {http://arxiv.org/abs/2010.15043} {arXiv:2010.15043 [nucl-th]} \BibitemShut
  {NoStop}%
\bibitem [{\citenamefont {Jiang}\ \emph {et~al.}(2021)\citenamefont {Jiang},
  \citenamefont {Wang},\ and\ \citenamefont {Zhou}}]{Jiang:2021gsw}%
  \BibitemOpen
  \bibfield  {author} {\bibinfo {author} {\bibfnamefont {L.}~\bibnamefont
  {Jiang}}, \bibinfo {author} {\bibfnamefont {L.}~\bibnamefont {Wang}}, \ and\
  \bibinfo {author} {\bibfnamefont {K.}~\bibnamefont {Zhou}},\ }\href {\doibase
  10.1103/PhysRevD.103.116023} {\bibfield  {journal} {\bibinfo  {journal}
  {Phys. Rev. D}\ }\textbf {\bibinfo {volume} {103}},\ \bibinfo {pages}
  {116023} (\bibinfo {year} {2021})},\ \Eprint
  {http://arxiv.org/abs/2103.04090} {arXiv:2103.04090 [nucl-th]} \BibitemShut
  {NoStop}%
\bibitem [{\citenamefont {Zhou}\ \emph {et~al.}(2019)\citenamefont {Zhou},
  \citenamefont {Endr\ifmmode~\mbox{\H{o}}\else \H{o}\fi{}di}, \citenamefont
  {Pang},\ and\ \citenamefont {St\"ocker}}]{PhysRevD.100.011501}%
  \BibitemOpen
  \bibfield  {author} {\bibinfo {author} {\bibfnamefont {K.}~\bibnamefont
  {Zhou}}, \bibinfo {author} {\bibfnamefont {G.}~\bibnamefont
  {Endr\ifmmode~\mbox{\H{o}}\else \H{o}\fi{}di}}, \bibinfo {author}
  {\bibfnamefont {L.-G.}\ \bibnamefont {Pang}}, \ and\ \bibinfo {author}
  {\bibfnamefont {H.}~\bibnamefont {St\"ocker}},\ }\href {\doibase
  10.1103/PhysRevD.100.011501} {\bibfield  {journal} {\bibinfo  {journal}
  {Phys. Rev. D}\ }\textbf {\bibinfo {volume} {100}},\ \bibinfo {pages}
  {011501} (\bibinfo {year} {2019})}\BibitemShut {NoStop}%
\bibitem [{\citenamefont {Boyda}\ \emph {et~al.}(2021)\citenamefont {Boyda},
  \citenamefont {Kanwar}, \citenamefont {Racani\`ere}, \citenamefont {Rezende},
  \citenamefont {Albergo}, \citenamefont {Cranmer}, \citenamefont {Hackett},\
  and\ \citenamefont {Shanahan}}]{Boyda:2020hsi}%
  \BibitemOpen
  \bibfield  {author} {\bibinfo {author} {\bibfnamefont {D.}~\bibnamefont
  {Boyda}}, \bibinfo {author} {\bibfnamefont {G.}~\bibnamefont {Kanwar}},
  \bibinfo {author} {\bibfnamefont {S.}~\bibnamefont {Racani\`ere}}, \bibinfo
  {author} {\bibfnamefont {D.~J.}\ \bibnamefont {Rezende}}, \bibinfo {author}
  {\bibfnamefont {M.~S.}\ \bibnamefont {Albergo}}, \bibinfo {author}
  {\bibfnamefont {K.}~\bibnamefont {Cranmer}}, \bibinfo {author} {\bibfnamefont
  {D.~C.}\ \bibnamefont {Hackett}}, \ and\ \bibinfo {author} {\bibfnamefont
  {P.~E.}\ \bibnamefont {Shanahan}},\ }\href {\doibase
  10.1103/PhysRevD.103.074504} {\bibfield  {journal} {\bibinfo  {journal}
  {Phys. Rev. D}\ }\textbf {\bibinfo {volume} {103}},\ \bibinfo {pages}
  {074504} (\bibinfo {year} {2021})},\ \Eprint
  {http://arxiv.org/abs/2008.05456} {arXiv:2008.05456 [hep-lat]} \BibitemShut
  {NoStop}%
\bibitem [{\citenamefont {Kanwar}\ \emph {et~al.}(2020)\citenamefont {Kanwar},
  \citenamefont {Albergo}, \citenamefont {Boyda}, \citenamefont {Cranmer},
  \citenamefont {Hackett}, \citenamefont {Racani\`ere}, \citenamefont
  {Rezende},\ and\ \citenamefont {Shanahan}}]{Kanwar:2020xzo}%
  \BibitemOpen
  \bibfield  {author} {\bibinfo {author} {\bibfnamefont {G.}~\bibnamefont
  {Kanwar}}, \bibinfo {author} {\bibfnamefont {M.~S.}\ \bibnamefont {Albergo}},
  \bibinfo {author} {\bibfnamefont {D.}~\bibnamefont {Boyda}}, \bibinfo
  {author} {\bibfnamefont {K.}~\bibnamefont {Cranmer}}, \bibinfo {author}
  {\bibfnamefont {D.~C.}\ \bibnamefont {Hackett}}, \bibinfo {author}
  {\bibfnamefont {S.}~\bibnamefont {Racani\`ere}}, \bibinfo {author}
  {\bibfnamefont {D.~J.}\ \bibnamefont {Rezende}}, \ and\ \bibinfo {author}
  {\bibfnamefont {P.~E.}\ \bibnamefont {Shanahan}},\ }\href {\doibase
  10.1103/PhysRevLett.125.121601} {\bibfield  {journal} {\bibinfo  {journal}
  {Phys. Rev. Lett.}\ }\textbf {\bibinfo {volume} {125}},\ \bibinfo {pages}
  {121601} (\bibinfo {year} {2020})},\ \Eprint
  {http://arxiv.org/abs/2003.06413} {arXiv:2003.06413 [hep-lat]} \BibitemShut
  {NoStop}%
\bibitem [{\citenamefont {Albergo}\ \emph {et~al.}(2019)\citenamefont
  {Albergo}, \citenamefont {Kanwar},\ and\ \citenamefont
  {Shanahan}}]{Albergo:2019eim}%
  \BibitemOpen
  \bibfield  {author} {\bibinfo {author} {\bibfnamefont {M.~S.}\ \bibnamefont
  {Albergo}}, \bibinfo {author} {\bibfnamefont {G.}~\bibnamefont {Kanwar}}, \
  and\ \bibinfo {author} {\bibfnamefont {P.~E.}\ \bibnamefont {Shanahan}},\
  }\href {\doibase 10.1103/PhysRevD.100.034515} {\bibfield  {journal} {\bibinfo
   {journal} {Phys. Rev. D}\ }\textbf {\bibinfo {volume} {100}},\ \bibinfo
  {pages} {034515} (\bibinfo {year} {2019})},\ \Eprint
  {http://arxiv.org/abs/1904.12072} {arXiv:1904.12072 [hep-lat]} \BibitemShut
  {NoStop}%
\bibitem [{\citenamefont {Omana~Kuttan}\ \emph {et~al.}(2020)\citenamefont
  {Omana~Kuttan}, \citenamefont {Steinheimer}, \citenamefont {Zhou},
  \citenamefont {Redelbach},\ and\ \citenamefont
  {Stoecker}}]{OmanaKuttan:2020brq}%
  \BibitemOpen
  \bibfield  {author} {\bibinfo {author} {\bibfnamefont {M.}~\bibnamefont
  {Omana~Kuttan}}, \bibinfo {author} {\bibfnamefont {J.}~\bibnamefont
  {Steinheimer}}, \bibinfo {author} {\bibfnamefont {K.}~\bibnamefont {Zhou}},
  \bibinfo {author} {\bibfnamefont {A.}~\bibnamefont {Redelbach}}, \ and\
  \bibinfo {author} {\bibfnamefont {H.}~\bibnamefont {Stoecker}},\ }\href
  {\doibase 10.1016/j.physletb.2020.135872} {\bibfield  {journal} {\bibinfo
  {journal} {Phys. Lett. B}\ }\textbf {\bibinfo {volume} {811}},\ \bibinfo
  {pages} {135872} (\bibinfo {year} {2020})},\ \Eprint
  {http://arxiv.org/abs/2009.01584} {arXiv:2009.01584 [hep-ph]} \BibitemShut
  {NoStop}%
\bibitem [{\citenamefont {Thaprasop}\ \emph {et~al.}(2021)\citenamefont
  {Thaprasop}, \citenamefont {Zhou}, \citenamefont {Steinheimer},\ and\
  \citenamefont {Herold}}]{Thaprasop:2020mzp}%
  \BibitemOpen
  \bibfield  {author} {\bibinfo {author} {\bibfnamefont {P.}~\bibnamefont
  {Thaprasop}}, \bibinfo {author} {\bibfnamefont {K.}~\bibnamefont {Zhou}},
  \bibinfo {author} {\bibfnamefont {J.}~\bibnamefont {Steinheimer}}, \ and\
  \bibinfo {author} {\bibfnamefont {C.}~\bibnamefont {Herold}},\ }\href
  {\doibase 10.1088/1402-4896/abf214} {\bibfield  {journal} {\bibinfo
  {journal} {Phys. Scripta}\ }\textbf {\bibinfo {volume} {96}},\ \bibinfo
  {pages} {064003} (\bibinfo {year} {2021})},\ \Eprint
  {http://arxiv.org/abs/2007.15830} {arXiv:2007.15830 [hep-ex]} \BibitemShut
  {NoStop}%
\bibitem [{\citenamefont {Li}\ \emph {et~al.}(2020)\citenamefont {Li},
  \citenamefont {Wang}, \citenamefont {L\"u}, \citenamefont {Li}, \citenamefont
  {Li},\ and\ \citenamefont {Liu}}]{Li:2020qqn}%
  \BibitemOpen
  \bibfield  {author} {\bibinfo {author} {\bibfnamefont {F.}~\bibnamefont
  {Li}}, \bibinfo {author} {\bibfnamefont {Y.}~\bibnamefont {Wang}}, \bibinfo
  {author} {\bibfnamefont {H.}~\bibnamefont {L\"u}}, \bibinfo {author}
  {\bibfnamefont {P.}~\bibnamefont {Li}}, \bibinfo {author} {\bibfnamefont
  {Q.}~\bibnamefont {Li}}, \ and\ \bibinfo {author} {\bibfnamefont
  {F.}~\bibnamefont {Liu}},\ }\href {\doibase 10.1088/1361-6471/abb1f9}
  {\bibfield  {journal} {\bibinfo  {journal} {J. Phys. G}\ }\textbf {\bibinfo
  {volume} {47}},\ \bibinfo {pages} {115104} (\bibinfo {year} {2020})},\
  \Eprint {http://arxiv.org/abs/2008.11540} {arXiv:2008.11540 [nucl-th]}
  \BibitemShut {NoStop}%
\bibitem [{\citenamefont {Andreassen}\ \emph {et~al.}(2021)\citenamefont
  {Andreassen}, \citenamefont {Hsu}, \citenamefont {Nachman}, \citenamefont
  {Suaysom},\ and\ \citenamefont {Suresh}}]{Andreassen:2020gtw}%
  \BibitemOpen
  \bibfield  {author} {\bibinfo {author} {\bibfnamefont {A.}~\bibnamefont
  {Andreassen}}, \bibinfo {author} {\bibfnamefont {S.-C.}\ \bibnamefont {Hsu}},
  \bibinfo {author} {\bibfnamefont {B.}~\bibnamefont {Nachman}}, \bibinfo
  {author} {\bibfnamefont {N.}~\bibnamefont {Suaysom}}, \ and\ \bibinfo
  {author} {\bibfnamefont {A.}~\bibnamefont {Suresh}},\ }\href {\doibase
  10.1103/PhysRevD.103.036001} {\bibfield  {journal} {\bibinfo  {journal}
  {Phys. Rev. D}\ }\textbf {\bibinfo {volume} {103}},\ \bibinfo {pages}
  {036001} (\bibinfo {year} {2021})},\ \Eprint
  {http://arxiv.org/abs/2010.03569} {arXiv:2010.03569 [hep-ph]} \BibitemShut
  {NoStop}%
\bibitem [{\citenamefont {Kuttan}\ \emph {et~al.}(2021)\citenamefont {Kuttan},
  \citenamefont {Zhou}, \citenamefont {Steinheimer}, \citenamefont
  {Redelbach},\ and\ \citenamefont {Stoecker}}]{Kuttan:2021npg}%
  \BibitemOpen
  \bibfield  {author} {\bibinfo {author} {\bibfnamefont {M.~O.}\ \bibnamefont
  {Kuttan}}, \bibinfo {author} {\bibfnamefont {K.}~\bibnamefont {Zhou}},
  \bibinfo {author} {\bibfnamefont {J.}~\bibnamefont {Steinheimer}}, \bibinfo
  {author} {\bibfnamefont {A.}~\bibnamefont {Redelbach}}, \ and\ \bibinfo
  {author} {\bibfnamefont {H.}~\bibnamefont {Stoecker}},\ }\href@noop {} {\
  (\bibinfo {year} {2021})},\ \Eprint {http://arxiv.org/abs/2107.05590}
  {arXiv:2107.05590 [hep-ph]} \BibitemShut {NoStop}%
\bibitem [{\citenamefont {Huang}\ \emph {et~al.}(2021)\citenamefont {Huang},
  \citenamefont {Xiao}, \citenamefont {Liu}, \citenamefont {Wu}, \citenamefont
  {Mu},\ and\ \citenamefont {Song}}]{Huang:2018fzn}%
  \BibitemOpen
  \bibfield  {author} {\bibinfo {author} {\bibfnamefont {H.}~\bibnamefont
  {Huang}}, \bibinfo {author} {\bibfnamefont {B.}~\bibnamefont {Xiao}},
  \bibinfo {author} {\bibfnamefont {Z.}~\bibnamefont {Liu}}, \bibinfo {author}
  {\bibfnamefont {Z.}~\bibnamefont {Wu}}, \bibinfo {author} {\bibfnamefont
  {Y.}~\bibnamefont {Mu}}, \ and\ \bibinfo {author} {\bibfnamefont
  {H.}~\bibnamefont {Song}},\ }\href {\doibase
  10.1103/PhysRevResearch.3.023256} {\bibfield  {journal} {\bibinfo  {journal}
  {Phys. Rev. Res.}\ }\textbf {\bibinfo {volume} {3}},\ \bibinfo {pages}
  {023256} (\bibinfo {year} {2021})},\ \Eprint
  {http://arxiv.org/abs/1801.03334} {arXiv:1801.03334 [nucl-th]} \BibitemShut
  {NoStop}%
\bibitem [{\citenamefont {Kingma}\ and\ \citenamefont
  {Ba}(2015)}]{2014arXiv1412.6980K}%
  \BibitemOpen
  \bibfield  {author} {\bibinfo {author} {\bibfnamefont {D.~P.}\ \bibnamefont
  {Kingma}}\ and\ \bibinfo {author} {\bibfnamefont {J.}~\bibnamefont {Ba}},\
  }in\ \href {http://arxiv.org/abs/1412.6980} {\emph {\bibinfo {booktitle} {3rd
  International Conference on Learning Representations, {ICLR} 2015, San Diego,
  CA, USA, May 7-9, 2015, Conference Track Proceedings}}},\ \bibinfo {editor}
  {edited by\ \bibinfo {editor} {\bibfnamefont {Y.}~\bibnamefont {Bengio}}\
  and\ \bibinfo {editor} {\bibfnamefont {Y.}~\bibnamefont {LeCun}}}\ (\bibinfo
  {year} {2015})\BibitemShut {NoStop}%
\bibitem [{\citenamefont {{Xie}}\ \emph {et~al.}(2020)\citenamefont {{Xie}},
  \citenamefont {{Liu}},\ and\ \citenamefont {{Wang}}}]{2020PhRvB.101x5139X}%
  \BibitemOpen
  \bibfield  {author} {\bibinfo {author} {\bibfnamefont {H.}~\bibnamefont
  {{Xie}}}, \bibinfo {author} {\bibfnamefont {J.-G.}\ \bibnamefont {{Liu}}}, \
  and\ \bibinfo {author} {\bibfnamefont {L.}~\bibnamefont {{Wang}}},\ }\href
  {\doibase 10.1103/PhysRevB.101.245139} {\bibfield  {journal} {\bibinfo
  {journal} {\prb}\ }\textbf {\bibinfo {volume} {101}},\ \bibinfo {eid}
  {245139} (\bibinfo {year} {2020})},\ \Eprint
  {http://arxiv.org/abs/2001.04121} {arXiv:2001.04121 [cond-mat.str-el]}
  \BibitemShut {NoStop}%
\bibitem [{\citenamefont {Rubinstein}\ and\ \citenamefont
  {Kroese}(2016)}]{rubinstein2016simulation}%
  \BibitemOpen
  \bibfield  {author} {\bibinfo {author} {\bibfnamefont {R.~Y.}\ \bibnamefont
  {Rubinstein}}\ and\ \bibinfo {author} {\bibfnamefont {D.~P.}\ \bibnamefont
  {Kroese}},\ }\href@noop {} {\emph {\bibinfo {title} {Simulation and the Monte
  Carlo method}}},\ Vol.~\bibinfo {volume} {10}\ (\bibinfo  {publisher} {John
  Wiley \& Sons},\ \bibinfo {year} {2016})\BibitemShut {NoStop}%
\bibitem [{\citenamefont {Graves}(2011)}]{graves2011practical}%
  \BibitemOpen
  \bibfield  {author} {\bibinfo {author} {\bibfnamefont {A.}~\bibnamefont
  {Graves}},\ }in\ \href@noop {} {\emph {\bibinfo {booktitle} {Advances in
  neural information processing systems}}}\ (\bibinfo {organization}
  {Citeseer},\ \bibinfo {year} {2011})\ pp.\ \bibinfo {pages}
  {2348--2356}\BibitemShut {NoStop}%
\bibitem [{\citenamefont {Blundell}\ \emph {et~al.}(2015)\citenamefont
  {Blundell}, \citenamefont {Cornebise}, \citenamefont {Kavukcuoglu},\ and\
  \citenamefont {Wierstra}}]{blundell2015weight}%
  \BibitemOpen
  \bibfield  {author} {\bibinfo {author} {\bibfnamefont {C.}~\bibnamefont
  {Blundell}}, \bibinfo {author} {\bibfnamefont {J.}~\bibnamefont {Cornebise}},
  \bibinfo {author} {\bibfnamefont {K.}~\bibnamefont {Kavukcuoglu}}, \ and\
  \bibinfo {author} {\bibfnamefont {D.}~\bibnamefont {Wierstra}},\ }in\
  \href@noop {} {\emph {\bibinfo {booktitle} {International Conference on
  Machine Learning}}}\ (\bibinfo {organization} {PMLR},\ \bibinfo {year}
  {2015})\ pp.\ \bibinfo {pages} {1613--1622}\BibitemShut {NoStop}%
\bibitem [{\citenamefont {Burnier}\ \emph {et~al.}(2010)\citenamefont
  {Burnier}, \citenamefont {Laine},\ and\ \citenamefont
  {Vepsalainen}}]{Burnier:2009bk}%
  \BibitemOpen
  \bibfield  {author} {\bibinfo {author} {\bibfnamefont {Y.}~\bibnamefont
  {Burnier}}, \bibinfo {author} {\bibfnamefont {M.}~\bibnamefont {Laine}}, \
  and\ \bibinfo {author} {\bibfnamefont {M.}~\bibnamefont {Vepsalainen}},\
  }\href {\doibase 10.1007/JHEP01(2010)054} {\bibfield  {journal} {\bibinfo
  {journal} {JHEP}\ }\textbf {\bibinfo {volume} {01}},\ \bibinfo {pages} {054}
  (\bibinfo {year} {2010})},\ \bibinfo {note} {[Erratum: JHEP 01, 180
  (2013)]},\ \Eprint {http://arxiv.org/abs/0911.3480} {arXiv:0911.3480
  [hep-ph]} \BibitemShut {NoStop}%
\bibitem [{\citenamefont {Lafferty}\ and\ \citenamefont
  {Rothkopf}(2020)}]{Lafferty:2019jpr}%
  \BibitemOpen
  \bibfield  {author} {\bibinfo {author} {\bibfnamefont {D.}~\bibnamefont
  {Lafferty}}\ and\ \bibinfo {author} {\bibfnamefont {A.}~\bibnamefont
  {Rothkopf}},\ }\href {\doibase 10.1103/PhysRevD.101.056010} {\bibfield
  {journal} {\bibinfo  {journal} {Phys. Rev. D}\ }\textbf {\bibinfo {volume}
  {101}},\ \bibinfo {pages} {056010} (\bibinfo {year} {2020})},\ \Eprint
  {http://arxiv.org/abs/1906.00035} {arXiv:1906.00035 [hep-ph]} \BibitemShut
  {NoStop}%
\bibitem [{\citenamefont {Fukushima}\ and\ \citenamefont
  {Su}(2013)}]{Fukushima:2013xsa}%
  \BibitemOpen
  \bibfield  {author} {\bibinfo {author} {\bibfnamefont {K.}~\bibnamefont
  {Fukushima}}\ and\ \bibinfo {author} {\bibfnamefont {N.}~\bibnamefont {Su}},\
  }\href {\doibase 10.1103/PhysRevD.88.076008} {\bibfield  {journal} {\bibinfo
  {journal} {Phys. Rev. D}\ }\textbf {\bibinfo {volume} {88}},\ \bibinfo
  {pages} {076008} (\bibinfo {year} {2013})},\ \Eprint
  {http://arxiv.org/abs/1304.8004} {arXiv:1304.8004 [hep-ph]} \BibitemShut
  {NoStop}%
\bibitem [{\citenamefont {Bala}\ \emph {et~al.}(2021)\citenamefont {Bala},
  \citenamefont {Kaczmarek}, \citenamefont {Larsen}, \citenamefont {Mukherjee},
  \citenamefont {Parkar}, \citenamefont {Petreczky}, \citenamefont {Rothkopf},\
  and\ \citenamefont {Weber}}]{Bala:2021fkm}%
  \BibitemOpen
  \bibfield  {author} {\bibinfo {author} {\bibfnamefont {D.}~\bibnamefont
  {Bala}}, \bibinfo {author} {\bibfnamefont {O.}~\bibnamefont {Kaczmarek}},
  \bibinfo {author} {\bibfnamefont {R.}~\bibnamefont {Larsen}}, \bibinfo
  {author} {\bibfnamefont {S.}~\bibnamefont {Mukherjee}}, \bibinfo {author}
  {\bibfnamefont {G.}~\bibnamefont {Parkar}}, \bibinfo {author} {\bibfnamefont
  {P.}~\bibnamefont {Petreczky}}, \bibinfo {author} {\bibfnamefont
  {A.}~\bibnamefont {Rothkopf}}, \ and\ \bibinfo {author} {\bibfnamefont
  {J.~H.}\ \bibnamefont {Weber}},\ }\href@noop {} {\  (\bibinfo {year}
  {2021})},\ \Eprint {http://arxiv.org/abs/2110.11659} {arXiv:2110.11659
  [hep-lat]} \BibitemShut {NoStop}%
\bibitem [{\citenamefont {Liu}\ and\ \citenamefont {Rapp}(2020)}]{Liu:2016ysz}%
  \BibitemOpen
  \bibfield  {author} {\bibinfo {author} {\bibfnamefont {S.~Y.~F.}\
  \bibnamefont {Liu}}\ and\ \bibinfo {author} {\bibfnamefont {R.}~\bibnamefont
  {Rapp}},\ }\href {\doibase 10.1140/epja/s10050-020-00024-z} {\bibfield
  {journal} {\bibinfo  {journal} {Eur. Phys. J. A}\ }\textbf {\bibinfo {volume}
  {56}},\ \bibinfo {pages} {44} (\bibinfo {year} {2020})},\ \Eprint
  {http://arxiv.org/abs/1612.09138} {arXiv:1612.09138 [nucl-th]} \BibitemShut
  {NoStop}%
\bibitem [{\citenamefont {Liu}\ and\ \citenamefont {Rapp}(2018)}]{Liu:2017qah}%
  \BibitemOpen
  \bibfield  {author} {\bibinfo {author} {\bibfnamefont {S.~Y.~F.}\
  \bibnamefont {Liu}}\ and\ \bibinfo {author} {\bibfnamefont {R.}~\bibnamefont
  {Rapp}},\ }\href {\doibase 10.1103/PhysRevC.97.034918} {\bibfield  {journal}
  {\bibinfo  {journal} {Phys. Rev. C}\ }\textbf {\bibinfo {volume} {97}},\
  \bibinfo {pages} {034918} (\bibinfo {year} {2018})},\ \Eprint
  {http://arxiv.org/abs/1711.03282} {arXiv:1711.03282 [nucl-th]} \BibitemShut
  {NoStop}%
\bibitem [{\citenamefont {Du}\ \emph {et~al.}(2017)\citenamefont {Du},
  \citenamefont {Rapp},\ and\ \citenamefont {He}}]{Du:2017qkv}%
  \BibitemOpen
  \bibfield  {author} {\bibinfo {author} {\bibfnamefont {X.}~\bibnamefont
  {Du}}, \bibinfo {author} {\bibfnamefont {R.}~\bibnamefont {Rapp}}, \ and\
  \bibinfo {author} {\bibfnamefont {M.}~\bibnamefont {He}},\ }\href {\doibase
  10.1103/PhysRevC.96.054901} {\bibfield  {journal} {\bibinfo  {journal} {Phys.
  Rev. C}\ }\textbf {\bibinfo {volume} {96}},\ \bibinfo {pages} {054901}
  (\bibinfo {year} {2017})},\ \Eprint {http://arxiv.org/abs/1706.08670}
  {arXiv:1706.08670 [hep-ph]} \BibitemShut {NoStop}%
\bibitem [{\citenamefont {Du}\ \emph {et~al.}(2019)\citenamefont {Du},
  \citenamefont {Liu},\ and\ \citenamefont {Rapp}}]{Du:2019tjf}%
  \BibitemOpen
  \bibfield  {author} {\bibinfo {author} {\bibfnamefont {X.}~\bibnamefont
  {Du}}, \bibinfo {author} {\bibfnamefont {S.~Y.~F.}\ \bibnamefont {Liu}}, \
  and\ \bibinfo {author} {\bibfnamefont {R.}~\bibnamefont {Rapp}},\ }\href
  {\doibase 10.1016/j.physletb.2019.07.032} {\bibfield  {journal} {\bibinfo
  {journal} {Phys. Lett. B}\ }\textbf {\bibinfo {volume} {796}},\ \bibinfo
  {pages} {20} (\bibinfo {year} {2019})},\ \Eprint
  {http://arxiv.org/abs/1904.00113} {arXiv:1904.00113 [nucl-th]} \BibitemShut
  {NoStop}%
\bibitem [{\citenamefont {Strickland}(2011)}]{Strickland:2011mw}%
  \BibitemOpen
  \bibfield  {author} {\bibinfo {author} {\bibfnamefont {M.}~\bibnamefont
  {Strickland}},\ }\href {\doibase 10.1103/PhysRevLett.107.132301} {\bibfield
  {journal} {\bibinfo  {journal} {Phys. Rev. Lett.}\ }\textbf {\bibinfo
  {volume} {107}},\ \bibinfo {pages} {132301} (\bibinfo {year} {2011})},\
  \Eprint {http://arxiv.org/abs/1106.2571} {arXiv:1106.2571 [hep-ph]}
  \BibitemShut {NoStop}%
\bibitem [{\citenamefont {Islam}\ and\ \citenamefont
  {Strickland}(2020{\natexlab{b}})}]{Islam:2020bnp}%
  \BibitemOpen
  \bibfield  {author} {\bibinfo {author} {\bibfnamefont {A.}~\bibnamefont
  {Islam}}\ and\ \bibinfo {author} {\bibfnamefont {M.}~\bibnamefont
  {Strickland}},\ }\href {\doibase 10.1007/JHEP03(2021)235} {\bibfield
  {journal} {\bibinfo  {journal} {JHEP}\ }\textbf {\bibinfo {volume} {21}},\
  \bibinfo {pages} {235} (\bibinfo {year} {2020}{\natexlab{b}})},\ \Eprint
  {http://arxiv.org/abs/2010.05457} {arXiv:2010.05457 [hep-ph]} \BibitemShut
  {NoStop}%
\bibitem [{\citenamefont {Mocsy}\ and\ \citenamefont
  {Petreczky}(2007)}]{Mocsy:2007jz}%
  \BibitemOpen
  \bibfield  {author} {\bibinfo {author} {\bibfnamefont {A.}~\bibnamefont
  {Mocsy}}\ and\ \bibinfo {author} {\bibfnamefont {P.}~\bibnamefont
  {Petreczky}},\ }\href {\doibase 10.1103/PhysRevLett.99.211602} {\bibfield
  {journal} {\bibinfo  {journal} {Phys. Rev. Lett.}\ }\textbf {\bibinfo
  {volume} {99}},\ \bibinfo {pages} {211602} (\bibinfo {year} {2007})},\
  \Eprint {http://arxiv.org/abs/0706.2183} {arXiv:0706.2183 [hep-ph]}
  \BibitemShut {NoStop}%
\bibitem [{\citenamefont {Burnier}\ \emph {et~al.}(2008)\citenamefont
  {Burnier}, \citenamefont {Laine},\ and\ \citenamefont
  {Vepsalainen}}]{Burnier:2007qm}%
  \BibitemOpen
  \bibfield  {author} {\bibinfo {author} {\bibfnamefont {Y.}~\bibnamefont
  {Burnier}}, \bibinfo {author} {\bibfnamefont {M.}~\bibnamefont {Laine}}, \
  and\ \bibinfo {author} {\bibfnamefont {M.}~\bibnamefont {Vepsalainen}},\
  }\href {\doibase 10.1088/1126-6708/2008/01/043} {\bibfield  {journal}
  {\bibinfo  {journal} {JHEP}\ }\textbf {\bibinfo {volume} {01}},\ \bibinfo
  {pages} {043} (\bibinfo {year} {2008})},\ \Eprint
  {http://arxiv.org/abs/0711.1743} {arXiv:0711.1743 [hep-ph]} \BibitemShut
  {NoStop}%
\bibitem [{\citenamefont {{Crater}}(1994)}]{1994JCoPh.115..470C}%
  \BibitemOpen
  \bibfield  {author} {\bibinfo {author} {\bibfnamefont {H.~W.}\ \bibnamefont
  {{Crater}}},\ }\href {\doibase 10.1006/jcph.1994.1211} {\bibfield  {journal}
  {\bibinfo  {journal} {Journal of Computational Physics}\ }\textbf {\bibinfo
  {volume} {115}},\ \bibinfo {pages} {470} (\bibinfo {year}
  {1994})}\BibitemShut {NoStop}%
\end{thebibliography}%

\end{document}